\documentclass[twocolumn,aps,showpacs,showkeys,prd,superscriptaddress,byrevtex, amsmath,nofootinbib]{revtex4-1}

\usepackage{amssymb}
\usepackage{amsmath}
\usepackage{verbatim}
\usepackage{mathrsfs}
\usepackage{amsfonts}
\usepackage{latexsym}
\usepackage{epsfig}
\usepackage{color}
\usepackage{graphicx,subfigure}
\usepackage{units}
\usepackage{aas_macros}
\usepackage{ulem}

\begin{document}


\renewcommand{\t}{\times}
\newcommand{\sg}[1]{\textcolor{blue}{SG: #1}}
\newcommand{\tf}[1]{\textcolor{red}{TF: #1}}
\newcommand{\ch}[1]{\textcolor{cyan}{#1}}

\long\def\symbolfootnote[#1]#2{\begingroup%
\def\thefootnote{\fnsymbol{footnote}}\footnote[#1]{#2}\endgroup}


\title{Magnetized accretion disks around Kerr black holes with scalar hair - Nonconstant angular momentum disks} 

\author{Sergio Gimeno-Soler}
\affiliation{Departamento de
  Astronom\'{\i}a y Astrof\'{\i}sica, Universitat de Val\`encia,
  C/ Dr. Moliner 50, 46100, Burjassot (Val\`encia), Spain}

\author{Jos\'e A. Font}
\affiliation{Departamento de
  Astronom\'{\i}a y Astrof\'{\i}sica, Universitat de Val\`encia,
  C/ Dr. Moliner 50, 46100, Burjassot (Val\`encia), Spain}
\affiliation{Observatori Astron\`omic, Universitat de Val\`encia, C/ Catedr\'atico 
  Jos\'e Beltr\'an 2, 46980, Paterna (Val\`encia), Spain}

    \author{Carlos Herdeiro}
\affiliation{Departamento  de  Matem\'{a}tica  da  Universidade  de  Aveiro  and  Centre  for  Research  and  Development in  Mathematics  and  Applications  (CIDMA),  Campus  de  Santiago,  3810-183  Aveiro,  Portugal} 
  
  \author{Eugen Radu}
\affiliation{Departamento  de  Matem\'{a}tica  da  Universidade  de  Aveiro  and  Centre  for  Research  and  Development in  Mathematics  and  Applications  (CIDMA),  Campus  de  Santiago,  3810-183  Aveiro,  Portugal} 


\date{\today}


\begin{abstract}
We present new equilibrium solutions of stationary models of magnetized thick disks (or tori) around Kerr black holes with synchronised scalar hair. The models reported here largely extend our previous results based on constant radial distributions of the specific angular momentum along the equatorial plane. We introduce a new way to prescribe the distribution of the disk's angular momentum based on a combination of two previous proposals and compute the angular momentum distribution outside the equatorial plane by resorting to the construction of von Zeipel cylinders. We find that the effect of the scalar hair on the black hole spacetime can yield significant differences in the disk morphology and properties compared to what is found if the spacetime is purely Kerr. Some of the tori built within the most extreme, background hairy black hole spacetime of our sample exhibit the appearance of two  maxima in the gravitational energy density which impacts the radial profile distributions of the disk's thermodynamical quantities. The models reported in this paper can be used as initial data for numerical evolutions with GRMHD codes to study their stability properties. Moreover, they can be employed as illuminating sources to build shadows of Kerr black holes with scalar hair which might help  further constrain the no-hair hypothesis as new observational data is collected.
\end{abstract}




\maketitle


\section{Introduction}

The Event Horizon Telescope (EHT) Collaboration has recently resolved the shadow of the supermassive dark compact object at the center of the giant elliptical galaxy M87~\cite{EHT1}. The image shows a remarkable similarity with the shadow a Kerr black hole from general relativity would produce. The observational capabilities offered by the EHT, thus, allow to measure strong-field lensing patterns from accretion disks which can be used to test the validity of the black hole hypothesis. Further evidences in support of such hypothesis are provided by the Advanced LIGO and Advanced Virgo observations of gravitational waves from compact binary coalescences~\cite{gwtc-1,gwtc-2} and by the study of orbital motions of stars near SgrA* at the center of the Milky Way~\cite{Ghez:2008,genzel:2010,gravity}. 

While the black hole hypothesis is thus far supported by current data, the available experimental efforts also place within observational reach the exploration of additional proposals that are collectively known as {\it exotic} compact objects (ECOs, see~\cite{Cardoso:2019rvt} and references therein). Indeed, recent examples have shown the intrinsic degeneracy between the prevailing Kerr black hole solutions of general relativity and bosonic star solutions, a class of horizonless, {dynamically robust} ECOs, using both actual gravitational-wave data~\cite{juan1} and electromagnetic data~\cite{imitation} (see also~\cite{olivares}). Likewise, testing the very nature of gravity in the strong-field regime is becoming increasingly possible using  gravitational-wave observations~\cite{testingGR1,testingGR2}. Moreover, proofs of concept of the feasibility of testing general relativity{, or even the existence of new particles} via EHT observations have been reported in{, $e.g.$~\cite{Mizuno18,Cunha:2019ikd,Cunha:2019dwb,Cruz-Osorio,Volkel:2020xlc,Psaltis:2020lvx,Kocherlakota:2021dcv}.} 

Those observational advances highly motivate the development of theoretical models to explain the available data. In particular, and in connection with the EHT observations, the establishing of sound theoretical descriptions of dark compact objects surrounded by accretion disks is much required. Disks act as illuminating sources leading through gravity to potentially observable strong-field lensing patterns and shadows. Indeed, a few proposals have recently discussed the observational appearance of the shadows of black holes and boson stars by analyzing the lensing patterns produced by a light source - an accretion disk - with identical morphology~\cite{Cunha:2015yba,Cunha:2016bjh,Vincent:2016,olivares}.
While boson star spacetimes lack an innermost stable circular orbit for timelike geodesics (which would prevent the occurence of the shadow as the disk can only terminate in the centre of the dark star) the general relativistic MHD simulations of~\cite{olivares} have shown the existence of an {\it effective} shadow at a given areal radius at which the angular velocity of the orbits attains a maximum. The intrinsic unstable nature of the spherical boson star model employed in~\cite{olivares} has been discussed in~\cite{imitation} who found that a  degenerate (effective) shadow comparable to that of a Schwarzschild black hole can exist for spherical {\it vector} (a.k.a.~Proca) boson stars. 


Despite the significance of the accretion disk model for the computation of lensing patterns as realistic as possible, existing studies are based on rigidly-rotating (geometrically thick) disks, assuming as an initial condition for the dynamical evolutions a constant radial profile of the specific angular momentum of the plasma.
In this paper we present stationary solutions of magnetized thick disks (or tori) whose angular momentum distribution deviates from a simplistic constant angular momentum law. We introduce a new way to prescribe the distribution of the disk's angular momentum based on a combination of two previous proposals~\cite{Daigne:2004,Qian:2009,Gimeno-Soler:2017} and compute the angular momentum distribution employing the so-called von Zeipel cylinders, i.e. the surfaces of constant specific angular momentum and constant angular velocity, which coincide for a barotropic equation of state. A major simplification of our approach is that the self-gravity of the disk is neglected and the models are built within the background spacetime provided by a particular class of ECO, namely the spacetime of a Kerr black hole with synchronised scalar hair. (We note in passing that building such disks around bosonic stars, extending the models of~\cite{Vincent:2016,olivares} would be straightforward in our approach.)
Kerr black holes with synchronised scalar hair (KBHsSH) result from minimally coupling Einstein’s gravity to bosonic matter fields~\cite{Herdeiro:2014a,Herdeiro:2015gia} and provide a sound counterexample to the no-hair conjecture~\cite{Herdeiro:2015_review}\footnote{{The solutions studied here are the fundamental states of the minimal Einstein-Klein-Gordon model without self-interactions. Different generalizations can be obtained, including charged~\cite{Delgado:2016jxq} and excited states in the same model~\cite{Wang:2018xhw,Delgado:2019prc}, as well as cousin solutions in different scalar~\cite{Herdeiro:2015tia,Herdeiro:2018daq,Herdeiro:2018djx,Brihaye:2018grv,Kunz:2019bhm,Collodel:2020gyp,Delgado:2020hwr}   or Proca  models~\cite{Herdeiro:2016,Santos:2020pmh}.}}. Such hairy black holes have been shown to form dynamically (in the vector case) as the end-product of the superradiant instability~\cite{East:2017} (but see also~\cite{sanchis-gual:2020} for an alternative formation channel through the post-merger dynamics of bosonic star binaries) and to be effectively stable  against superradiance in regions of the parameter space~\cite{Degollado:2018}. 
As we show below, the effect of the scalar hair on the black hole spacetime can introduce significant differences in the properties and morphology of the disks compared to what is found in a purely Kerr spacetime. The models discussed in this paper can be used as initial data for general-relativistic MHD codes and employed as illuminating sources to compute shadows of KBHsSH that might be confronted with prospective new observational data.

The organization of this paper is as follows: Section~\ref{framework} briefly describes the spacetime properties of KBHsSH, the combined approaches we employ to prescribe the angular momentum distribution in the disk, along with the way the magnetic field is incorporated in the models. The numerical procedure to build the tori and our choice of parameter space is discussed in Section~\ref{procedure}. The equilibrium models are presented and analyzed in Section~\ref{results}. This section also contains the discussion of the morphological features of the disks along with potential astrophysical implications of our models. Finally, our conclusions are summarized in Section~\ref{conclusions}. Geometrized units ($G=c=1$) are used throughout the paper. 

\section{Framework}
\label{framework}

\subsection{Spacetime metric and KBHsSH models}

As in~\cite{paper1} (hereafter Paper I) we use the KBHsSH models built using the procedure described in~\cite{Herdeiro:2015b} where the interested reader is addressed for further details. In the following we briefly review their basic properties.

KBHsSH are asymptotically flat, stationary and axisymmetric solutions of the Einstein-(complex)Klein-Gordon (EKG) field equations 
\begin{eqnarray}
R_{ab} - \frac{1}{2}R g _{ab} = 8 \pi (T_{\mathrm{SF}})_{ab}\,,
\end{eqnarray}
describing a massive, complex scalar field $\Psi$ minimally coupled to Einstein gravity. The metric and the scalar field can be written using the ansatz (see~\cite{Herdeiro:2014a})
\begin{eqnarray}
\mathrm{d}s^2 &=& e^{2F_1}\left(\frac{\mathrm{d}r^2}{N} + r^2\mathrm{d}\theta^2\right) +  e^{2F_2}r^2\sin^2 \theta(\mathrm{d}\phi-W\mathrm{d}t)^2 
\nonumber \\ 
&-&  e^{2F_0}N\mathrm{d}t^2\,,
\label{metric}
\end{eqnarray}
\begin{eqnarray}
\Psi = \varphi(r, \theta) e^{\mathrm{i}(m\phi - \omega t)} \,,
\end{eqnarray}
where $W$, $F_1$, $F_2$, $F_0$ are functions of $r$ and $\theta$, $\omega$ is the scalar field frequency, and $m$ is the azimuthal harmonic index. The latter two are related through $\omega/m = \Omega_{\mathrm{H}}$, where $\Omega_{\mathrm{H}}$ is the angular velocity of the event horizon. Moreover $N = 1 - r_{\rm H}/r$, where $r_{\rm H}$ is the radius of the event horizon of the black hole. The energy-momentum tensor acting as a source of the EKG equations can be written as
\begin{eqnarray}\label{eq:e-m_scalar_field}
(T_{\mathrm{SF}})_{ab} &=& \partial_a \Psi^* \partial_b \Psi + \partial_b \Psi^* \partial_a \Psi 
 \\ 
&-& g_{ab} \left(\frac{1}{2} g^{cd}(\partial_c \Psi^* \partial_d \Psi + \partial_d \Psi^* \partial_c \Psi) + \mu^2 \Psi^* \Psi \right), \nonumber
\end{eqnarray}
where $\mu$ is the mass of the scalar field and superscript $(^*)$ denotes complex conjugation.

\begin{figure}
\centering
\includegraphics[scale=0.21]{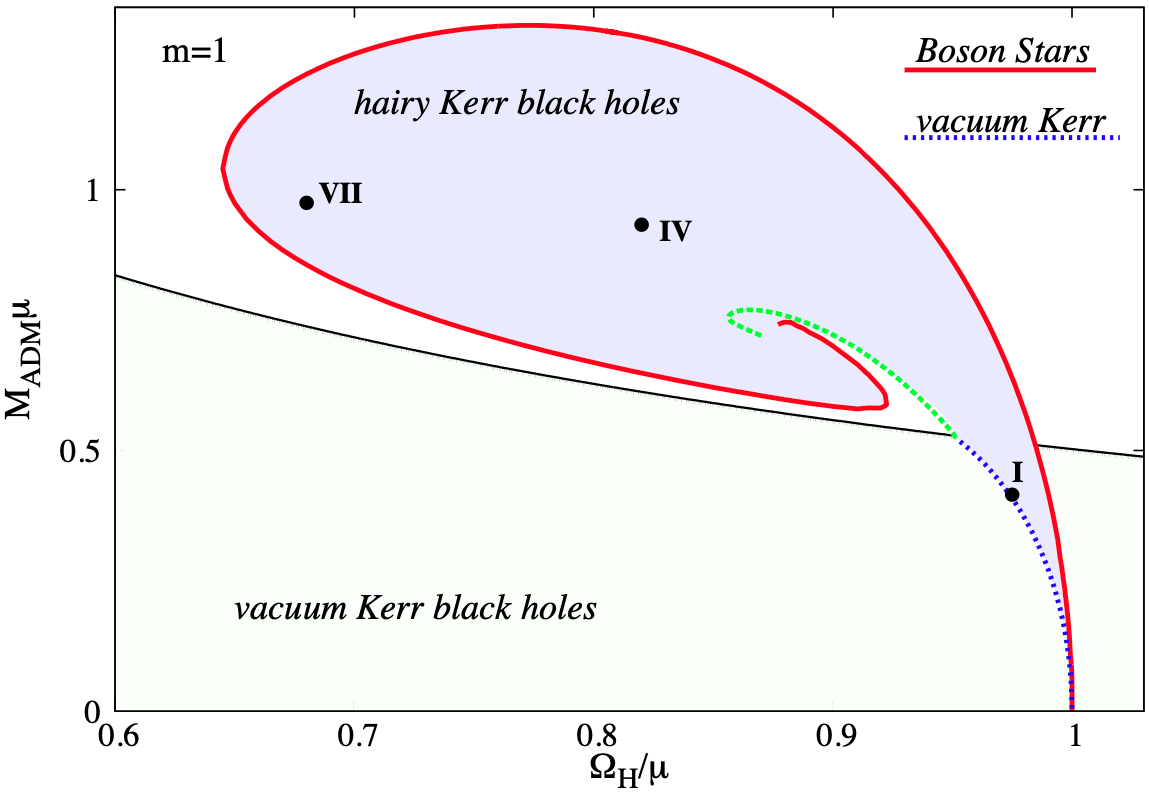}
\caption{Domain of existence for KBHsSH (shaded blue area) in an ADM mass versus scalar
field frequency diagram. The three solutions to be studied herein are highlighted in this diagram.}
\label{existence}
\end{figure}

Table~\ref{models_list} reports the properties of the three KBHsSH models we use in this work. The corresponding models are plotted in Fig.~\ref{existence} within the domain of existence of KBHsSH in an ADM mass versus scalar field frequency diagram. As we consider a subset of the models we used in Paper I we keep the same labels so that the comparison with the previous results for constant angular momentum disks is easier to do. In particular, model I corresponds to a Kerr-like model, with almost all the mass and angular momentum stored in the BH (namely, $94.7\%$ of the total mass and $87.2\%$ of the total angular momentum of the spacetime are stored in the BH), while model VII corresponds to a hairy Kerr BH with almost all the mass (98.15\%) and angular momentum (99.76\%) stored in the scalar field.  Moreover, it is worth noting that, even though KBHsSH can violate the Kerr bound in terms of the horizon quantities (i.e.~the normalized spin of the BH $a_{\mathrm{H}}$ can be greater than one), this fact does not have the same implications as in Kerr spacetime. In particular, the linear velocity of the horizon, $v_{\mathrm{H}}$, never exceeds the speed of light~\citep{Herdeiro:2015c}.

\begin{table*}[t]
\caption{List of models of KBHsSH used in this work. From left to right the columns report the name of the model, the ADM mass, $M_{\mathrm{ADM}}$, the ADM angular momentum, $J_{\mathrm{ADM}}$, the horizon mass, $M_{\mathrm{H}}$, the horizon angular momentum, $J_{\mathrm{H}}$, the mass of the scalar field, $M_{\mathrm{SF}}$, the angular momentum of the scalar field, $J_{\mathrm{SF}}$, the radius of the event horizon, $r_{\mathrm{H}}$, the values of the normalized spin parameter for the ADM quantities, $a_{\mathrm{ADM}}$, and for the BH horizon quantities, $a_{\mathrm{H}}$, the horizon linear velocity, $v_{\mathrm{H}}$, the spin parameter corresponding to a Kerr BH with a linear velocity equal to $v_{\mathrm{H}}$, $a_{\mathrm{H_{eq}}}$, and the sphericity of the horizon as defined in~\cite{Delgado:2018}, $\mathfrak{s}$. Here $\mu=1$.}        
\label{models_list}      
\centering          
\begin{tabular}{c c c c  c c c c   c c c c c}
\hline\hline       
 Model & $M_{\mathrm{ADM}}$ & $J_{\mathrm{ADM}}$ & $M_{\mathrm{H}}$ &  $J_{\mathrm{H}}$ & $M_{\mathrm{SF}}$ & $J_{\mathrm{SF}}$ & $r_{\mathrm{H}}$ & $a_{\mathrm{ADM}}$ & $a_{\mathrm{H}}$ & $v_{\mathrm{H}}$ & $a_{\mathrm{H_{eq}}}$ & $\mathfrak{s}$\\ 
\hline           
I & $0.415$ & $0.172$ & $0.393$ &  $0.150$  & $0.022$ & $0.022$ & $0.200$ & $0.9987$ & $0.971$ & $0.7685$ & $0.9663$ & $1.404$\\ 
IV & $0.933$ & $0.739$ & $0.234$ &  $0.114$  & $0.699$ & $0.625$ & $0.100$ & $0.8489$ & $2.082$ & $0.5635$ & $0.8554$ & $1.425$ \\ 
VII & $0.975$ & $0.850$ & $0.018$ &  $0.002$  & $0.957$ & $0.848$ & $0.040$ & $0.8941$ & $6.173$ & $0.0973$ & $0.1928$ & $1.039$ \\ 
\hline\hline
\end{tabular}
\end{table*}

\subsection{Angular momentum distributions in the disk}

As in Paper I equilibrium solutions of thick accretion disks are built assuming stationarity and axisymmetry in both the spacetime and in the matter fields (i.e.~$\partial_{t} f(r,\theta) = \partial_{\phi} f(r,\theta) = 0$ when $f(r,\theta)$ is a fluid quantity). We use the standard definitions of the specific angular momentum $l = -u_{\phi}/u_t$ and of the angular velocity $\Omega = u^{\phi}/u^{t}$, where we further assume circular motion, i.e.~the 4-velocity of the fluid is given by $u^{\mu} = (u^t, 0, 0, u^{\phi})$. It is straightforward to obtain the relationship between $l$ and $\Omega$,
\begin{equation}\label{eq:ang_mon_ang_vel}
l = - \frac{\Omega g_{\phi\phi} + g_{t\phi}}{\Omega g_{t\phi} + g_{tt}}, \;\;\; \Omega = - \frac{l g_{tt} + g_{t\phi}}{l g_{t\phi} + g_{\phi\phi}}.
\end{equation}

\begin{figure*}[t]
\includegraphics[scale=0.36]{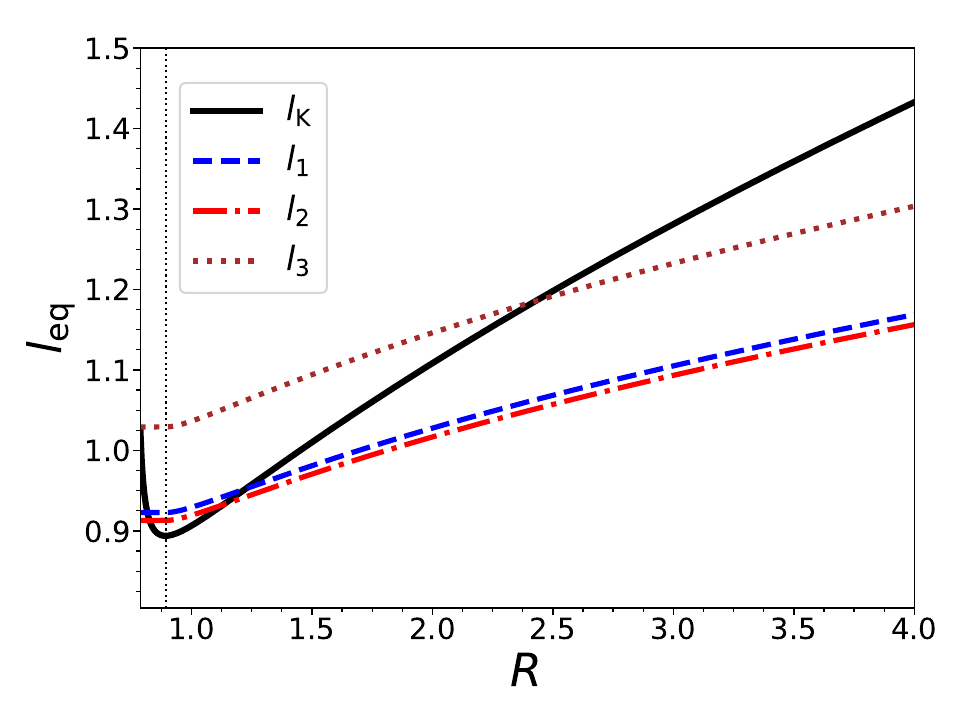}
\hspace{-0.3cm}
\includegraphics[scale=0.36]{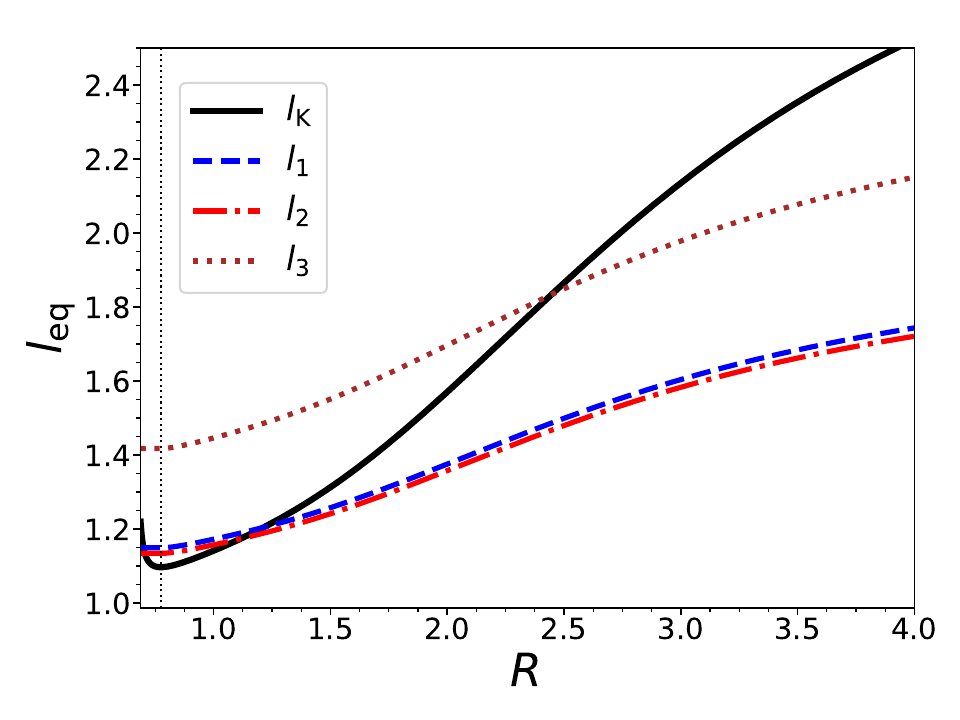}
\hspace{-0.2cm}
\includegraphics[scale=0.36]{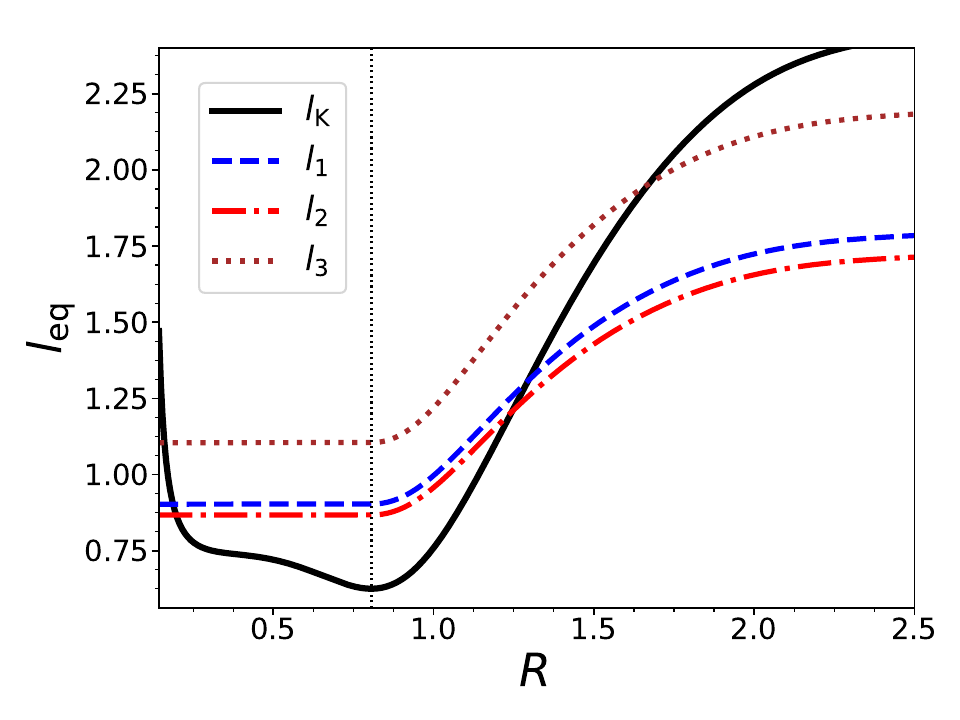}
\\
\vspace{0.0cm}
\includegraphics[scale=0.36]{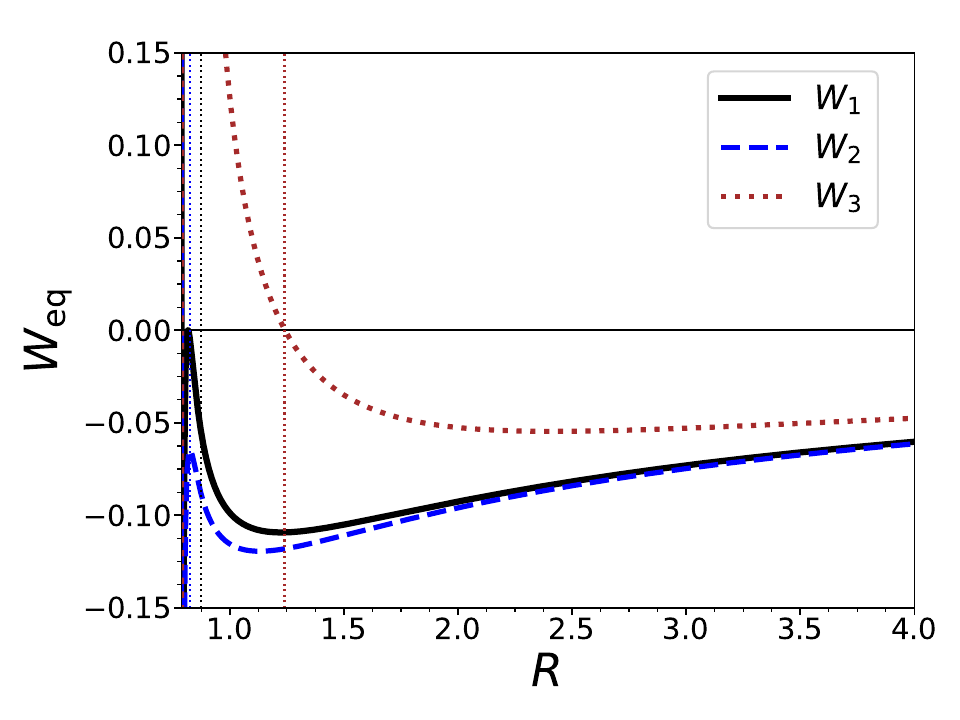}
\hspace{-0.3cm}
\includegraphics[scale=0.36]{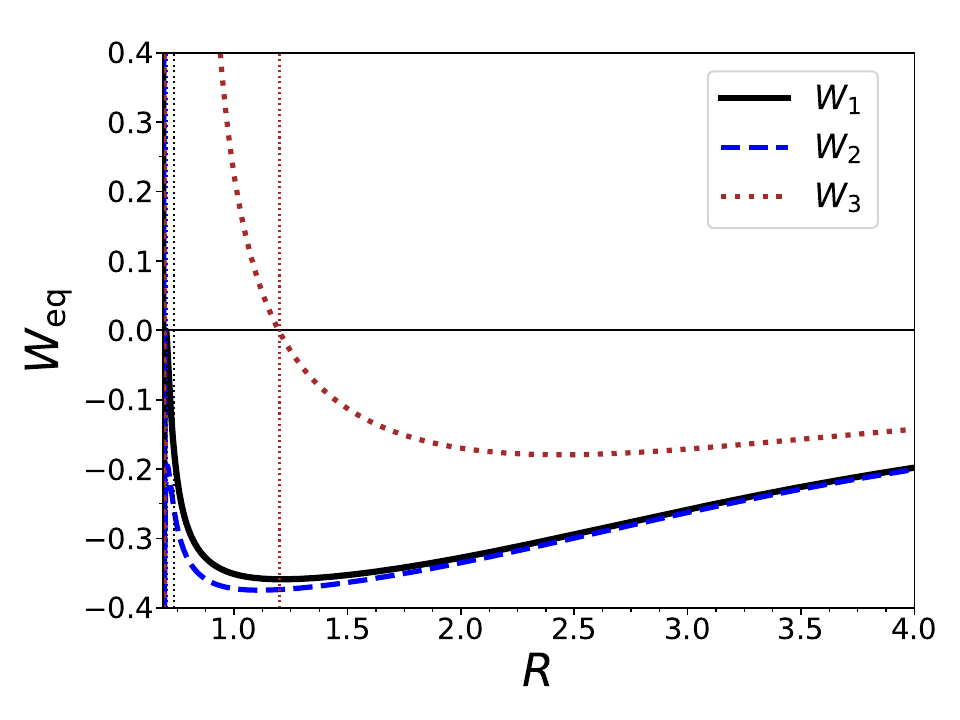}
\hspace{-0.2cm}
\includegraphics[scale=0.36]{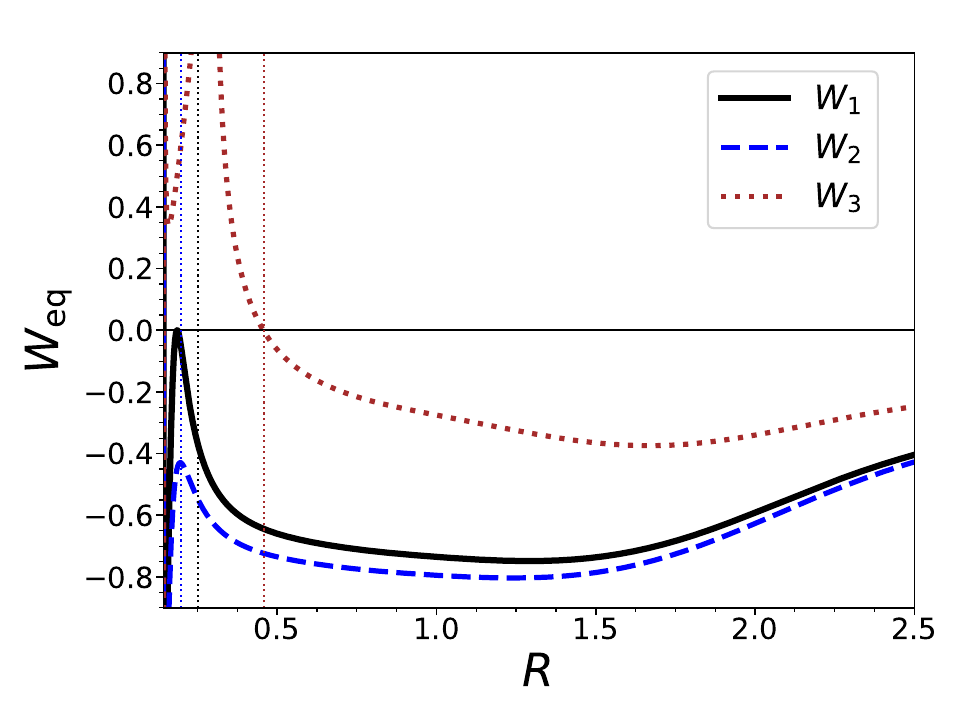}
\caption{Radial distributions of the angular momentum and of the potential at the equatorial plane for $\alpha = 0.5$. Each column corresponds to our three KBHsSH spacetimes (I, IV and VII from left to right). In the first row we show the distribution of angular momentum at the equatorial plane for the three different criteria discussed in Section III.A (namely, a blue dashed line for criterion 1, a red dash-dotted line for criterion 2 and a brown dotted line for criterion 3). The Keplerian angular momentum is also shown in a solid black line. The location of $r_{\mathrm{ms}}$ is displayed with a vertical dotted black line. In the second row we show the potential profiles corresponding to each angular momentum distribution displayed in the first row. They are shown by a solid black line, a blue dashed line and a brown dotted line, for each of the three criteria. The vertical dotted lines  indicate the location of $r_{\mathrm{in}}$ for each case. Note that all the panels use the perimeteral radius $R$.}
\label{radial_angmom_potential}
\end{figure*}

In this work we depart from Paper I by introducing a non-constant distribution of specific angular momentum in the disk. This new prescription is the result of combining two different approaches: one to formulate the angular momentum distribution in the equatorial plane, and another one to do so outside the equatorial plane. The reason for this split will be explained below.

\subsubsection{Angular momentum distribution in the equatorial plane}

To obtain the specific angular momentum distribution in the equatorial plane, $\theta=\pi/2$, we consider the following procedure
\begin{equation}
l \left(r,\frac{\pi}{2}\right) = \left\{ \label{eq:ansatz} 
  \begin{array}{lr}
    l_0 \left(\frac{l_{\mathrm{K}}(r)}{l_{\mathrm{K}}(r_{\mathrm{ms}})}\right)^{\alpha} &  \text{for } r \geq r_{\mathrm{ms}}\\
    l_0 & \text{for } r < r_{\mathrm{ms}}
  \end{array}
\right.
\end{equation}
where $l_0$ is a constant, $l_{\mathrm{K}}(r)$ is the Keplerian specific angular momentum, $r_{\mathrm{ms}}$ is the radius of the innermost stable circular orbit (ISCO) and the exponent $\alpha$ (where $0 \leq \alpha < 1$) is a parameter which controls how {\it Keplerian} the angular momentum profile on the equatorial plane is. The value $\alpha = 0$ would produce a constant profile and $\alpha = 1$ would produce a Keplerian profile. This prescription, extended outside the equatorial plane, was first introduced for so-called Polish-doughnuts in~\cite{Qian:2009}.
We also used this recipe in the context of magnetized accretion disks around Kerr black holes in~\cite{Gimeno-Soler:2017}. 

\begin{table*}[t]
\caption{Values of the relevant physical magnitudes of our results for model I. From left to right, the columns correspond to: the constant part of the specific angular momentum distribution $l_0$, the exponent of the angular momentum distribution $\alpha$, the depth of the potential well at the center $\Delta W_{\mathrm{c}}$, the values of the perimeteral radial coordinate at the inner edge of the disk $R_{\mathrm{in}}$, at the outer edge of the disk $R_{\mathrm{out}}$ and at the center of the disk $R_{\mathrm{c}}$, the value of the magnetization parameter at the center of the disk $\beta_{\mathrm{m,c}}$, the maximum values of the rest-mass density $\rho_{\mathrm{max}}$, specific enthalpy $h_{\mathrm{max}}$, fluid pressure $p_{\mathrm{max}}$ and magnetic pressure $p_{\mathrm{m,max}}$, the location of the maximum of the fluid pressure and the magnetic pressure $R_{\mathrm{max}}$, $R_{\mathrm{m, max}}$.}        
\label{results_I}
\centering          
\begin{tabular}{c c c c  c c c c   c c c c c}
\hline\hline       
 $l_0$ & $\alpha$ & $\Delta W_{\mathrm{c}}$ &  $R_{\mathrm{in}}$ & $R_{\mathrm{out}}$ & $R_{\mathrm{c}}$ & $\beta_{\mathrm{m,c}}$ & $\rho_{\mathrm{max}}$ & $h_{\mathrm{max}}$ & $p_{\mathrm{max}}$ & $p_{\mathrm{m,max}}$ & $R_{\mathrm{max}}$ & $R_{\mathrm{m, max}}$\\ 
\hline           
$0.934$ & $0$ & $-9.37 \times 10^{-2}$ &  $0.858$  & $3.56$ & $1.14$ & $10^{10}$ & $1.00$ & $1.10$ & $2.46 \times 10^{-2}$ & $2.54 \times 10^{-12}$ & $1.14$ & $1.20$ \\ 
 &  &  &  &  &  & $1$ & $1.09$ & $1.05$ & $1.37 \times 10^{-2}$ & $1.25 \times 10^{-2}$ & $1.06$ & $1.09$\\ 
 &  &  &  &  &  & $10^{-10}$ & $1.43$ & $1.00$ & $3.76 \times 10^{-12}$ & $3.09 \times 10^{-2}$ & $1.00$ & $1.03$ \\ 
$0.922$ & $0$ & $-9.37 \times 10^{-2}$ &  $0.818$  & $3.08$ & $1.08$ & $10^{10}$ & $1.00$ & $1.10$ & $2.46 \times 10^{-2}$ & $2.55 \times 10^{-12}$ & $1.08$ & $1.13$ \\ 
 &  &  &  &  &  & $1$ & $1.10$ & $1.05$ & $1.38 \times 10^{-2}$ & $1.26 \times 10^{-2}$ & $1.01$ & $1.04$ \\ 
 &  &  &  &  &  & $10^{-10}$ & $1.48$ & $1.00$ & $3.93 \times 10^{-12}$ & $3.17 \times 10^{-2}$ & $0.953$ & $0.975$ \\ 
 $1.14$ & $0$ & $-9.37 \times 10^{-2}$ &  $1.24$  & $\infty$ & $2.17$ & $10^{10}$ & $1.00$ & $1.10$ & $2.46 \times 10^{-2}$ & $2.56 \times 10^{-12}$ & $2.17$ & $2.40$ \\ 
 & & &  &  & & $1$ & $1.10$ & $1.05$ & $1.38 \times 10^{-2}$ & $1.26 \times 10^{-2}$ & $1.89$ & $2.01$ \\ 
 & & &  &  & & $10^{-10}$ & $1.44$ & $1.00$ & $3.81 \times 10^{-12}$ & $3.14 \times 10^{-2}$ & $1.72$ & $1.79$ \\
 $0.930$ & $0.25$ & $-7.57 \times 10^{-2}$ &  $0.864$  & $3.94$ & $1.17$ & $10^{10}$ & $1.00$ & $1.08$ & $1.97 \times 10^{-2}$ & $2.04 \times 10^{-12}$ & $1.17$ & $1.24$ \\ 
 &  &  &  &  &  & $1$ & $1.10$ & $1.04$ & $1.10 \times 10^{-2}$ & $1.01 \times 10^{-2}$ & $1.08$ & $1.12$ \\
 &  &  &  &  &  & $10^{-10}$ & $1.45$ & $1.00$ & $3.10 \times 10^{-12}$ & $2.53 \times 10^{-2}$ & $1.02$ & $1.05$ \\
 $0.918$ & $0.25$ & $-7.57 \times 10^{-2}$ &  $0.821$  & $3.32$ & $1.10$ & $10^{10}$ & $1.00$ & $1.08$ & $1.97 \times 10^{-2}$ & $2.04 \times 10^{-12}$ & $1.10$ & $1.16$ \\ 
 & & &  &  &  & $1$ & $1.11$ & $1.04$ & $1.11 \times 10^{-2}$ & $1.01 \times 10^{-2}$ & $1.02$ & $1.05$ \\ 
 & & &  &  &  & $10^{-10}$ & $1.50$ & $1.00$ & $3.25 \times 10^{-2}$ & $2.60 \times 10^{-2}$ & $0.963$ & $0.986$ \\ $1.09$ & $0.25$ & $-7.57 \times 10^{-2}$ &  $1.24$  & $\infty$ & $2.27$ & $10^{10}$ & $1.00$ & $1.08$ & $1.97 \times 10^{-2}$ & $2.06 \times 10^{-12}$ & $2.27$ & $2.56$ \\
 & & & & & & $1$ & $1.11$ & $1.04$ & $1.12 \times 10^{-2}$ & $1.01 \times 10^{-2}$ & $1.95$ & $2.08$ \\ 
 & & & & & & $10^{-10}$ & $1.49$ & $1.00$ & $3.22 \times 10^{-12}$ & $2.61 \times 10^{-2}$ & $1.75$ & $1.83$ \\ 
 $0.923$ & $0.5$ & $-5.46 \times 10^{-2}$ & $0.874$ & $4.52$ & $1.23$ & $10^{10}$ & $1.00$ & $1.06$ & $1.40 \times 10^{-2}$ & $1.46 \times 10^{-12}$ & $1.23$ & $1.31$ \\
 & & & & & & $1$ & $1.11$ & $1.03$ & $7.97\times 10^{-3}$ & $7.22 \times 10^{-3}$ & $1.12$ & $1.16$ \\
 & & & & & & $10^{-10}$ & $1.48$ & $1.00$ & $2.30\times 10^{-12}$ & $1.86 \times 10^{-2}$ & $1.05$ & $1.08$ \\
 $0.913$ & $0.5$ & $-5.46 \times 10^{-2}$ & $0.826$ & $3.65$ & $1.13$ & $10^{10}$ & $1.00$ & $1.06$ & $1.40 \times 10^{-2}$ & $1.46 \times 10^{-12}$ & $1.13$ & $1.46$ \\
  & &  &  &  &  & $1$ & $1.11$ & $1.03$ & $8.03 \times 10^{-3}$ & $7.24 \times 10^{-3}$ & $1.04$ & $1.08$ \\
& &  &  &  &  & $10^{-10}$ & $1.53$ & $1.00$ & $2.40 \times 10^{-12}$ & $1.90 \times 10^{-2}$ & $0.977$ & $1.00$ \\
$1.03$ & $0.5$ & $-5.46 \times 10^{-2}$ & $1.24$ & $\infty$ & $2.42$ & $10^{10}$ & $1.00$ & $1.06$ & $1.40 \times 10^{-2}$ & $1.48 \times 10^{-12}$ & $2.42$ & $2.79$ \\
 &  &  &  &  &  & $1$ & $1.13$ & $1.03$ & $8.17 \times 10^{-3}$ & $7.28 \times 10^{-3}$ & $2.03$ & $2.19$ \\
 &  &  &  &  &  & $10^{-10}$ & $1.56$ & $1.00$ & $2.47 \times 10^{-12}$ & $1.95 \times 10^{-2}$ & $1.81$ & $1.89$ \\
 $0.913$ & $0.75$ & $-2.96 \times 10^{-2}$ & $0.892$ & $5.51$ & $1.31$ & $10^{10}$ & $1.00$ & $1.03$ & $7.50 \times 10^{-3}$ & $7.83 \times 10^{-13}$ & $1.31$ & $1.41$ \\
 &  &  &  & &  & $1$ & $1.11$ & $1.02$ & $4.31 \times 10^{-3}$ & $3.88 \times 10^{-3}$ & $1.18$ & $1.24$ \\
 &  &  &  & & & $10^{-10}$ & $1.51$ & $1.00$ & $1.29 \times 10^{-12}$ & $1.03 \times 10^{-3}$ & $1.09$ & $1.13$ \\
 $0.906$ & $0.75$ & $-2.96 \times 10^{-2}$ & $0.836$ & $4.14$ & $1.18$ & $10^{10}$ & $1.00$ & $1.03$ & $7.50 \times 10^{-3}$ & $7.84 \times 10^{-13}$ & $1.18$ & $1.26$ \\
 & &  & &  &  & $1$ & $1.12$ & $1.02$ & $4.33 \times 10^{-3}$ & $3.88 \times 10^{-3}$ & $1.07$ & $1.12$ \\
 & &  & &  &  & $10^{-10}$ & $1.55$ & $1.00$ & $1.32 \times 10^{-12}$ & $1.04 \times 10^{-2}$ & $1.00$ & $1.03$ \\
 $0.967$ & $0.75$ & $-2.96 \times 10^{-2}$ & $1.26$ & $\infty$ & $2.65$ & $10^{10}$ & $1.00$ & $1.03$ & $7.50 \times 10^{-3}$ & $8.00 \times 10^{-13}$ & $2.65$ & $3.15$ \\
 & & & & & & $1$ & $1.15$ & $1.02$ & $4.48 \times 10^{-3}$ & $3.93 \times 10^{-3}$ & $2.17$ & $2.36$ \\
 & & & & & & $10^{-10}$ & $1.66$ & $1.00$ & $1.45 \times 10^{-12}$ & $1.11 \times 10^{-2}$ & $1.89$ & $2.00$ \\
\hline\hline
\end{tabular}
\end{table*}

In contrast with the Kerr case, for KBHsSH spacetimes we do not have a simple expression for the Keplerian angular momentum distribution $l_{\mathrm{K}}(r)$ or for the radius of the ISCO $r_{\mathrm{ms}}$. However, it can be shown (see, for instance~\citep{Dyba:2020}) that  in a stationary and axisymmetric spacetime, the Keplerian angular momentum (for prograde motion) takes the general form 
\begin{equation}\label{eq:keplerian_ang_mom}
l_{\mathrm{K}}(r) = -\frac{\mathcal{B} g_{\phi\phi} + \partial_r g_{\phi\phi} g_{t\phi}}{\mathcal{B} g_{t\phi} + \partial_r g_{\phi\phi} g_{tt}} \  ,
\end{equation}
where $\mathcal{B}$ is defined as
\begin{equation}
\mathcal{B} = -\partial_r g_{t\phi} + \sqrt{(\partial_r g_{t\phi})^2-\partial_r g_{tt}\partial_r g_{\phi\phi}}.
\end{equation}
It can also be seen that for most BH spacetimes $l_{\mathrm{K}}(r)$ only has one minimum outside the event horizon and this minimum coincides with the location of the ISCO. Examples where this condition is not fulfilled are discussed in~\cite{Dyba:2020, Dyba:2021} in the context of self-gravitating accretion disks. 

\begin{figure*}[t]
\includegraphics[scale=0.175]{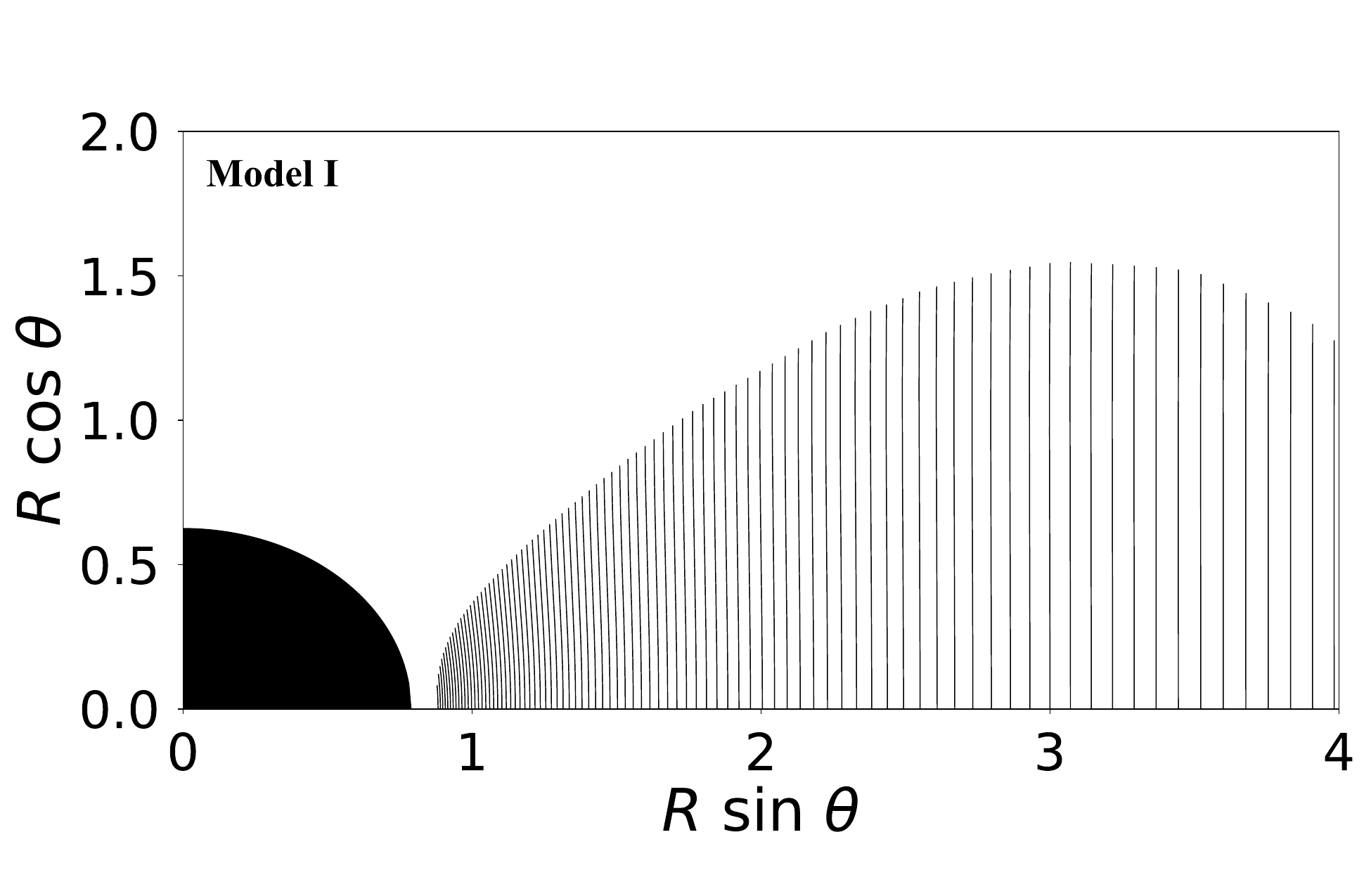}
\hspace{-0.3cm}
\includegraphics[scale=0.175]{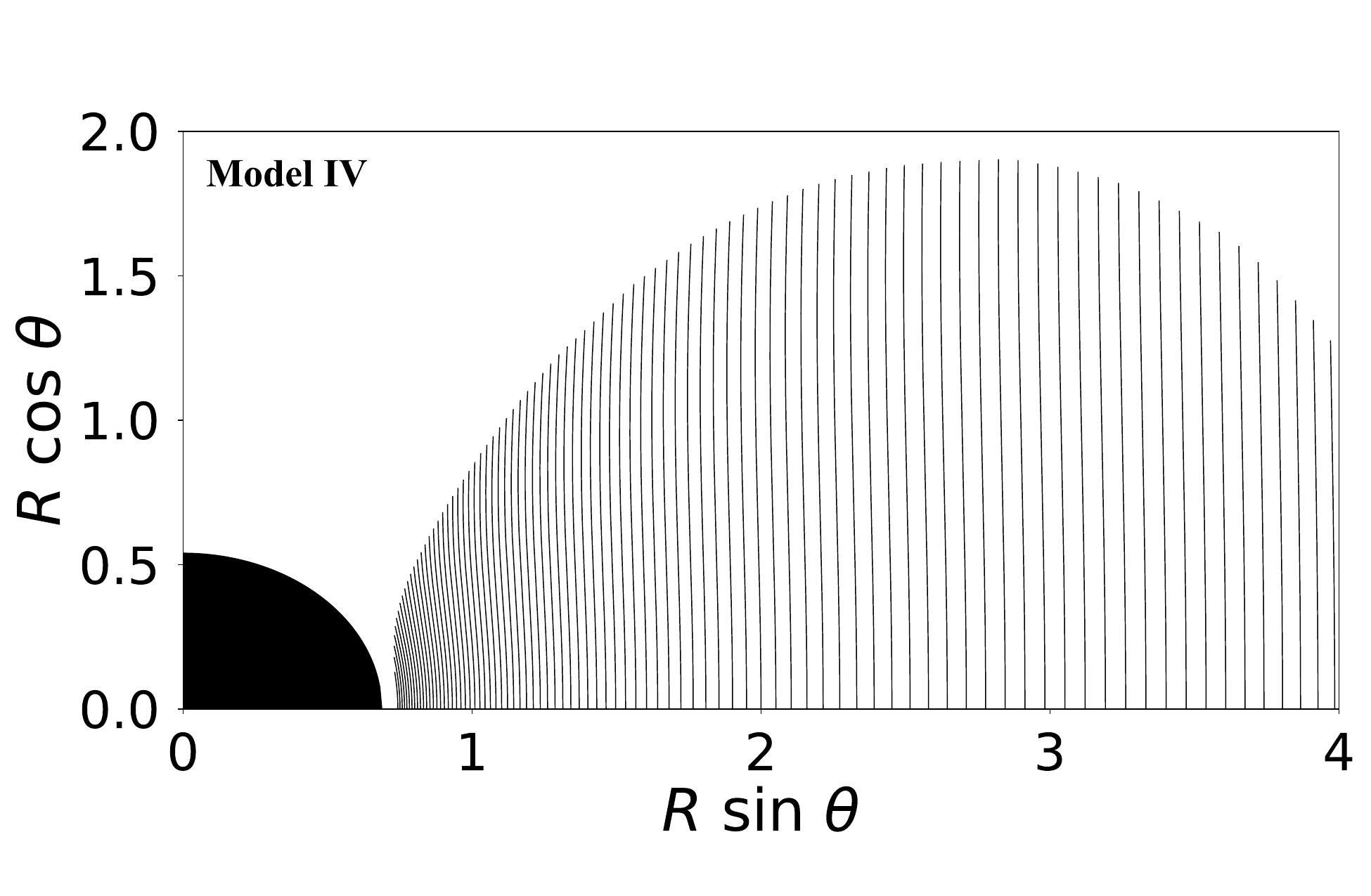}
\hspace{-0.2cm}
\includegraphics[scale=0.18]{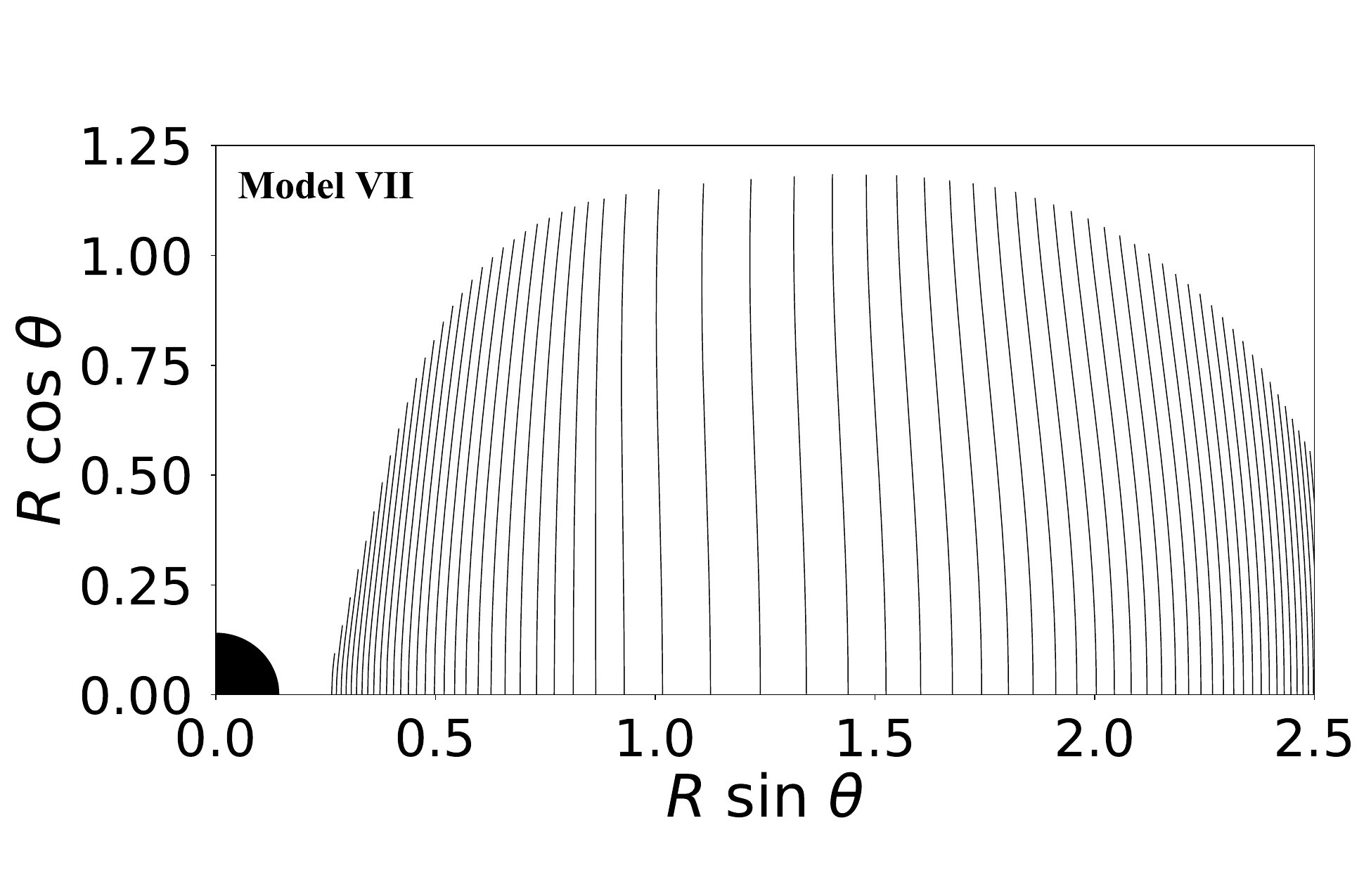}
\caption{Structure of the von Zeipel cylinders (i.e.~the surfaces of constant angular velocity $\Omega$ that correspond to surfaces of constant specific angular momentum~\cite{Daigne:2004}) for the three KBHsSH  spacetimes we consider. The particular angular momentum distribution used in this figure  corresponds to $l_0$ computed using criterion 1 in Section III.A and $\alpha = 0.5$. Note that the cylinders are only shown in the region occupied by the disks.}
\label{vonZeipel_cylinders}
\end{figure*}

\begin{table*}[t]
\caption{Same as in Table~\ref{results_I} but for model IV.}        
\label{results_IV}
\centering          
\begin{tabular}{c c c c  c c c c   c c c c c}
\hline\hline       
 $l_0$ & $\alpha$ & $\Delta W_{\mathrm{c}}$ &  $R_{\mathrm{in}}$ & $R_{\mathrm{out}}$ & $R_{\mathrm{c}}$ & $\beta_{\mathrm{m,c}}$ & $\rho_{\mathrm{max}}$ & $h_{\mathrm{max}}$ & $p_{\mathrm{max}}$ & $p_{\mathrm{m,max}}$ & $R_{\mathrm{max}}$ & $R_{\mathrm{m, max}}$\\ 
\hline           
$1.16$ & $0$ & $-0.273$ &  $0.725$  & $3.85$ & $1.06$ & $10^{10}$ & $1.00$ & $1.31$ & $7.86 \times 10^{-2}$ & $8.52 \times 10^{-12}$ & $1.06$ & $1.21$ \\ 
&  &  &  &  &  & $1$ & $1.22$ & $1.16$ & $4.95 \times 10^{-2}$ & $4.10 \times 10^{-2}$ & $0.908$ & $0.967$ \\ 
&  &  &  &  &  & $10^{-10}$ & $2.21$ & $1.00$ & $1.97 \times 10^{-11}$ & $0.130$ & $0.827$ & $0.852$ \\ 
$1.14$ & $0$ & $-0.273$ & $0.701$ & $3.70$ & $1.01$ & $10^{10}$ & $1.00$ & $1.31$ & $7.86 \times 10^{-2}$ & $8.56 \times 10^{-12}$ & $1.01$ & $1.15$ \\
&  &  &  &  &  & $1$ & $1.24$ & $1.16$ & $5.04 \times 10^{-2}$ & $4.13 \times 10^{-2}$ & $0.865$ & $0.920$ \\
&  &  &  &  &  & $10^{-10}$ & $2.38$ & $1.00$ & $2.17 \times 10^{-11}$ & $0.138$ & $0.792$ & $0.815$ \\
$1.83$ & $0$ & $-0.273$ & $1.42$ & $\infty$ & $2.44$ & $10^{10}$ & $1.00$ & $1.31$ & $7.86 \times 10^{-2}$ & $8.13 \times 10^{-12}$ & $2.44$ & $2.61$ \\
&  &  &  &  &  & $1$ & $1.09$ & $1.16$ & $4.29 \times 10^{-2}$ & $3.94 \times 10^{-2}$ & $2.17$ & $2.29$ \\
&  &  &  &  &  & $10^{-10}$ & $1.43$ & $1.00$ & $1.10 \times 10^{-11}$ & $9.08 \times 10^{-2}$ & $1.98$ & $2.06$ \\
$1.16$ & $0.25$ & $-0.233$ & $0.729$ & $4.06$ & $1.12$ & $10^{10}$ & $1.00$ & $1.26$ & $6.57 \times 10^{-2}$ & $7.17 \times 10^{-12}$ & $1.12$ & $1.29$ \\ 
&  &  &  &  &  & $1$ & $1.24$ & $1.14$ & $4.23 \times 10^{-2}$ & $3.46 \times 10^{-2}$ & $0.935$ & $1.01$ \\
&  &  &  &  &  & $10^{-10}$ & $2.33$ & $1.00$ & $1.80 \times 10^{-11}$ & $0.116$ & $0.842$ & $0.870$ \\
$1.14$ & $0.25$ & $-0.233$ & $0.703$ & $3.86$ & $1.05$ & $10^{10}$ & $1.00$ & $1.26$ & $6.57 \times 10^{-2}$ & $7.22 \times 10^{-12}$ & $1.05$ & $1.22$ \\ 
&  &  &  &  &  & $1$ & $1.26$ & $1.14$ & $4.33 \times 10^{-2}$ & $3.49 \times 10^{-2}$ & $0.881$ & $0.944$ \\
&  &  &  &  &  & $10^{-10}$ & $2.52$ & $1.00$ & $2.00 \times 10^{-11}$ & $0.124$ & $0.800$ & $0.825$ \\
$1.61$ & $0.25$ & $-0.233$ & $1.31$ & $\infty$ & $2.44$ & $10^{10}$ & $1.00$ & $1.26$ & $6.57 \times 10^{-2}$ & $6.83 \times 10^{-12}$ & $2.44$ & $2.64$ \\
&  &  &  &  &  & $1$ & $1.11$ & $1.13$ & $3.67 \times 10^{-2}$ & $3.32 \times 10^{-2}$ & $2.13$ & $2.26$ \\ 
&  &  &  &  &  & $10^{-10}$ & $1.52$ & $1.00$ & $1.02 \times 10^{-11}$ & $8.13 \times 10^{-2}$ & $1.91$ & $1.99$ \\
$1.15$ & $0.5$ & $-0.179$ & $0.736$ & $4.35$ & $1.21$ & $10^{10}$ & $1.00$ & $1.20$ & $4.91 \times 10^{-2}$ & $5.38 \times 10^{-12}$ & $1.21$ & $1.42$ \\
&  &  &  &  &  & $1$ & $1.25$ & $1.10$ & $3.24 \times 10^{-2}$ & $2.61 \times 10^{-2}$ & $0.980$ & $1.07$ \\
&  &  &  &  &  & $10^{-10}$ & $2.46$ & $1.00$ & $1.49 \times 10^{-11}$ & $9.35 \times 10^{-2}$ & $0.867$ & $ 0.901$ \\
$1.13$ & $0.5$ & $-0.179$ & $0.705$ & $4.07$ & $1.11$ & $10^{10}$ & $1.00$ & $1.20$ & $4.91 \times 10^{-2}$ & $5.45 \times 10^{-12}$ & $1.11$ & $1.32$ \\
&  &  &  &  &  & $1$ & $1.28$ & $1.10$ & $3.33 \times 10^{-2}$ & $2.63 \times 10^{-2}$ & $0.907$ & $0.983$ \\
&  &  &  &  &  & $10^{-10}$ & $2.68$ & $1.00$ & $1.67 \times 10^{-11}$ & $0.100$ & $0.812$ & $0.840$ \\
$1.42$ & $0.5$ & $-0.179$ & $1.20$ & $\infty$ & $2.44$ & $10^{10}$ & $1.00$ & $1.20$ & $4.91 \times 10^{-2}$ & $5.14 \times 10^{-12}$ & $2.44$ & $2.68$ \\
&  &  &  &  &  & $1$ & $1.13$ & $1.10$ & $2.83 \times 10^{-2}$ & $2.51 \times 10^{-2}$ & $2.08$ & $2.24$ \\
&  &  &  &  &  & $10^{-10}$ & $1.64$ & $1.00$ & $8.71 \times 10^{-12}$ & $6.65 \times 10^{-2}$ & $1.82$ & $1.92$ \\
$1.14$ & $0.75$ & $-0.103$ & $0.753$ & $4.79$ & $1.37$ & $10^{10}$ & $1.00$ & $1.11$ & $2.72 \times 10^{-2}$ & $2.98 \times 10^{-12}$ & $1.37$ & $1.62$ \\
&  &  &  &  &  & $1$ & $1.26$ & $1.06$ & $1.83 \times 10^{-2}$ & $1.46 \times 10^{-2}$ & $1.07$ & $1.19$ \\
&  &  &  &  &  & $10^{-10}$ & $2.55$ & $1.00$ & $9.03 \times 10^{-12}$ & $5.53 \times 10^{-2}$ & $0.922$ & $0.968$ \\
$1.12$ & $0.75$ & $-0.103$ & $0.712$ & $4.29$ & $1.21$ & $10^{10}$ & $1.00$ & $1.11$ & $2.72 \times 10^{-2}$ & $3.03 \times 10^{-12}$ & $1.21$ & $1.47$ \\
&  &  &  &  &  & $1$ & $1.30$ & $1.06$ & $1.90 \times 10^{-2}$ & $1.48 \times 10^{-2}$ & $0.956$ & $1.05$ \\
&  &  &  &  &  & $10^{-10}$ & $2.84$ & $1.00$ & $1.04 \times 10^{-11}$ & $6.06 \times 10^{-2}$ & $0.839$ & $0.873$ \\
$1.25$ & $0.75$ & $-0.103$ & $1.10$ & $\infty$ & $2.48$ & $10^{10}$ & $1.00$ & $1.11$ & $2.72 \times 10^{-2}$ & $2.88 \times 10^{-12}$ & $2.48$ & $2.78$ \\
&  &  &  &  &  & $1$ & $1.16$ & $1.06$ & $1.63 \times 10^{-2}$ & $1.41 \times 10^{-2}$ & $2.05$ & $2.24$ \\
&  &  &  &  &  & $10^{-10}$ & $1.82$ & $1.00$ & $5.75 \times 10^{-12}$ & $4.14 \times 10^{-2}$ & $1.73$ & $1.85$ \\
\hline\hline
\end{tabular}
\end{table*}

Our ansatz for the angular momentum law brings some advantages when compared to a simpler choice. For instance, one could consider a power-law radial dependence like the one discussed in~\cite{Daigne:2004}
\begin{equation}\label{eq:power_law}
l(r) = k r^{\alpha}.
\end{equation}
Due to the explicit dependence on the radial coordinate in Eq.~\eqref{eq:power_law} it is apparent that this functional form is not coordinate independent. This fact should be no more than a minor inconvenience when dealing with solutions of the Kerr family where  algebraic coordinate transformations exist. However, this becomes an insurmountable problem in our case, as there is no way of {\it translating} a specific choice of angular momentum distribution to a different spacetime in such a way that the physical meaning of Eq.~\eqref{eq:power_law} is preserved (e.g.~from KBHsSH in our coordinate ansatz to a Kerr BH in Boyer-Lindquist coordinates). The angular momentum ansatz used in
this work, Eq.~\eqref{eq:ansatz}, could be seen as a power law in the same way as Eq.~\eqref{eq:power_law} (for $r \geq r_{\mathrm{ms}}$) if we consider that $k = l_0/l_{\mathrm{K}}(r_{\mathrm{ms}})^{\alpha}$ and $l_{\mathrm{K}}(r)$ plays the role of the radial coordinate. This choice is particularly good as $l_{\mathrm{K}}(r)$ captures the relevant physical information about circular orbits and it is strictly increasing with $r$, as a well-chosen radial coordinate should be. Furthermore, if $l_0$ is expressed in terms of quantities determined by the kinematics of the disk, one specific choice of angular momentum will have the same physical meaning irrespective of the particular spacetime we considered.

\begin{table*}[t]
\caption{Same as in Table~\ref{results_I} but for model VII.}        
\label{results_VII}
\centering          
\begin{tabular}{c c c c  c c c c   c c c c c}
\hline\hline       
 $l_0$ & $\alpha$ & $\Delta W_{\mathrm{c}}$ &  $R_{\mathrm{in}}$ & $R_{\mathrm{out}}$ & $R_{\mathrm{c}}$ & $\beta_{\mathrm{m,c}}$ & $\rho_{\mathrm{max}}$ & $h_{\mathrm{max}}$ & $p_{\mathrm{max}}$ & $p_{\mathrm{m,max}}$ & $R_{\mathrm{max}}$ & $R_{\mathrm{m, max}}$\\ 
\hline           
$0.920$ & $0$ & $-0.618$ & $0.227$ & $2.51$ & $1.10$ & $10^{10}$ & $1.00$ & $1.85$ & $0.214$ & $2.23 \times 10^{-11}$ & $1.10$ & $1.21$ \\ 
&  &  &  &  &  & $1$ & $1.32$ & $1.44$ & $0.144$ & $0.108$ & $0.574$ & $0.863$ \\
&  &  &  &  &  & $10^{-10}$ & $4.77$ & $1.00$ & $1.24 \times 10^{-10}$ & $0.555$ & $0.363$ & $0.400$ \\
$0.895$ & $0$ & $-0.618$ & $0.189$ & $2.50$ & $1.08$ & $10^{10}$ & $1.00$ & $1.85$ & $0.214$ & $2.23 \times 10^{-11}$ & $1.08$ & $1.20$ \\ 
&  &  &  &  & & $1$ & $1.39$ & $1.44$ & $0.153$ & $0.109$ & $0.519$ & $0.820$ \\ 
&  &  &  &  & & $10^{-10}$ & $5.84$ & $1.00$ & $1.62 \times 10^{-10}$ & $0.670$ & $0.326$ & $0.360$ \\
$1.92$ & $0$ & $-0.618$ & $1.01$ & $\infty$ & $1.64$ & $10^{10}$ & $1.00$ & $1.85$ & $0.214$ & $2.18 \times 10^{-11}$ & $1.64$ & $1.71$ \\
&  &  &  &  &  & $1$ & $1.08$ & $1.41$ & $0.109$ & $0.102$ & $1.52$ & $1.57$ \\
&  &  &  &  &  & $10^{-10}$ & $1.34$ & $1.00$ & $2.29 \times 10^{-11}$ & $0.196$ & $1.41$ & $1.45$ \\
$0.916$ & $0.25$ & $-0.531$ & $0.234$ & $2.54$ & $1.16$ & $10^{10}$ & $1.00$ & $1.70$ & $0.175$ & $1.82 \times 10^{-11}$ & $1.16$ & $1.27$ \\
&  &  &  &  &  & $1$ & $1.30$ & $1.36$ & $0.116$ & $8.85 \times 10^{-2}$ & $0.629$ & $0.931$ \\
&  &  &  &  &  & $10^{-10}$ & $4.57$ & $1.00$ & $1.01 \times 10^{-10}$ & $0.453$ & $0.383$ & $0.425$ \\
$0.887$ & $0.25$ & $-0.531$ & $0.192$ & $2.50$ & $1.14$ & $10^{10}$ & $1.00$ & $1.70$ & $0.175$ & $1.82 \times 10^{-11}$ & $1.14$ & $1.25$ \\
&  &  &  &  &  & $1$ & $1.37$ & $1.36$ & $0.125$ & $8.96 \times 10^{-2}$ & $0.552$ & $0.876$ \\
&  &  &  &  &  & $10^{-10}$ & $5.75$ & $1.00$ & $1.37 \times 10^{-10}$ & $0.561$ & $0.339$ & $0.375$ \\
$1.44$ & $0.25$ & $-0.531$ & $0.672$ & $\infty$ & $1.63$ & $10^{10}$ & $1.00$ & $1.70$ & $0.175$ & $1.80 \times 10^{-11}$ & $1.63$ & $1.71$ \\
&  &  &  &  &  & $1$ & $1.09$ & $1.34$ & $9.23 \times 10^{-2}$ & $8.51 \times 10^{-2}$ & $1.47$ & $1.55$ \\
&  &  &  &  &  & $10^{-10}$ & $1.46$ & $1.00$ & $2.20 \times 10^{-11}$ & $0.178$ & $1.31$ & $1.38$ \\
$0.903$ & $0.5$ & $-0.374$ & $0.253$ & $2.59$ & $1.29$ & $10^{10}$ & $1.00$ & $1.45$ & $0.113$ & $1.17 \times 10^{-11}$ & $1.29$ & $1.38$ \\
&  &  &  &  &  & $1$ & $1.23$ & $1.23$ & $7.15 \times 10^{-2}$ & $5.72 \times 10^{-2}$ & $0.785$ & $1.12$ \\
&  &  &  &  &  & $10^{-10}$ & $3.93$ & $1.00$ & $5.80 \times 10^{-11}$ & $0.268$ & $0.438$ & $0.493$ \\
$0.867$ & $0.5$ & $-0.374$ & $0.198$ & $2.48$ & $1.24$ & $10^{10}$ & $1.00$ & $1.45$ & $0.113$ & $1.18 \times 10^{-11}$ & $1.24$ & $1.34$ \\
&  &  &  &  &  & $1$ & $1.30$ & $1.24$ & $7.65 \times 10^{-2}$ & $5.78 \times 10^{-2}$ & $0.663$ & $1.03$ \\
&  &  &  &  &  & $10^{-10}$ & $5.18$ & $1.00$ & $8.39 \times 10^{-11}$ & $0.346$ & $0.367$ & $0.412$ \\
$1.10$ & $0.5$ & $-0.374$ & $0.460$ & $\infty$ & $1.67$ & $10^{10}$ & $1.00$ & $1.45$ & $0.113$ & $1.17 \times 10^{-11}$ & $1.67$ & $1.75$ \\
&  &  &  &  &  & $1$ & $1.11$ & $1.22$ & $6.22 \times 10^{-2}$ & $5.64 \times 10^{-2}$ & $1.48$ & $1.57$ \\
&  &  &  &  &  & $10^{-10}$ & $1.66$ & $1.00$ & $1.84 \times 10^{-11}$ & $0.134$ & $0.973$ & $1.34$ \\
$0.829$ & $0.75$ & $-0.111$ & $0.380$ & $3.07$ & $1.65$ & $10^{10}$ & $1.00$ & $1.12$ & $2.95 \times 10^{-2}$ & $3.02 \times 10^{-12}$ & $1.65$ & $1.71$ \\
&  &  &  &  &  & $1$ & $1.07$ & $1.06$ & $1.60 \times 10^{-2}$ & $1.49 \times 10^{-2}$ & $1.53$ & $1.59$ \\
&  &  &  &  &  & $10^{-10}$ & $1.37$ & $1.00$ & $4.24 \times 10^{-12}$ & $3.50 \times 10^{-2}$ & $1.34$ & $1.44$ \\
$0.795$ & $0.75$ & $-0.111$ & $0.233$ & $2.29$ & $1.47$ & $10^{10}$ & $1.00$ & $1.12$ & $2.95 \times 10^{-2}$ & $3.01 \times 10^{-12}$ & $1.47$ & $1.52$ \\
&  &  &  &  &  & $1$ & $1.08$ & $1.06$ & $1.60 \times 10^{-2}$ & $1.49 \times 10^{-2}$ & $1.31$ & $1.40$ \\
&  &  &  &  &  & $10^{-10}$ & $1.92$ & $1.00$ & $6.62 \times 10^{-12}$ & $3.85 \times 10^{-2}$ & $0.594$ & $0.843$ \\
$0.861$ & $0.75$ & $-0.111$ & $0.643$ & $\infty$ & $1.95$ & $10^{10}$ & $1.00$ & $1.12$ & $2.95 \times 10^{-2}$ & $3.08 \times 10^{-12}$ & $1.95$ & $2.06$ \\
&  &  &  &  &  & $1$ & $1.10$ & $1.06$ & $1.66 \times 10^{-2}$ & $1.51 \times 10^{-2}$ & $1.81$ & $1.87$ \\
&  &  &  &  &  & $10^{-10}$ & $1.48$ & $1.00$ & $4.69 \times 10^{-12}$ & $3.79 \times 10^{-2}$ & $1.67$ & $1.73$ \\
\hline\hline
\end{tabular}
\end{table*}

\begin{figure*}[t]
\includegraphics[scale=0.17]{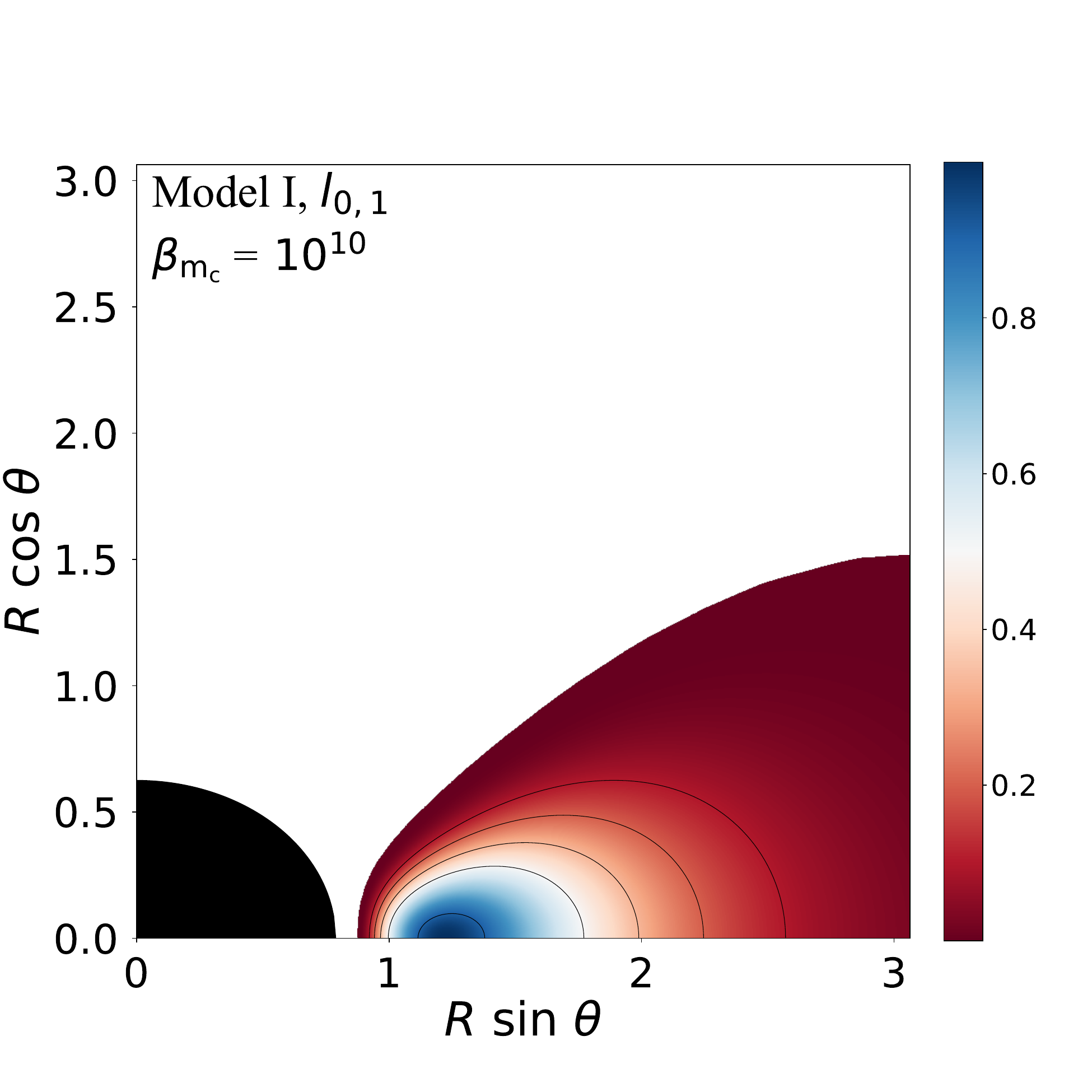}
\hspace{-0.3cm}
\includegraphics[scale=0.17]{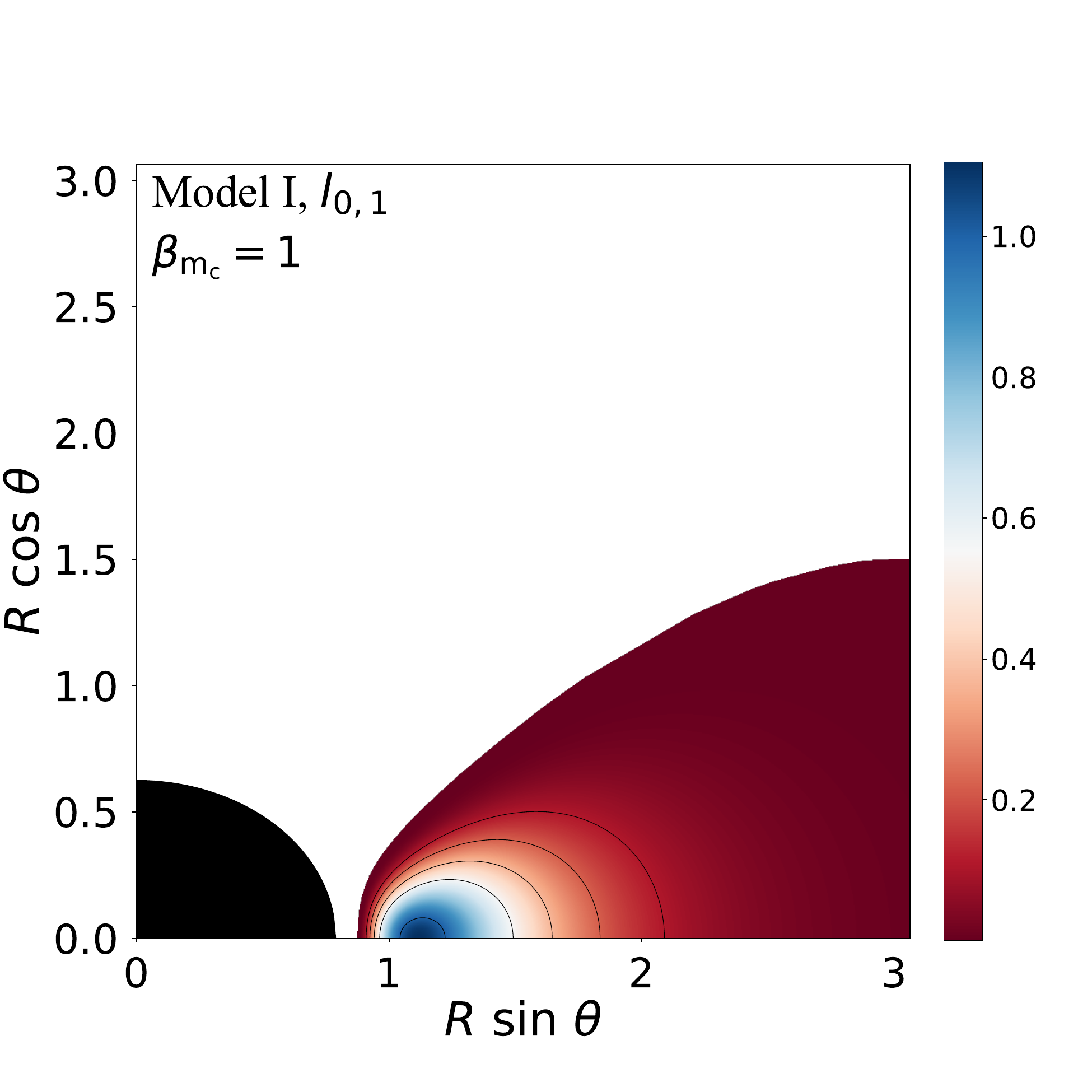}
\hspace{-0.2cm}
\includegraphics[scale=0.17]{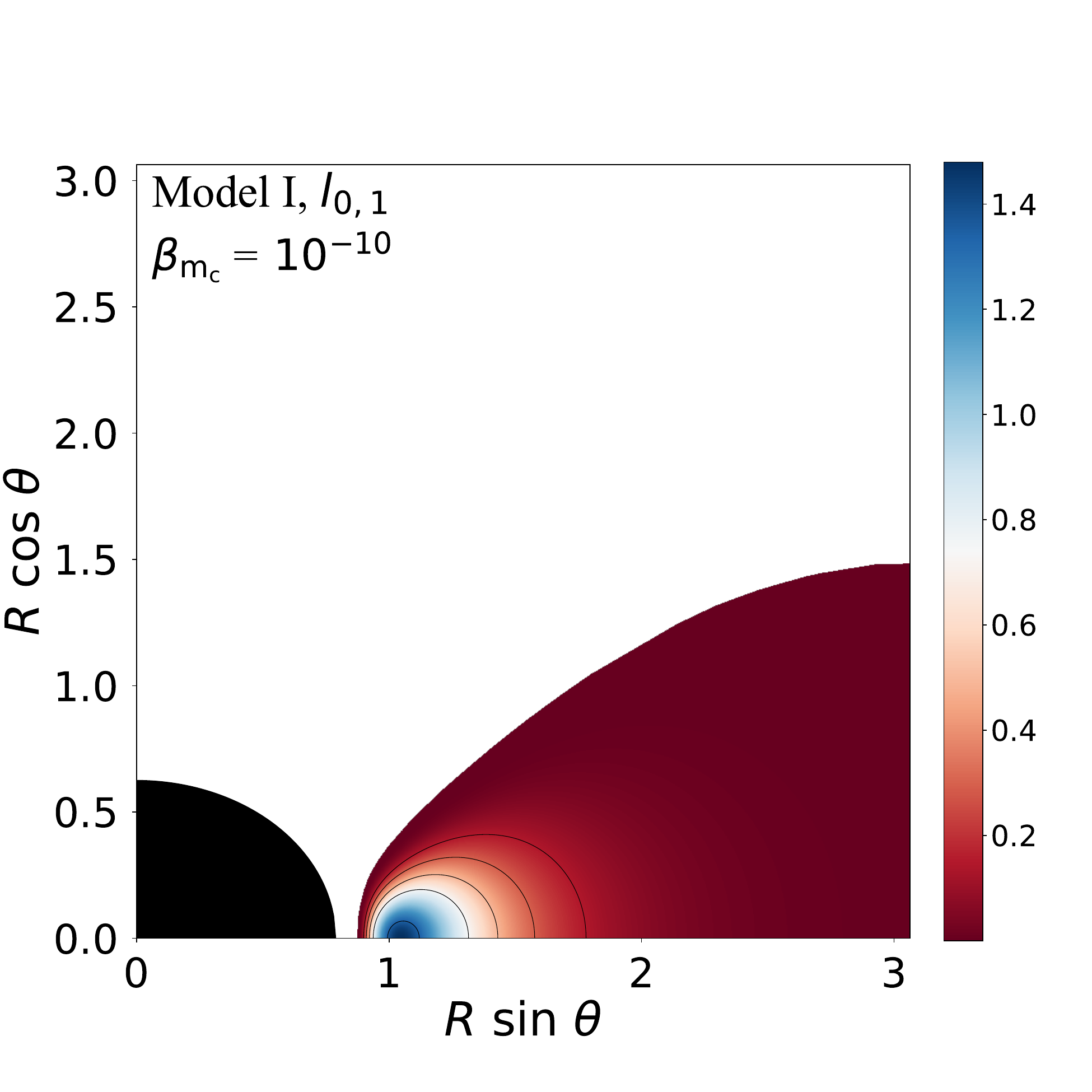}
\\
\vspace{-0.5cm}
\includegraphics[scale=0.17]{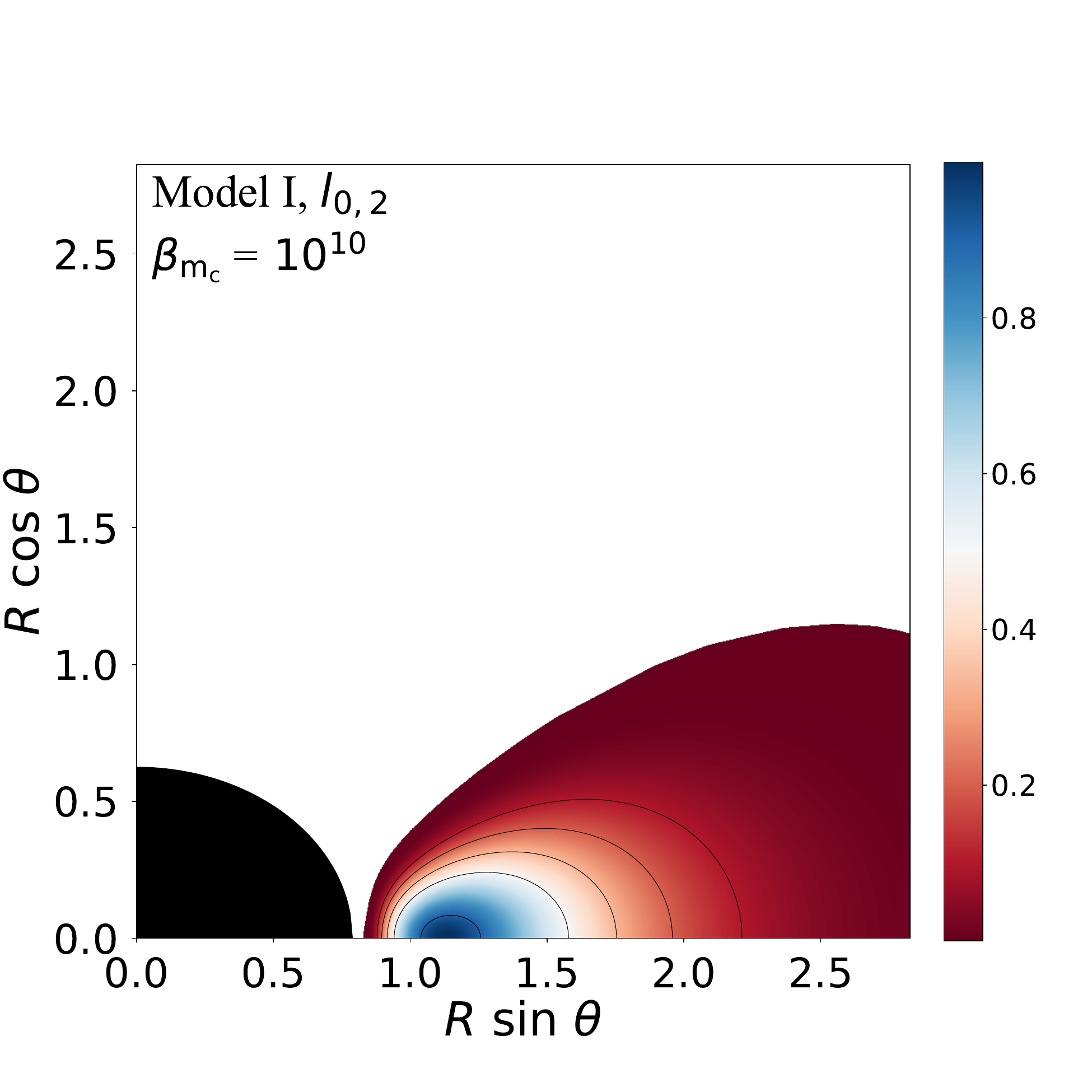}
\hspace{-0.3cm}
\includegraphics[scale=0.17]{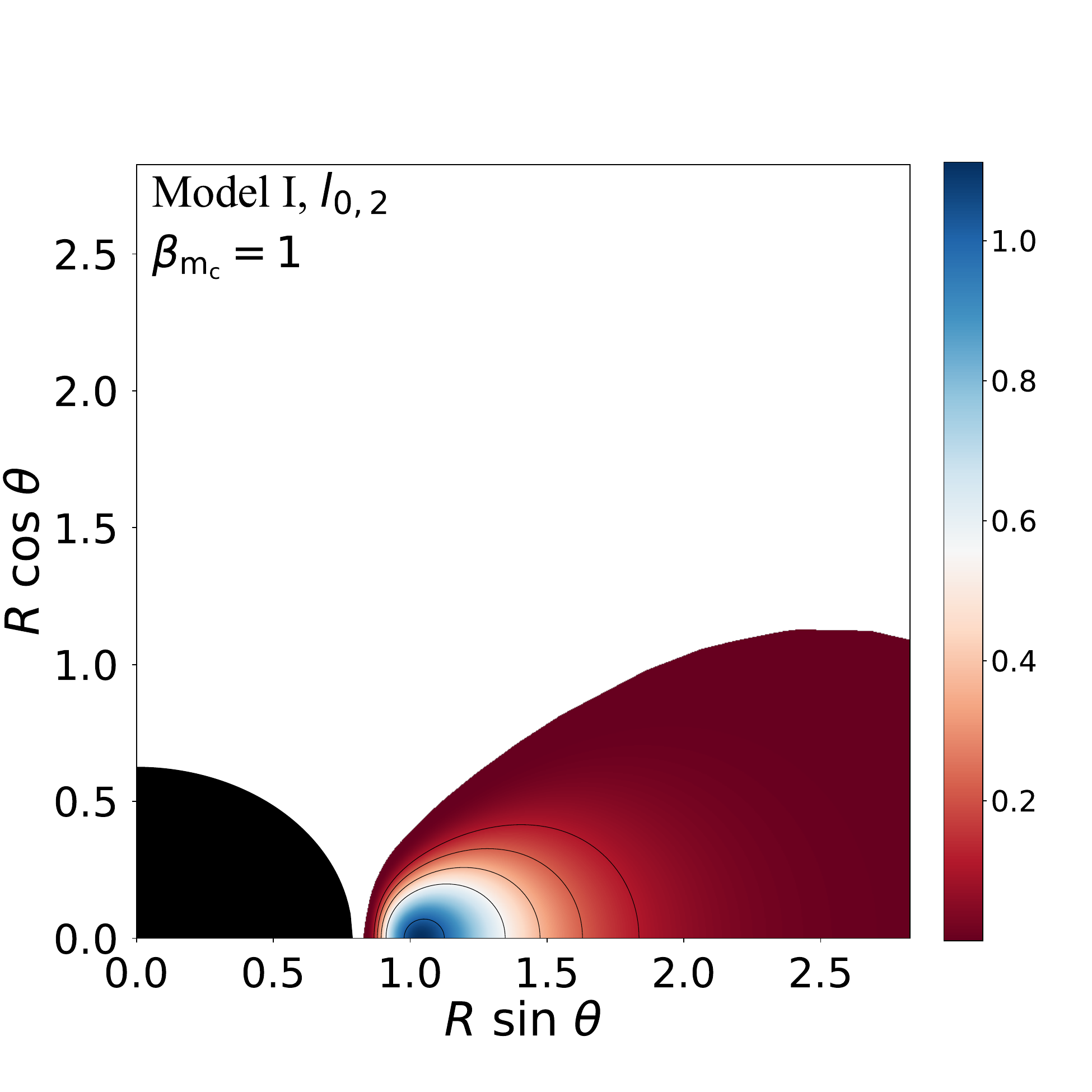}
\hspace{-0.2cm}
\includegraphics[scale=0.17]{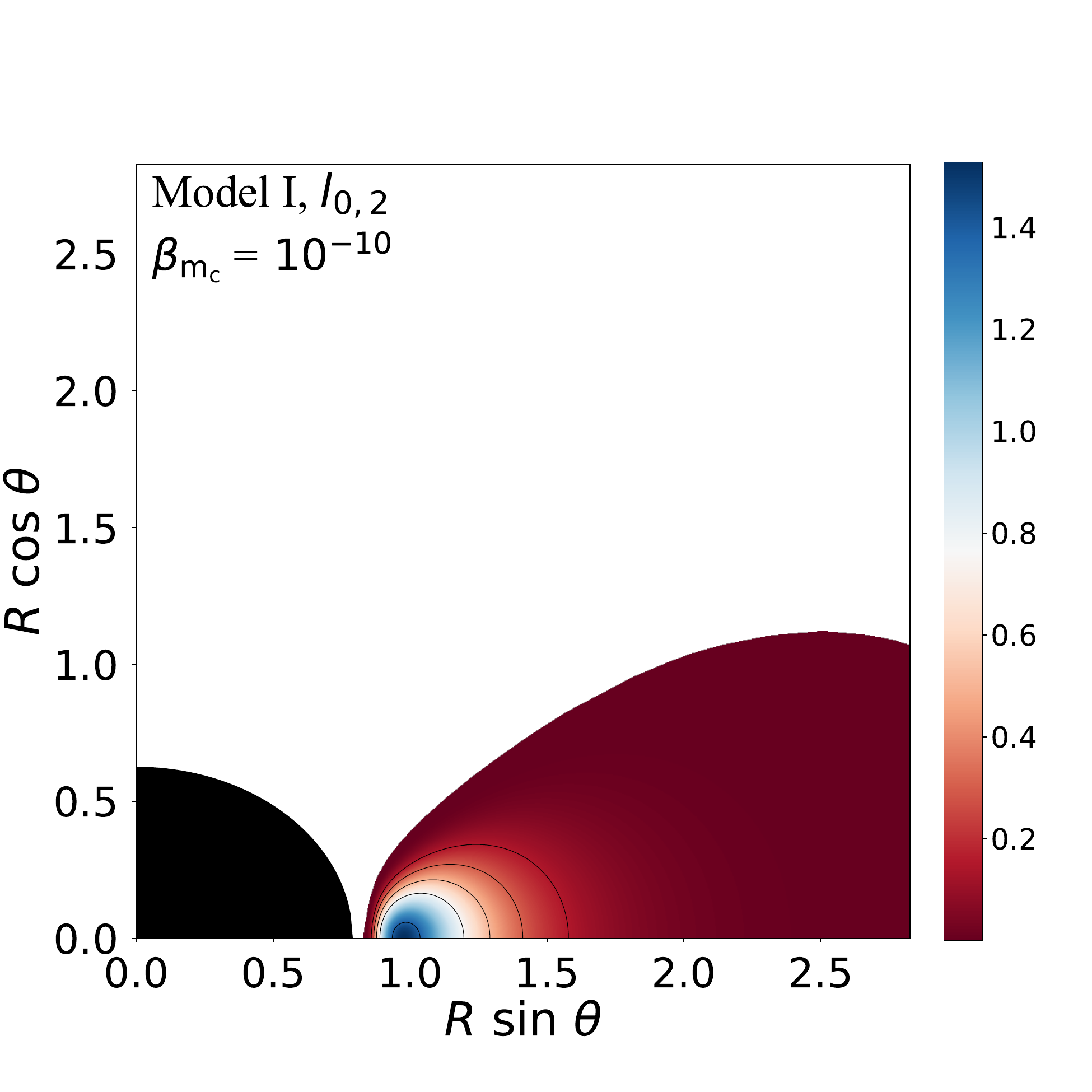}
\\
\vspace{-0.5cm}
\includegraphics[scale=0.17]{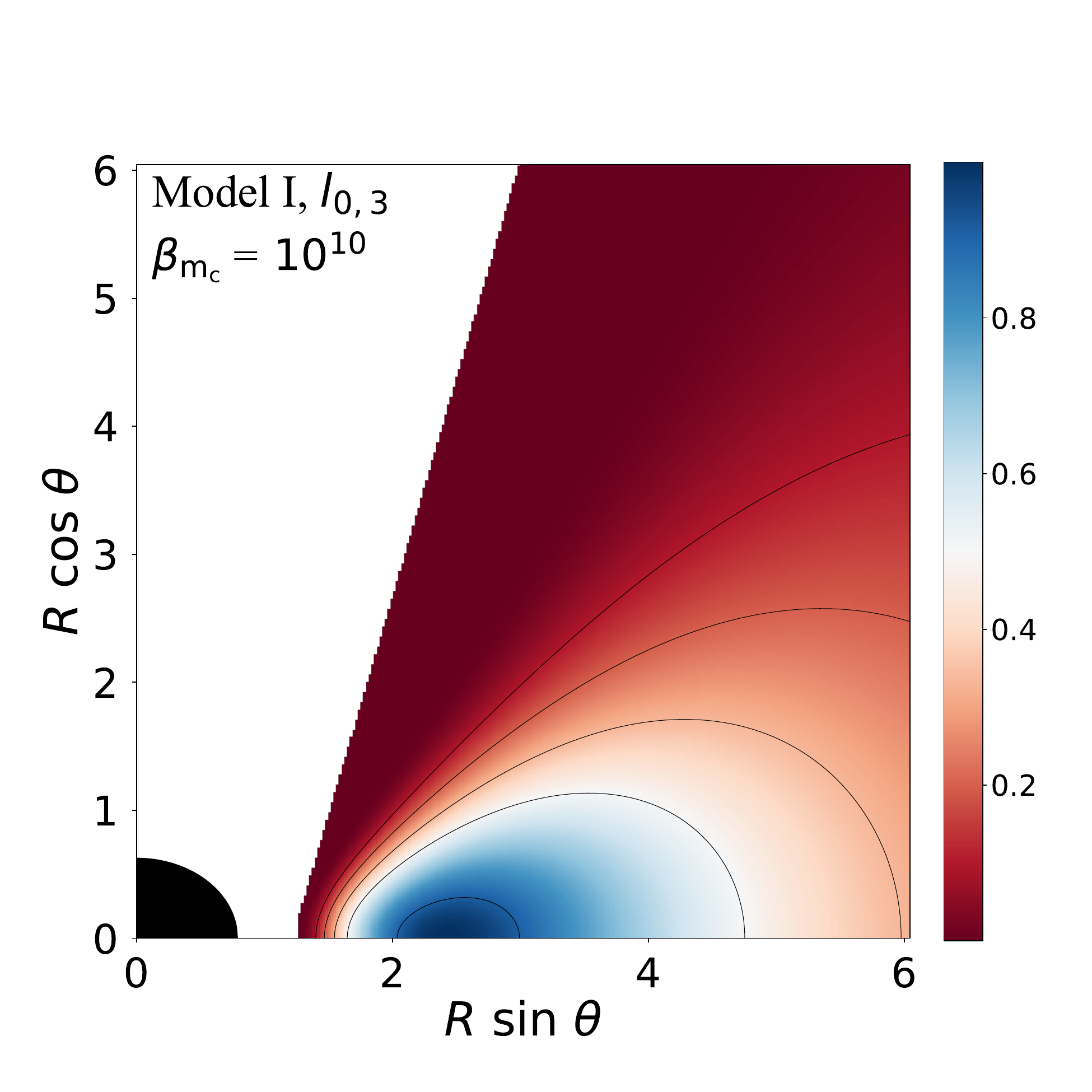}
\hspace{-0.3cm}
\includegraphics[scale=0.17]{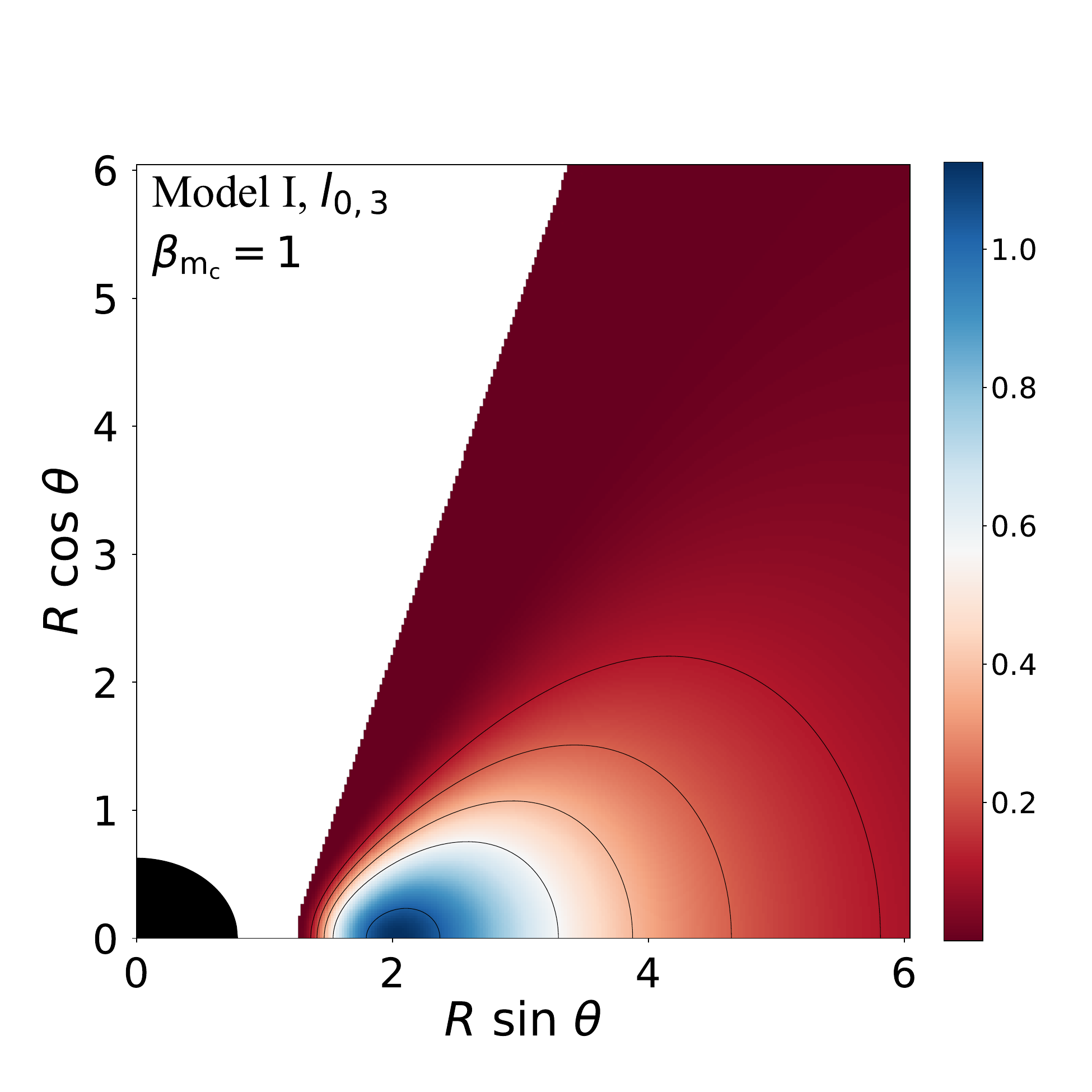}
\hspace{-0.2cm}
\includegraphics[scale=0.17]{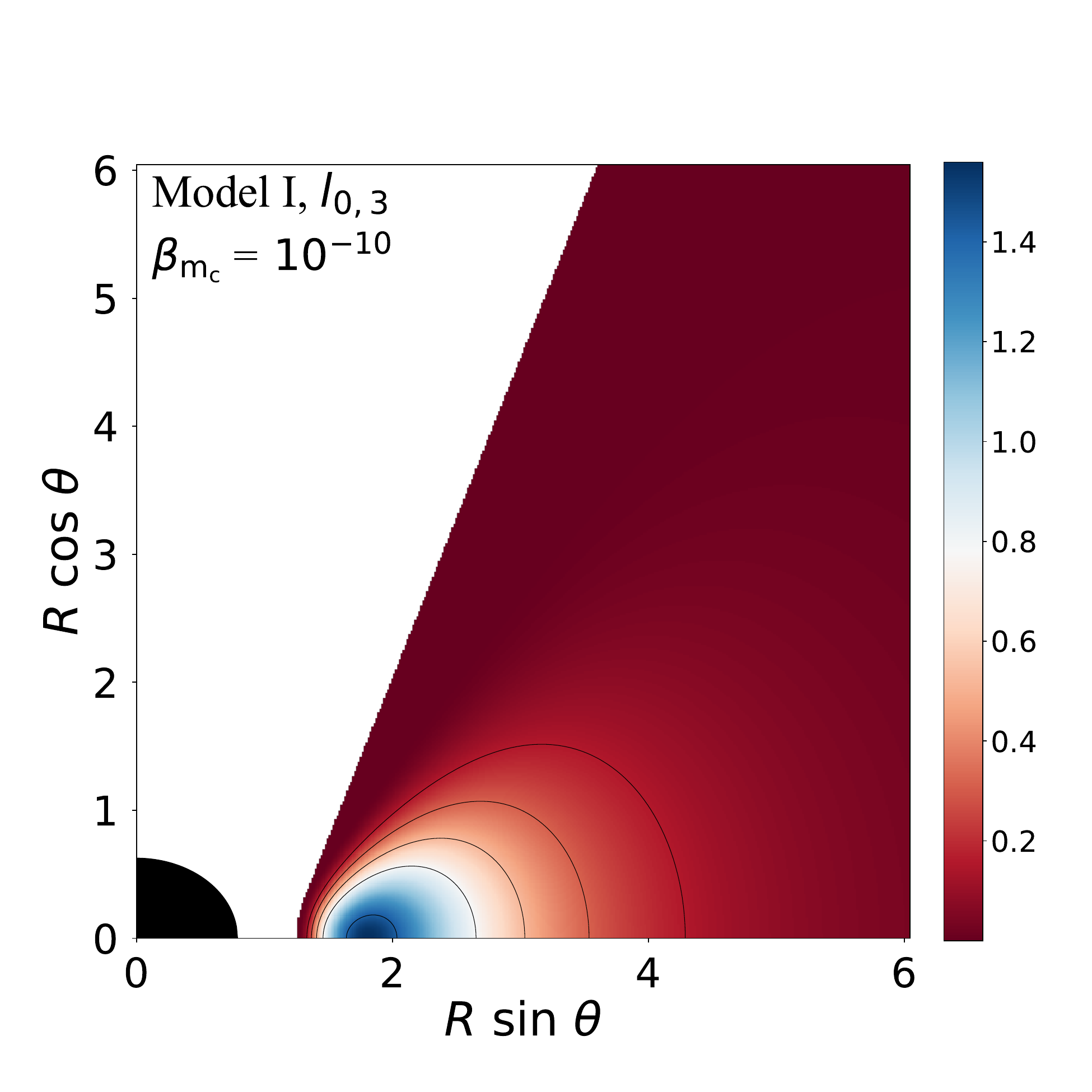}
\caption{Distribution of the rest-mass density $\rho$ in the $(x,y)=(R\sin\theta,R\cos\theta)$ plane (in terms of the perimeteral coordinate $R$) for spacetime I and $\alpha = 0.5$. From top to bottom the rows correspond to different values of the constant part of the specific angular momentum $l_0$, namely $l_{0, 1}$, $l_{0,2}$ and $l_{0,3}$. From left to right, the columns correspond to different values of the magnetization parameter at the center $\beta_{\mathrm{m,c}}$, namely $10^{10}$, $1$ and $10^{-10}$. The black quarter-circle in the bottom-left corner of each plot marks the position of the black hole. The black curves represent rest-mass density isocontours, corresponding to the values $\rho = \rho_{\mathrm{max}}/x$, where $x = \{10, 5, 3, 2, 1.1\}$.}
\label{Model_I_2Dplots}
\end{figure*}

\begin{figure*}[t]
\includegraphics[scale=0.17]{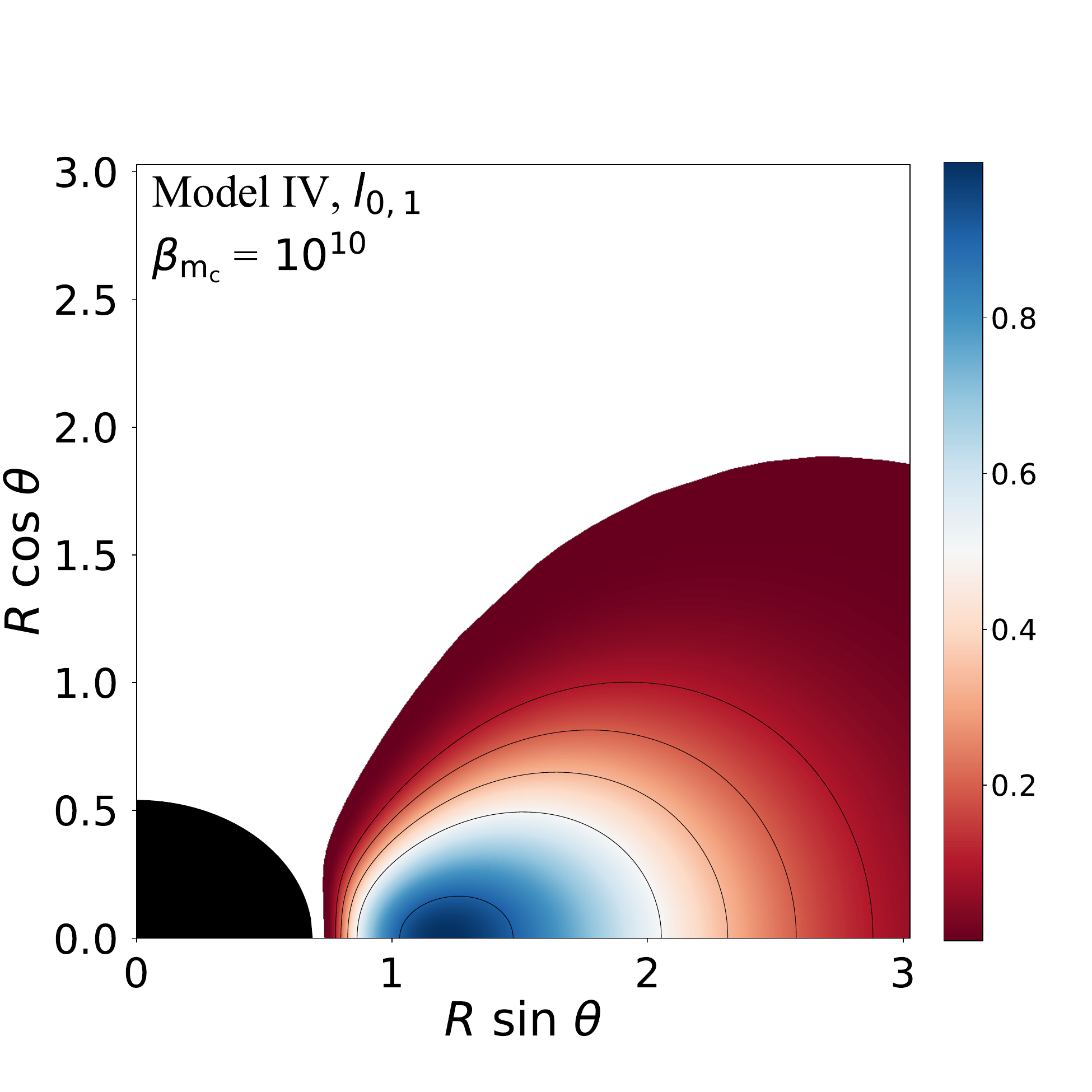}
\hspace{-0.3cm}
\includegraphics[scale=0.17]{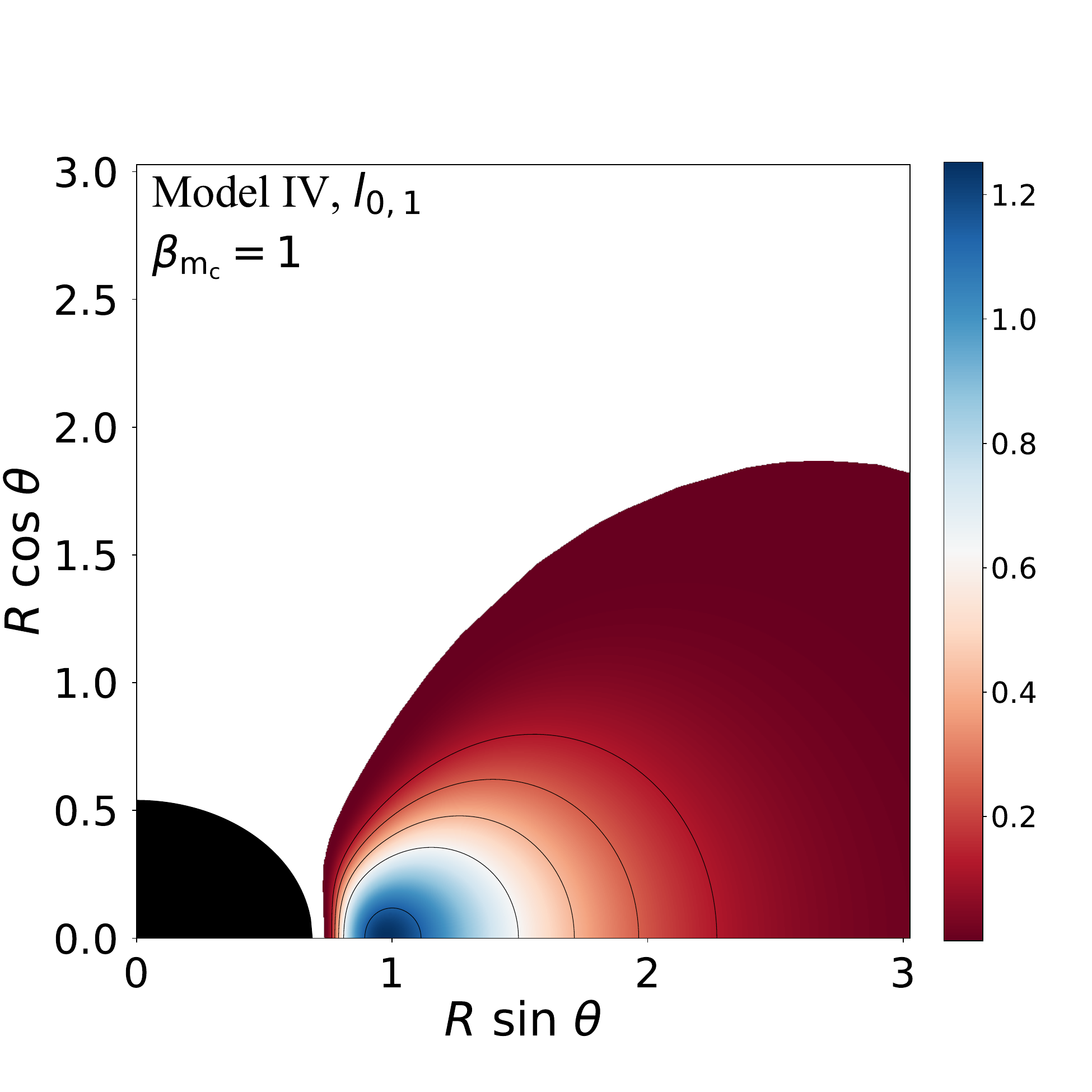}
\hspace{-0.2cm}
\includegraphics[scale=0.17]{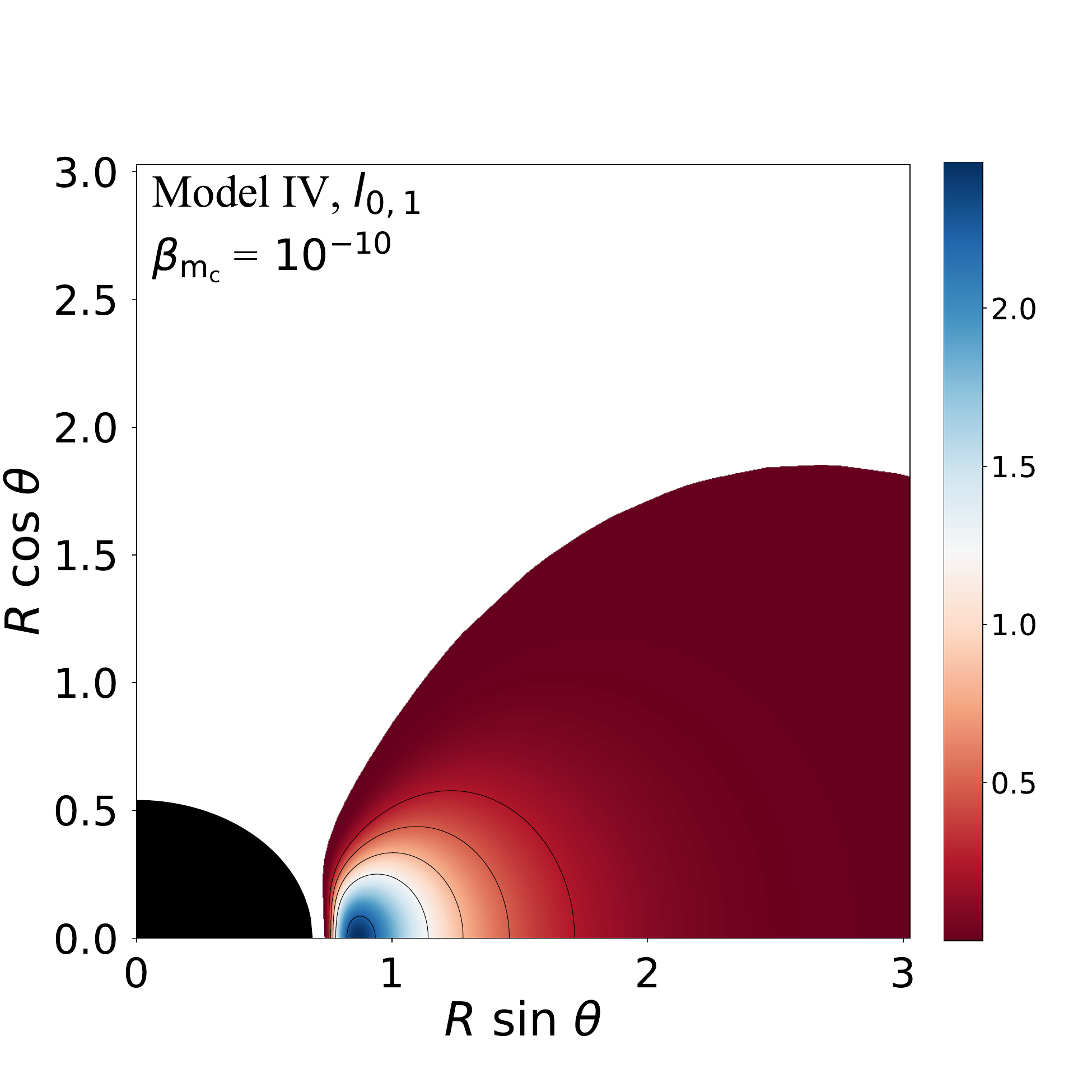}
\\
\vspace{-0.5cm}
\includegraphics[scale=0.17]{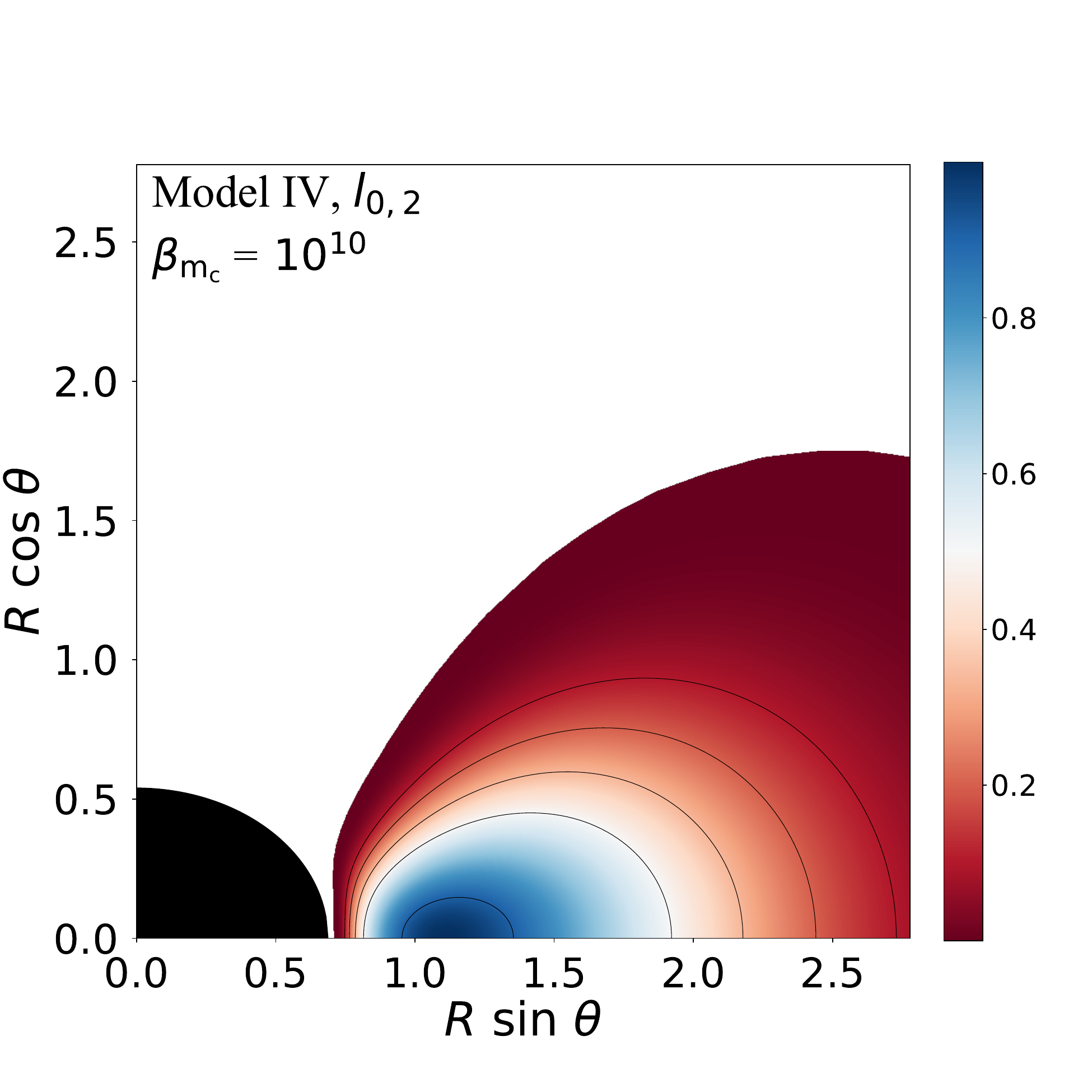}
\hspace{-0.3cm}
\includegraphics[scale=0.17]{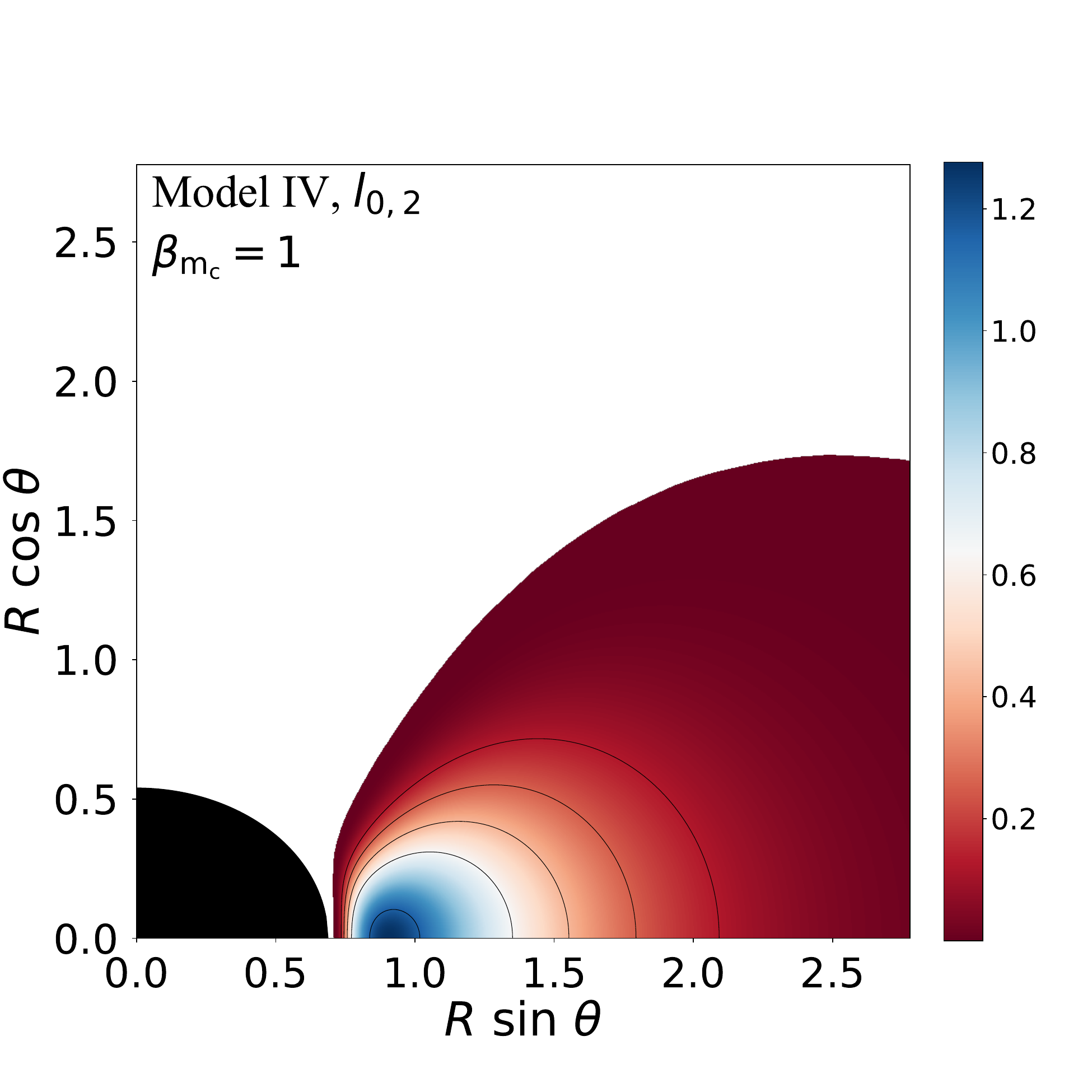}
\hspace{-0.2cm}
\includegraphics[scale=0.17]{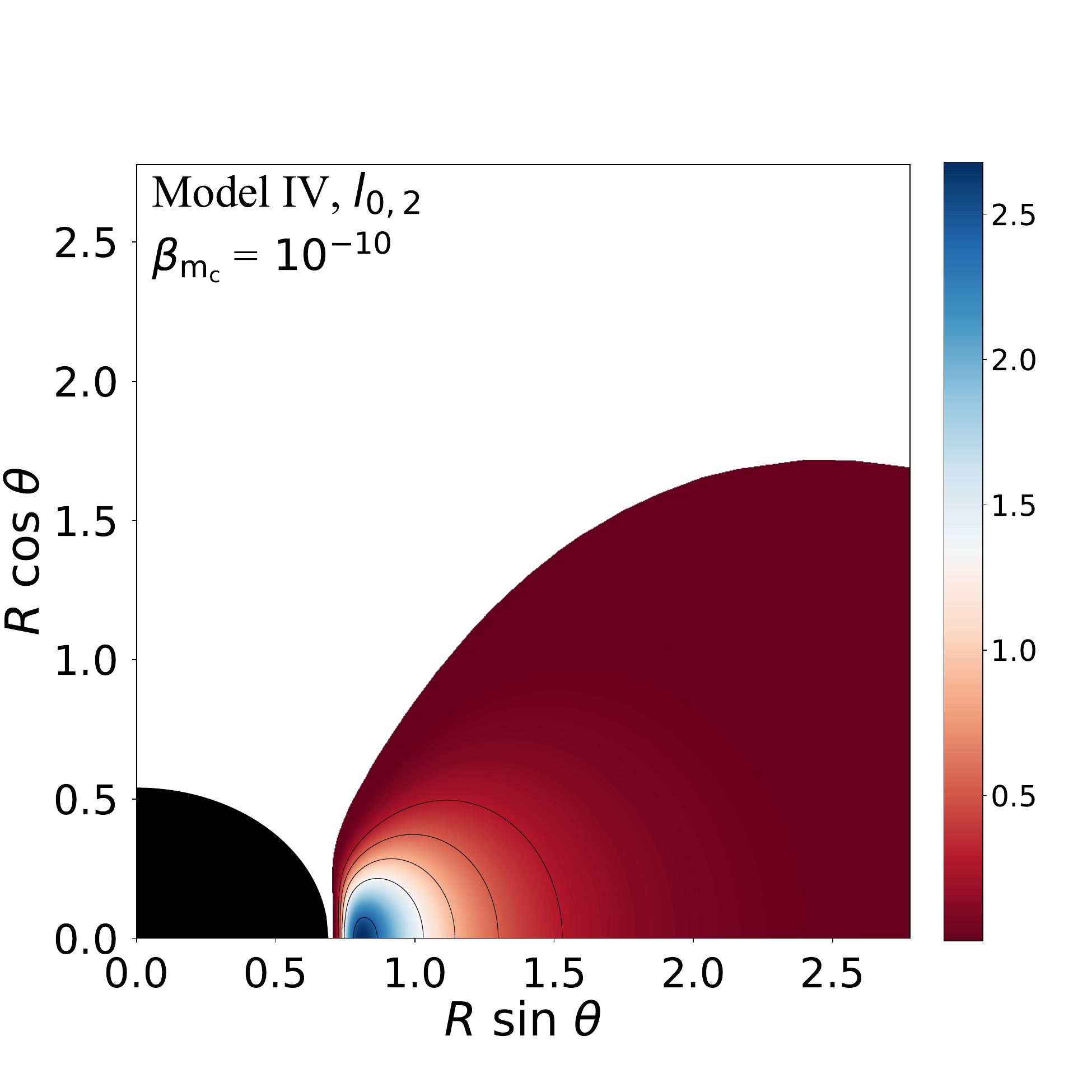}
\\
\vspace{-0.5cm}
\includegraphics[scale=0.17]{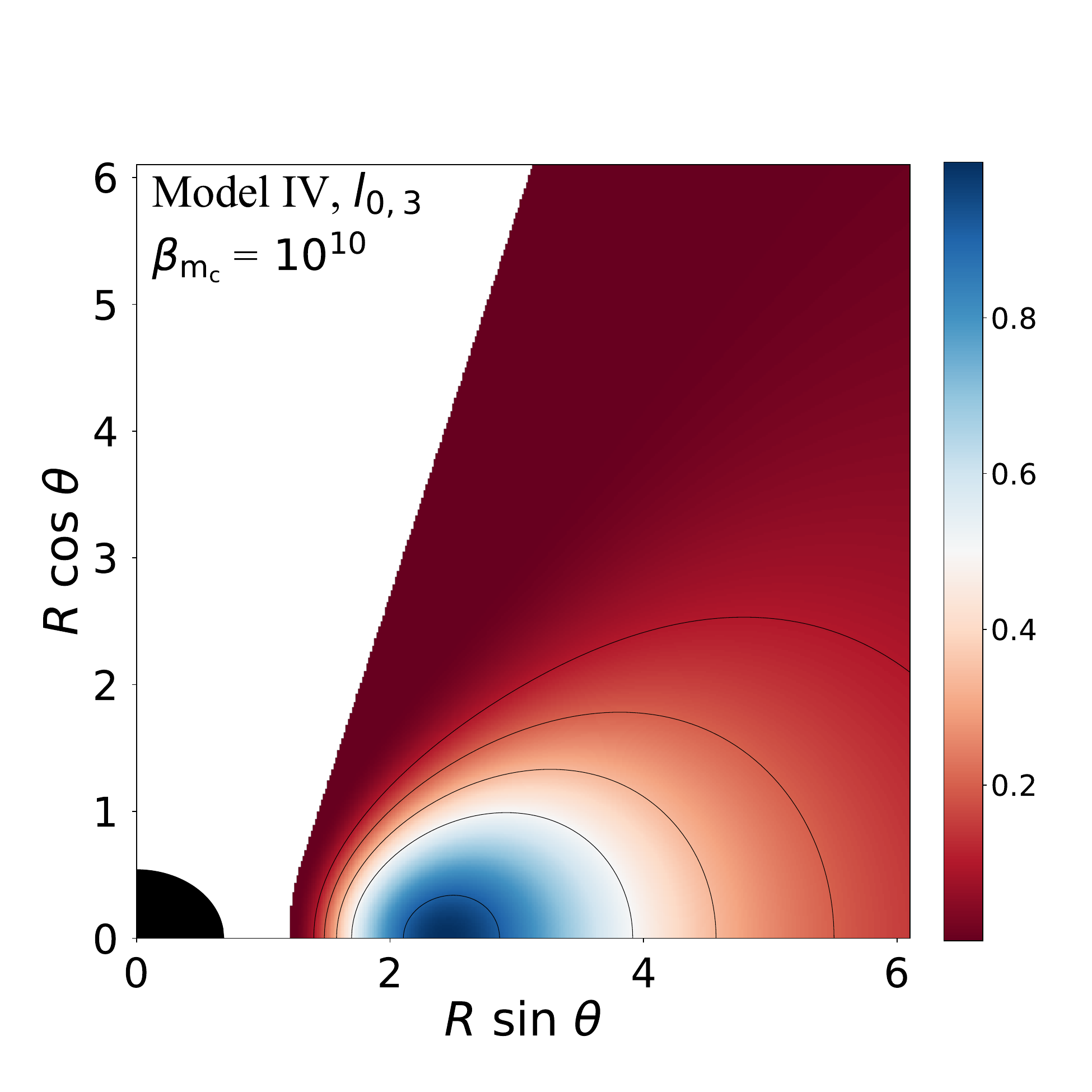}
\hspace{-0.3cm}
\includegraphics[scale=0.17]{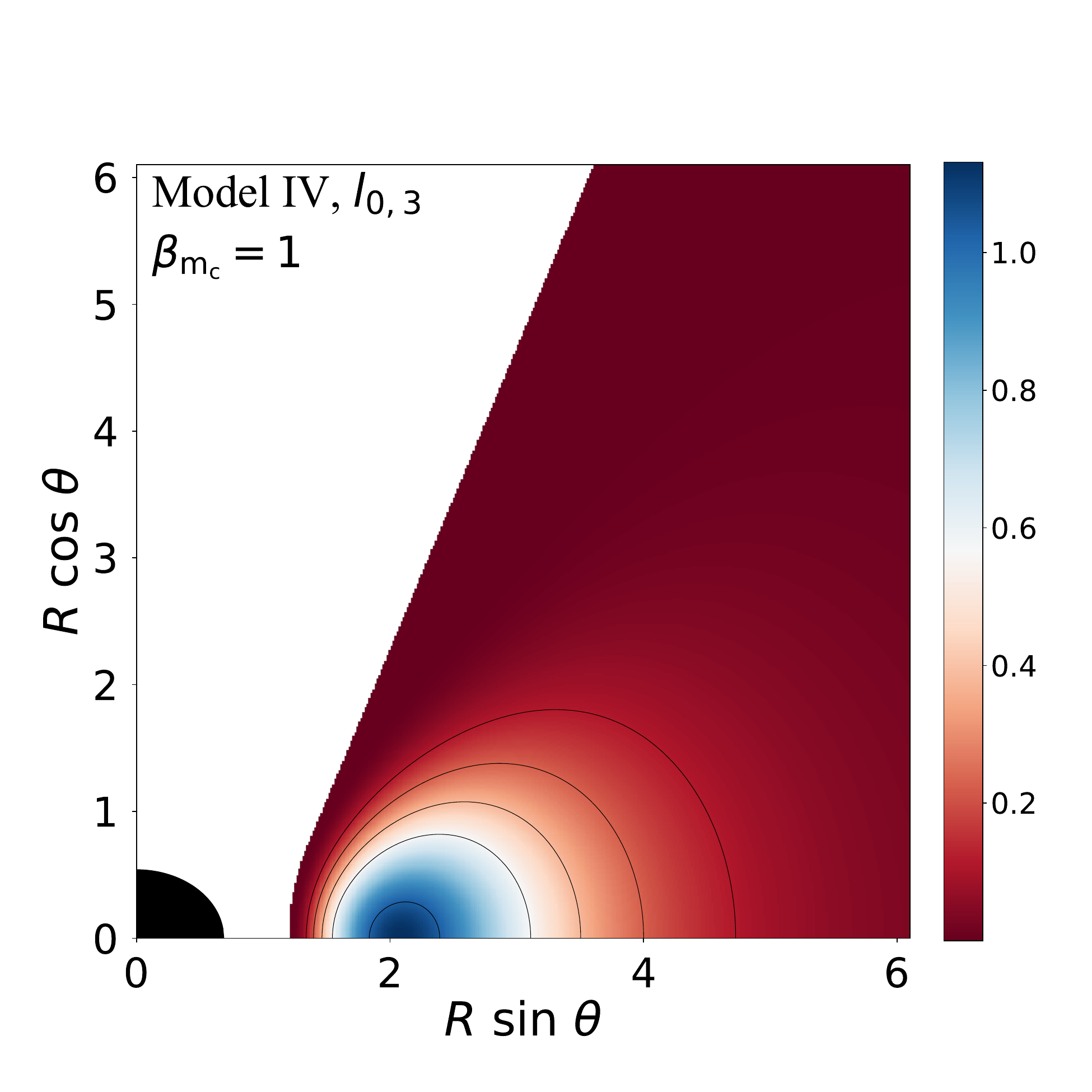}
\hspace{-0.2cm}
\includegraphics[scale=0.17]{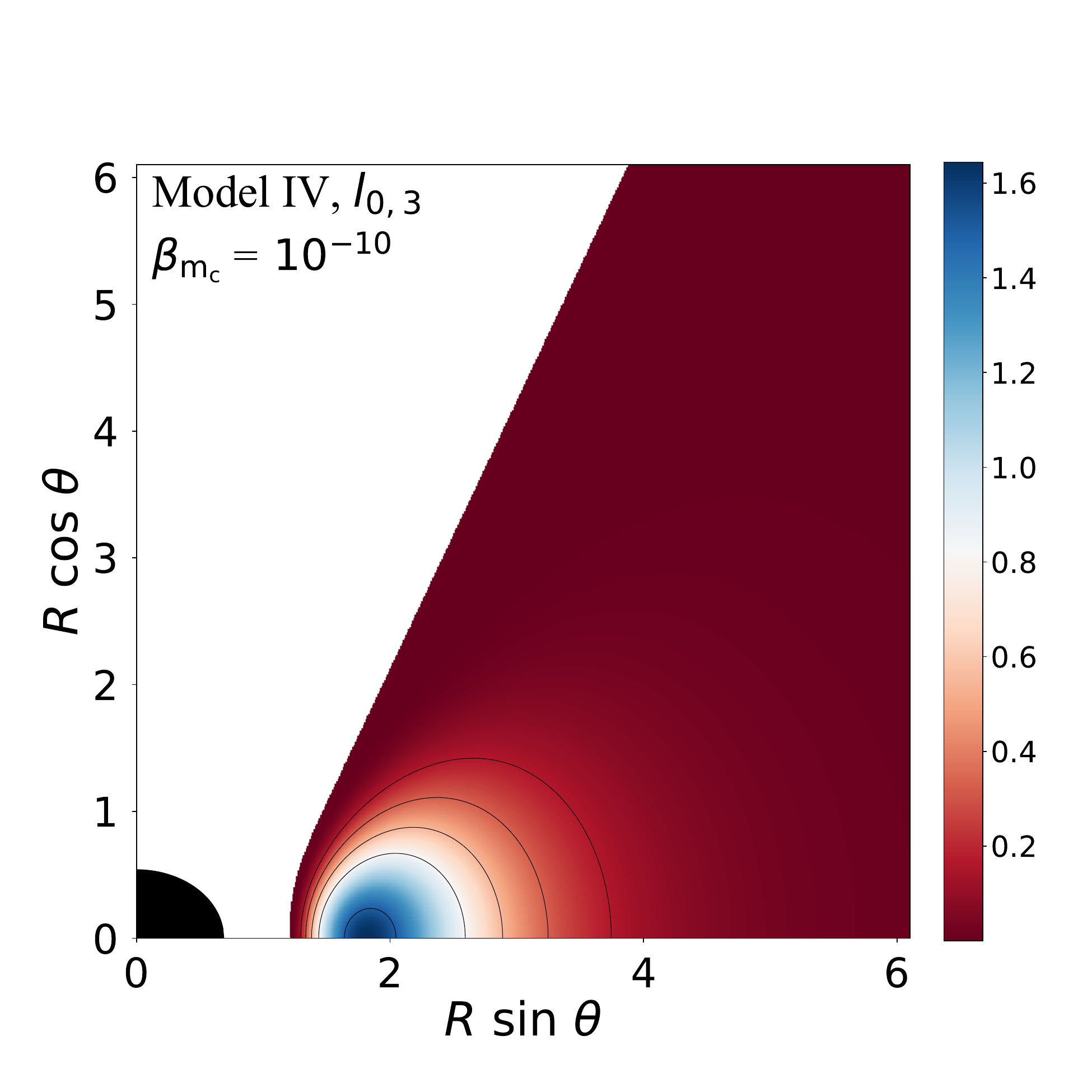}
\caption{Same as Fig.~\ref{Model_I_2Dplots} but for spacetime IV.}
\label{Model_IV_2Dplots}
\end{figure*}	 

\begin{figure*}[t]
\includegraphics[scale=0.17]{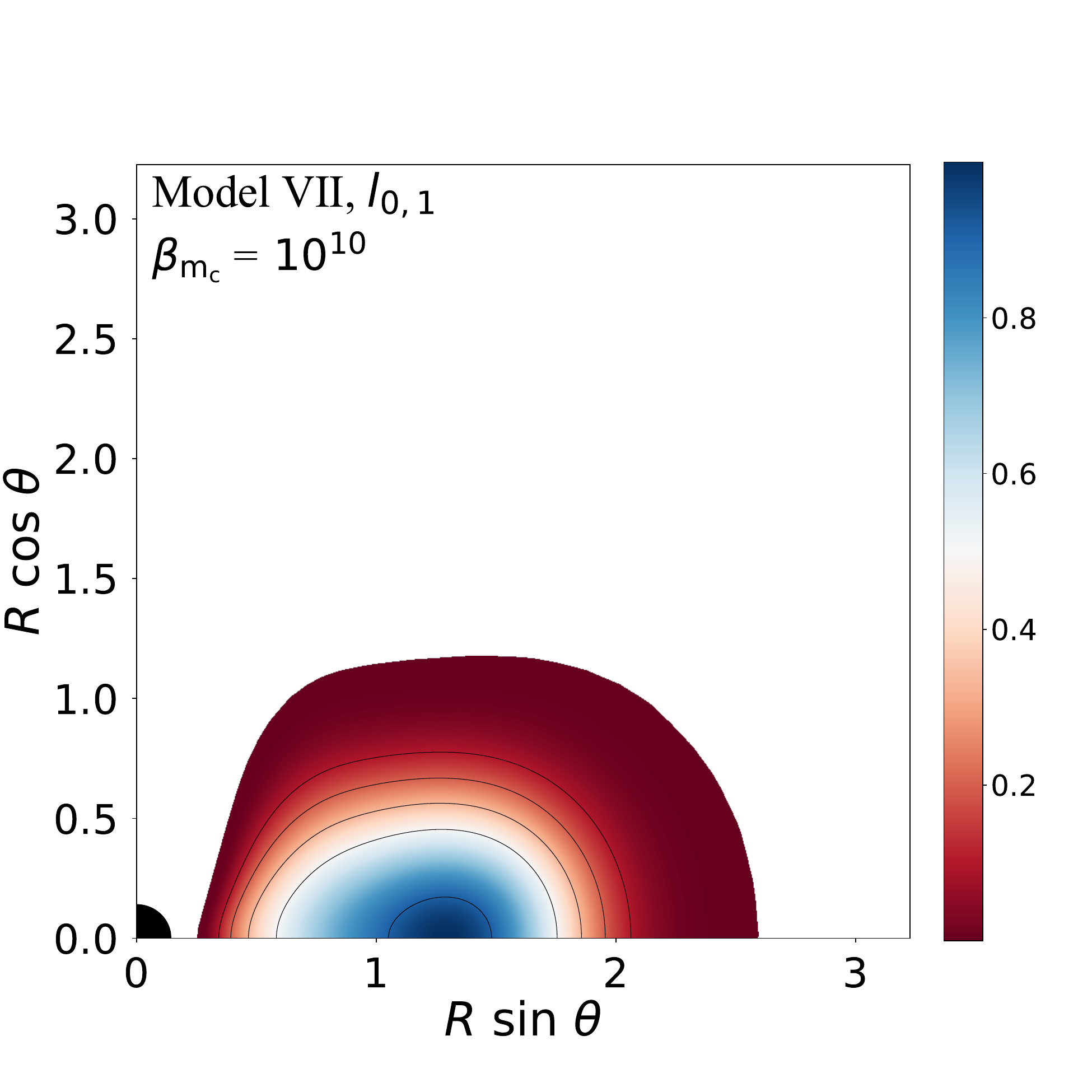}
\hspace{-0.3cm}
\includegraphics[scale=0.17]{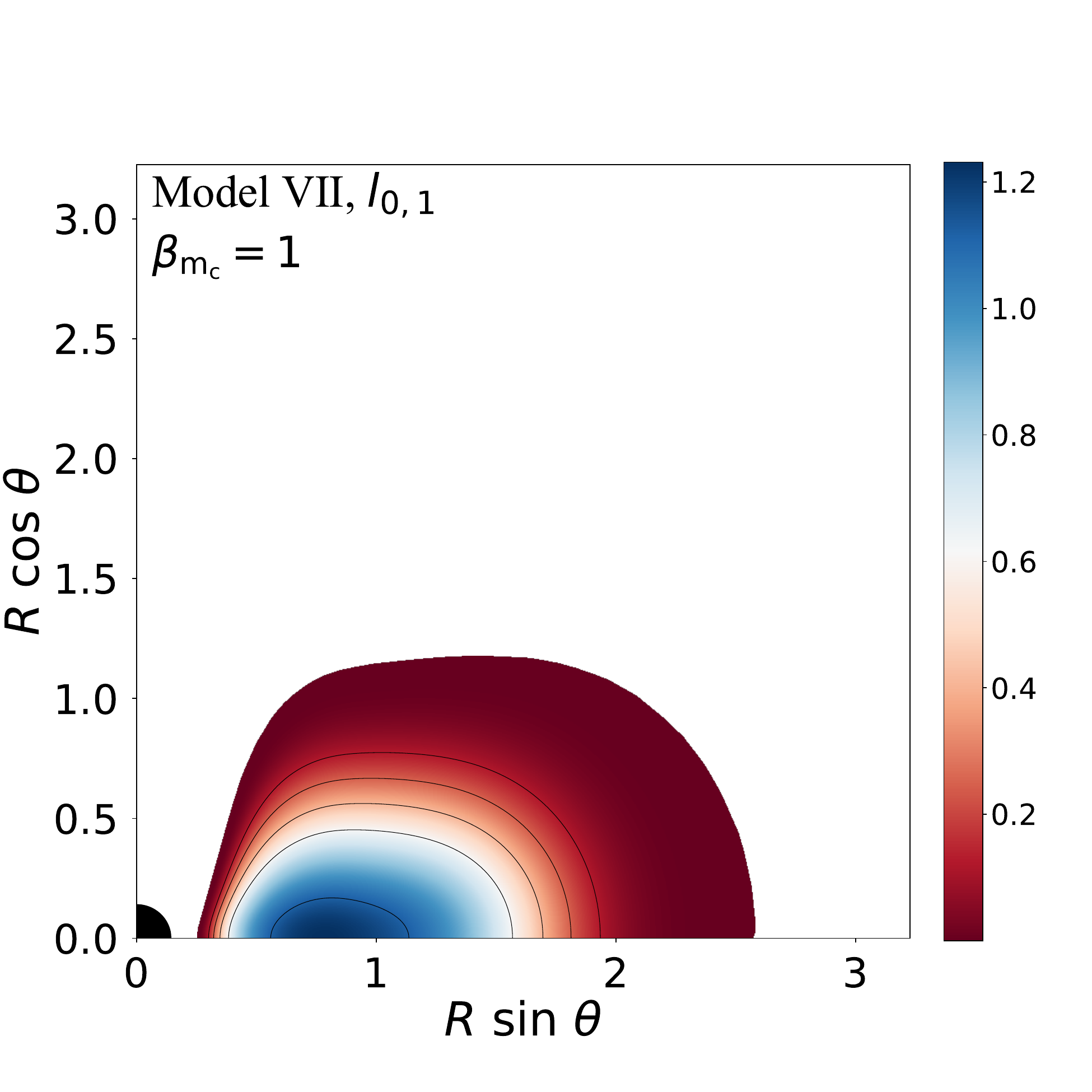}
\hspace{-0.2cm}
\includegraphics[scale=0.17]{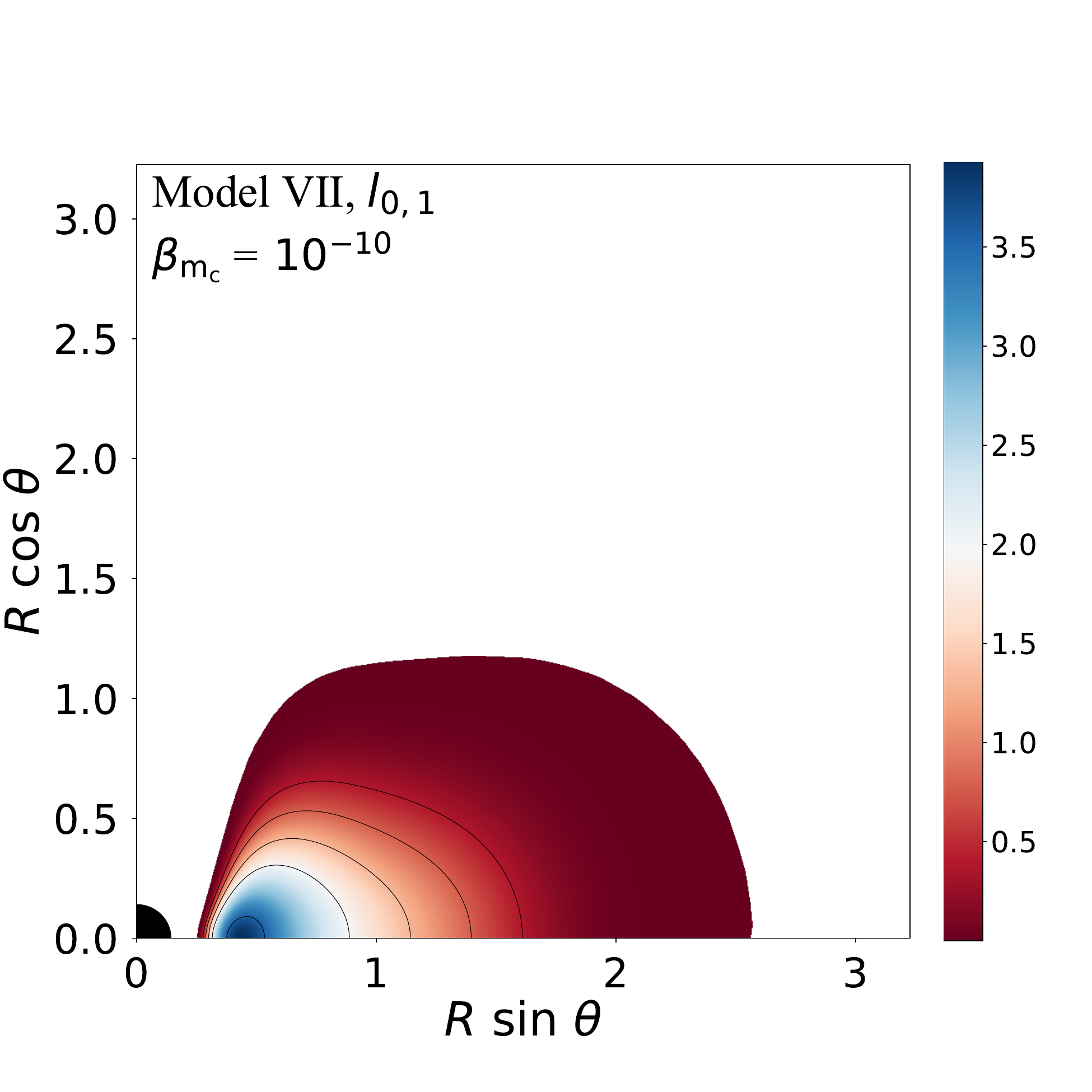}
\\
\vspace{-0.5cm}
\includegraphics[scale=0.17]{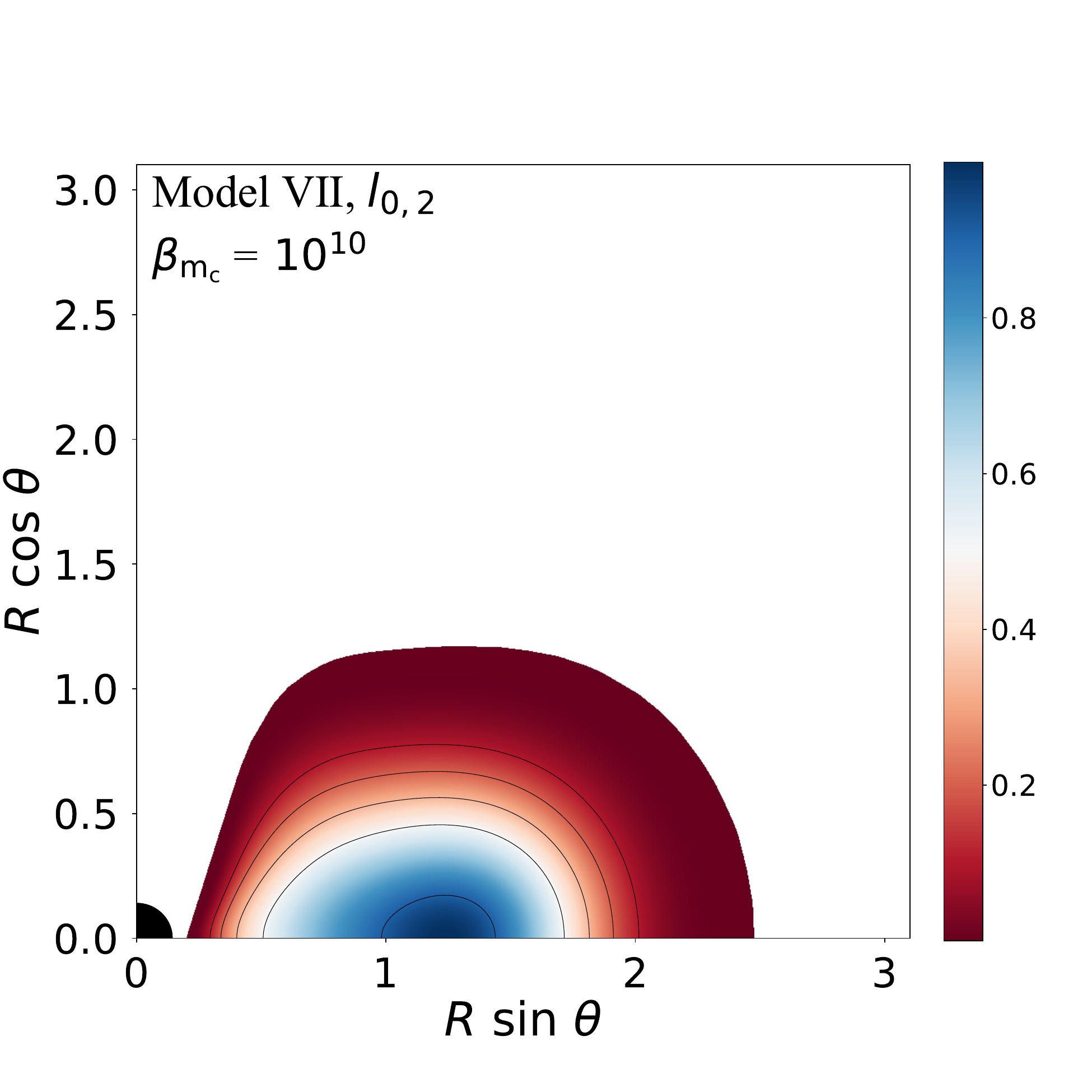}
\hspace{-0.3cm}
\includegraphics[scale=0.17]{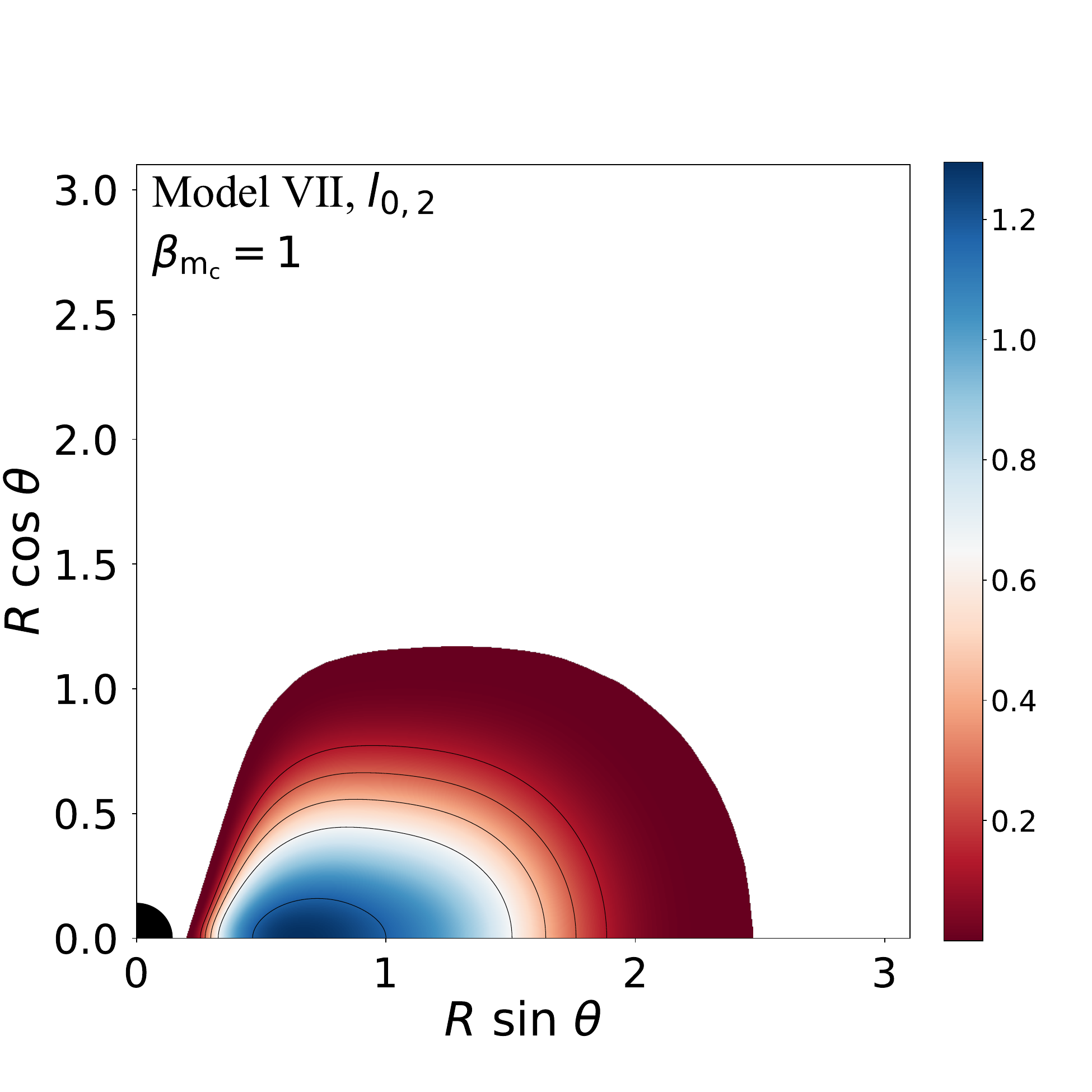}
\hspace{-0.2cm}
\includegraphics[scale=0.17]{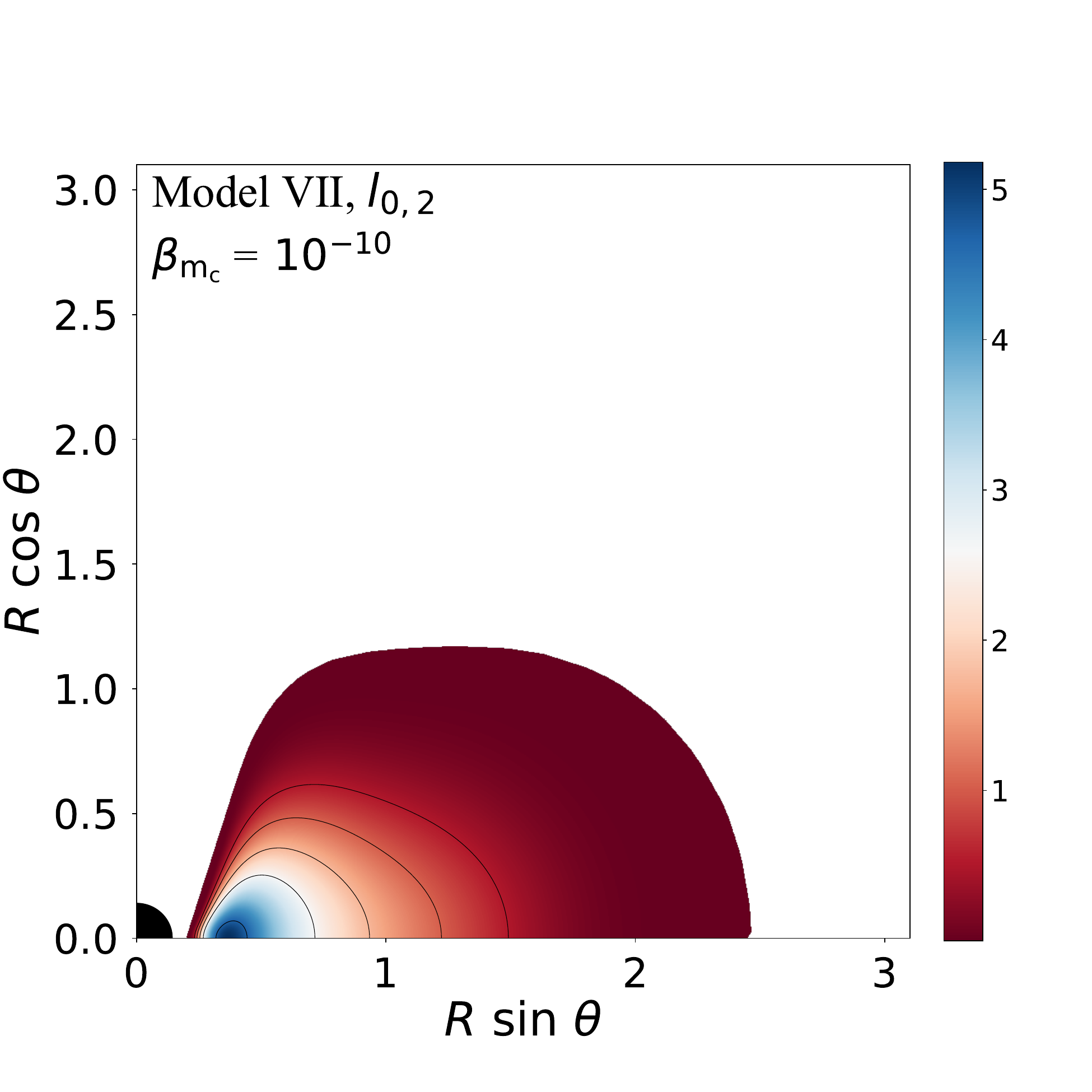}
\\
\vspace{-0.5cm}
\includegraphics[scale=0.17]{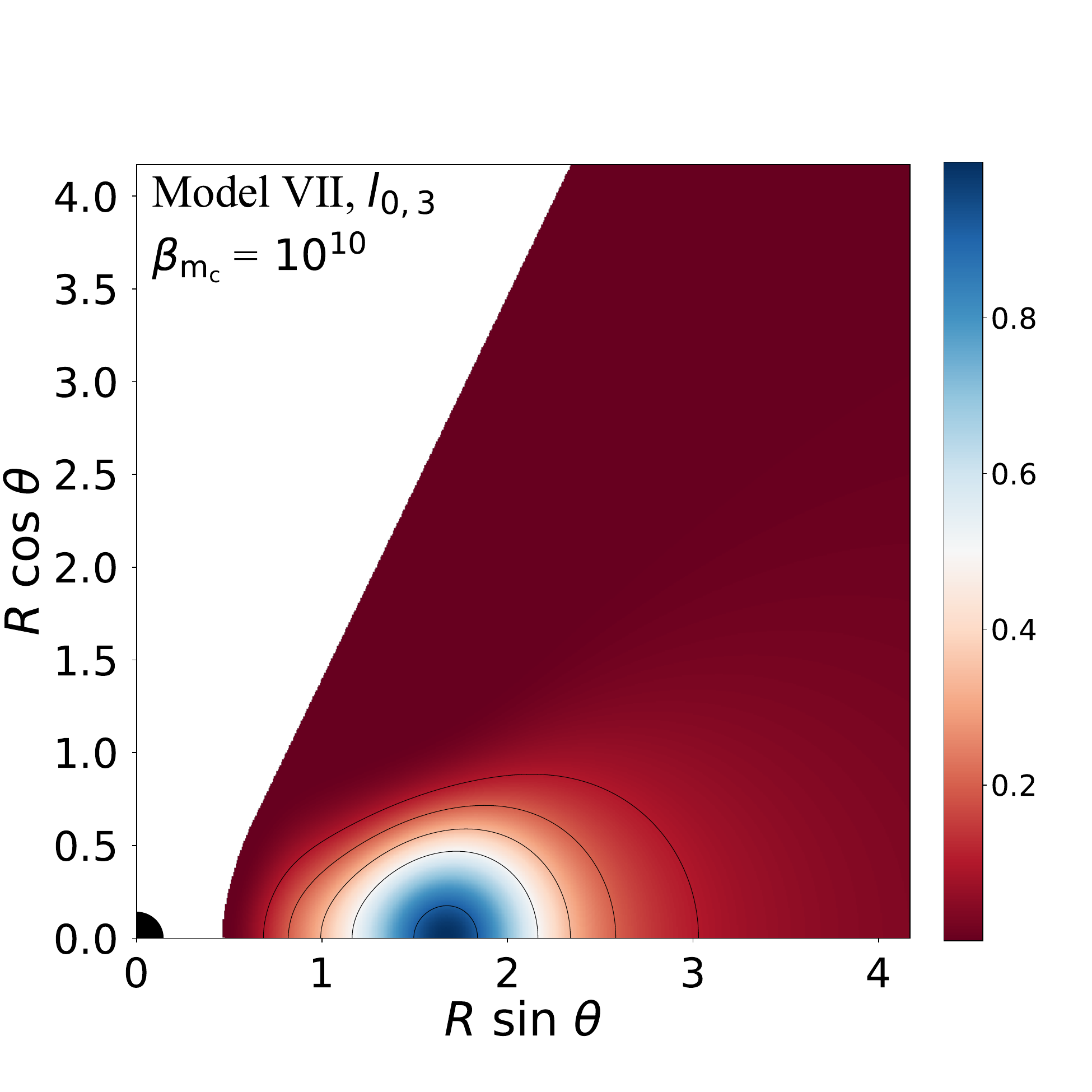}
\hspace{-0.3cm}
\includegraphics[scale=0.17]{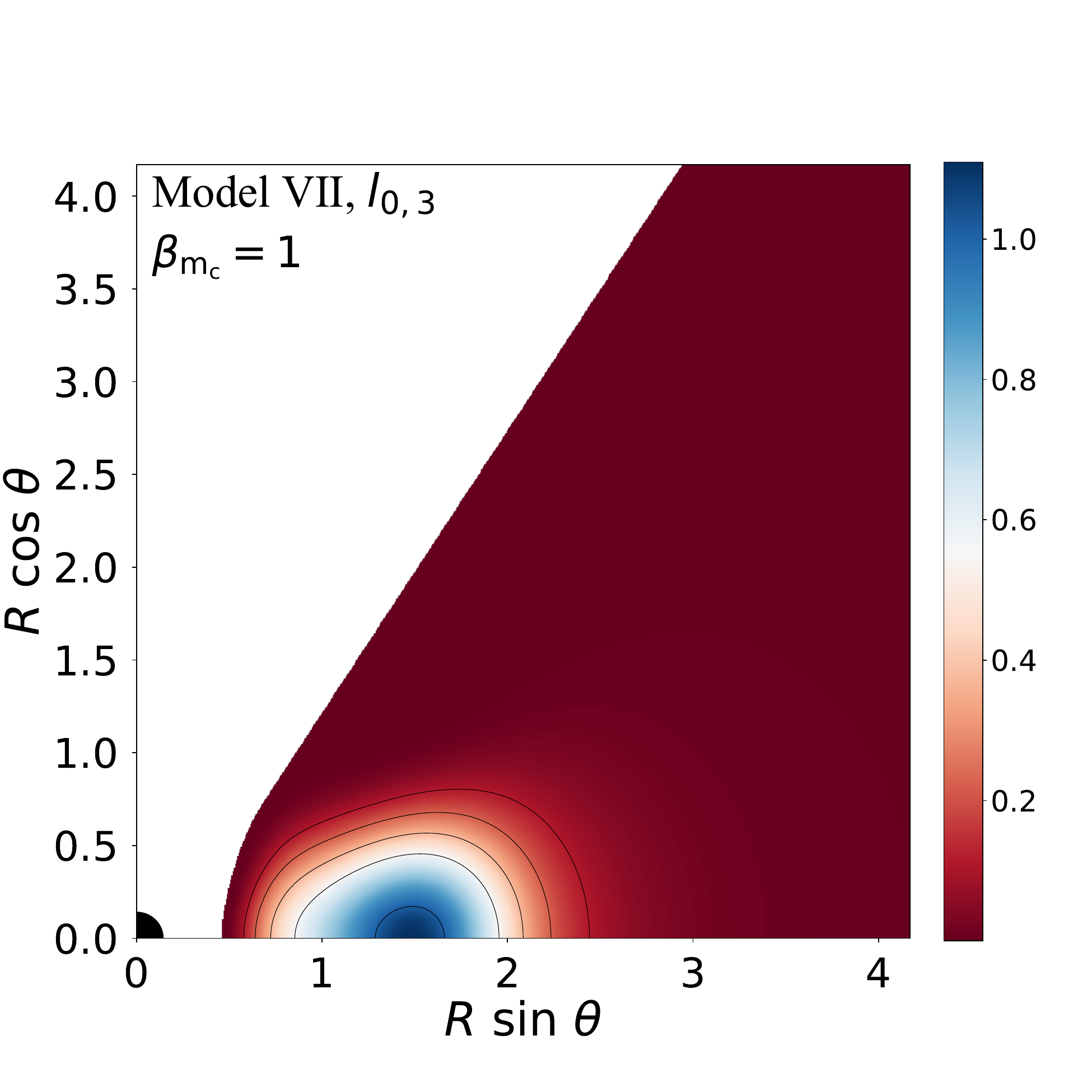}
\hspace{-0.2cm}
\includegraphics[scale=0.17]{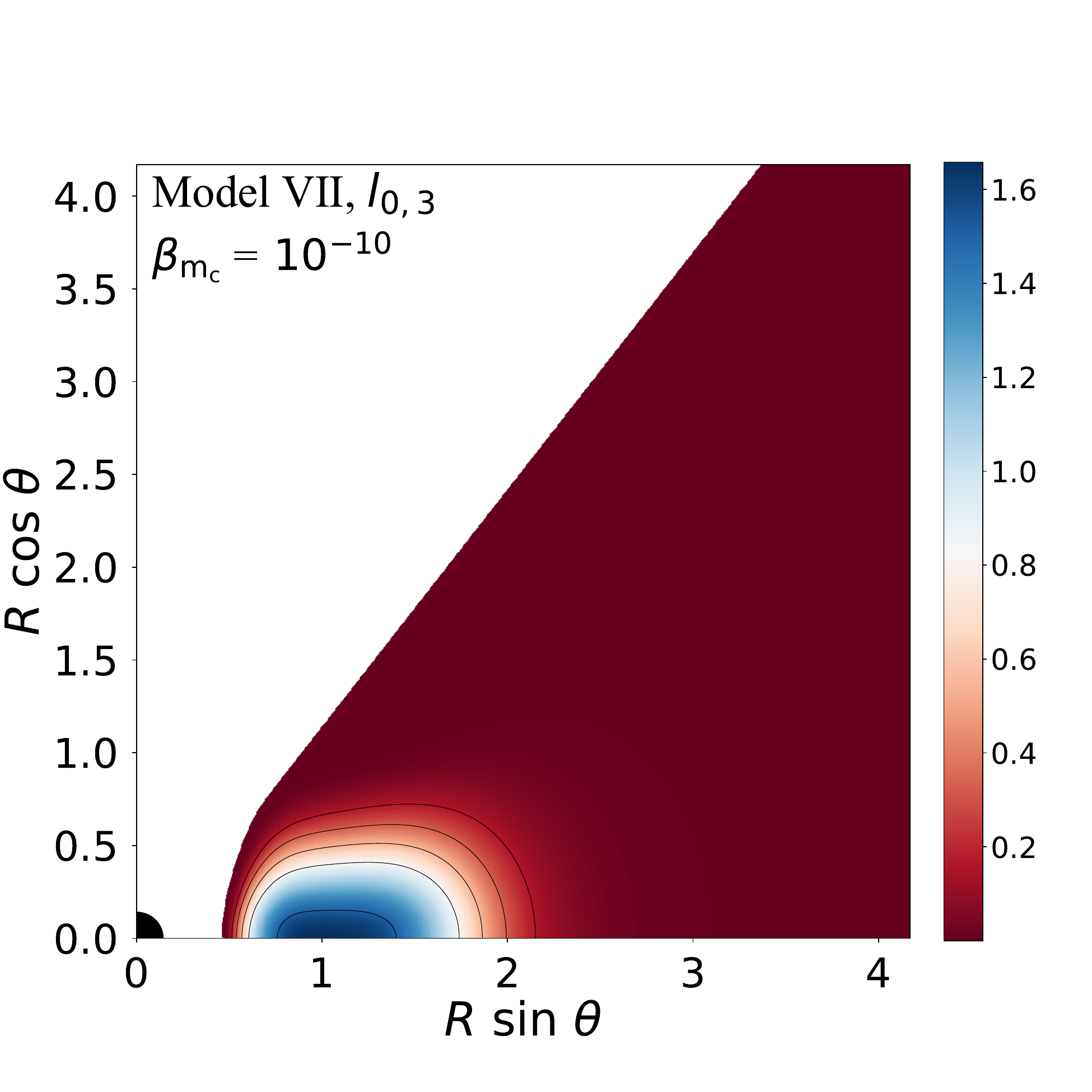}
\caption{Same as Fig.~\ref{Model_I_2Dplots} but for spacetime VII}
\label{Model_VII_2Dplots}
\end{figure*}

\begin{figure*}[t]
\includegraphics[scale=0.14]{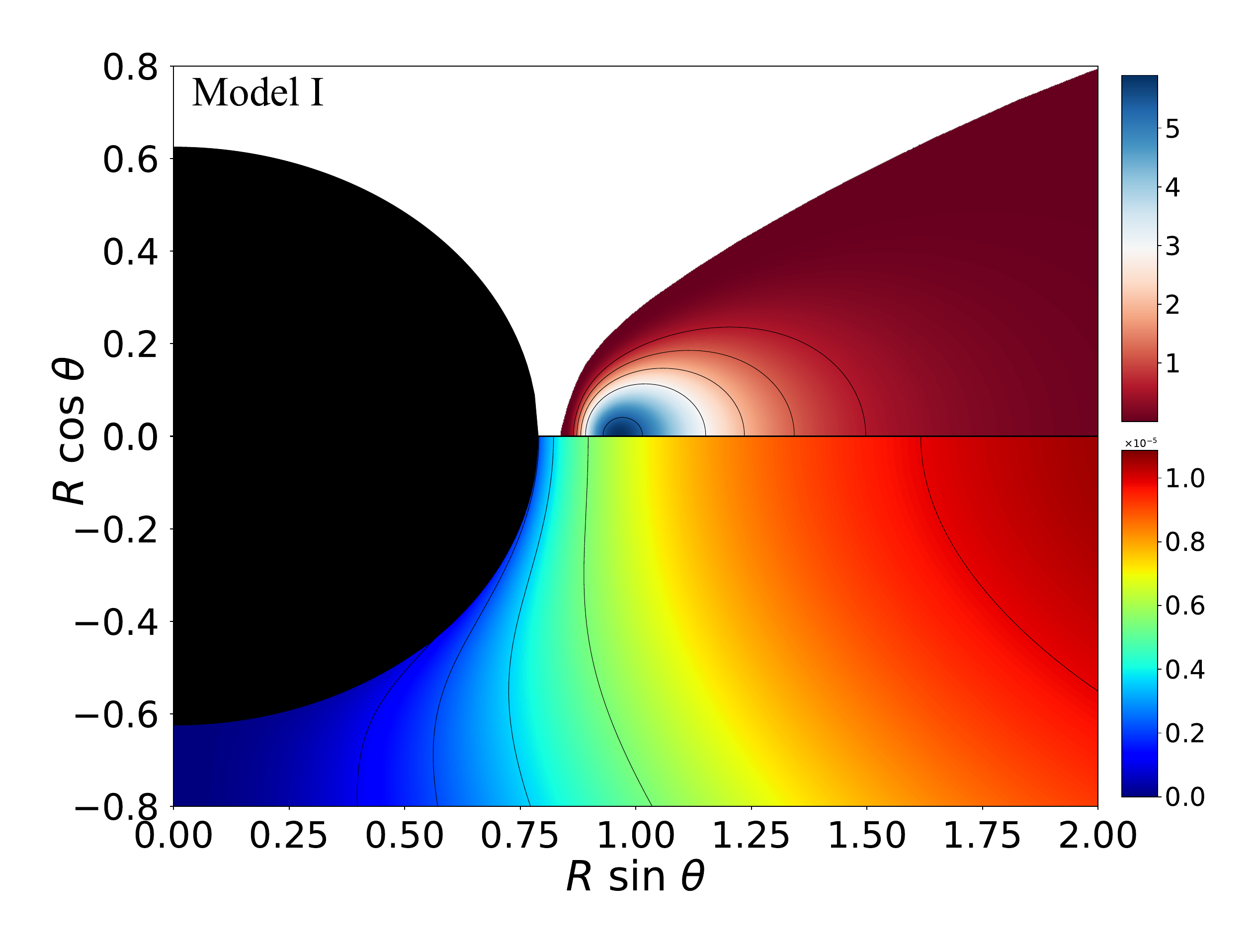}
\hspace{-0.45cm}
\includegraphics[scale=0.14]{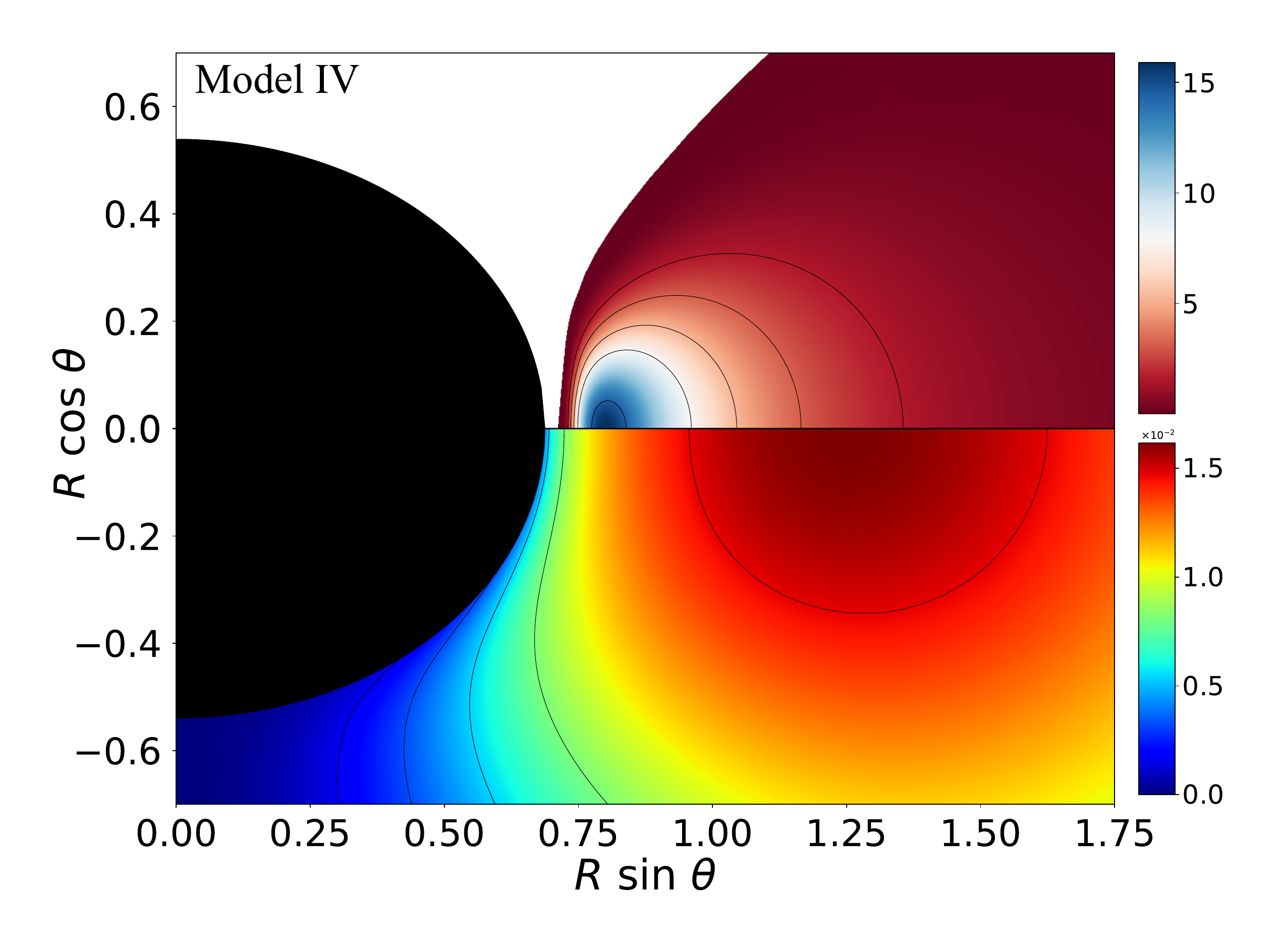}
\hspace{-0.45cm}
\includegraphics[scale=0.14]{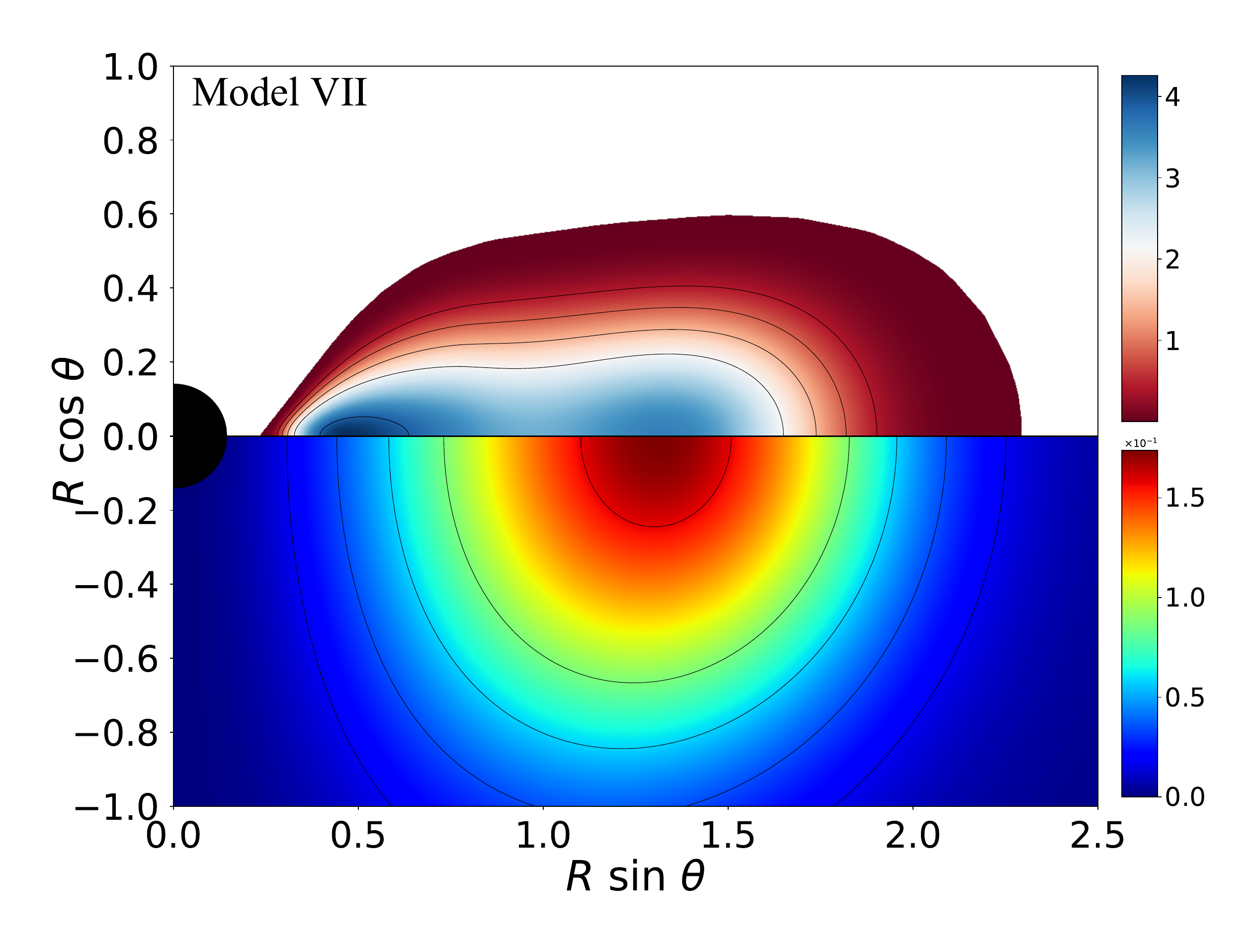}
\\
\vspace{-0.1cm}
\includegraphics[scale=0.14]{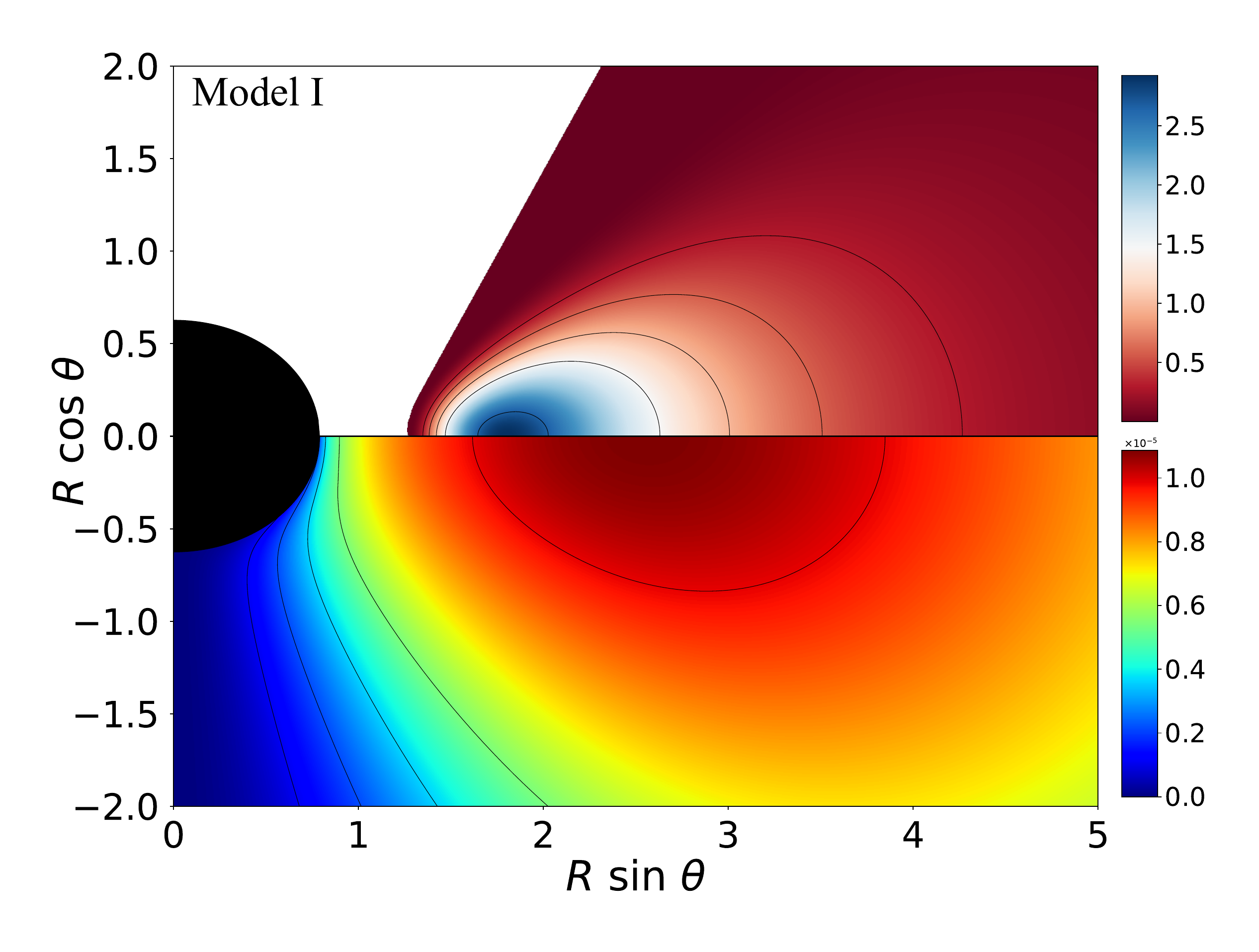}
\hspace{-0.45cm}
\includegraphics[scale=0.14]{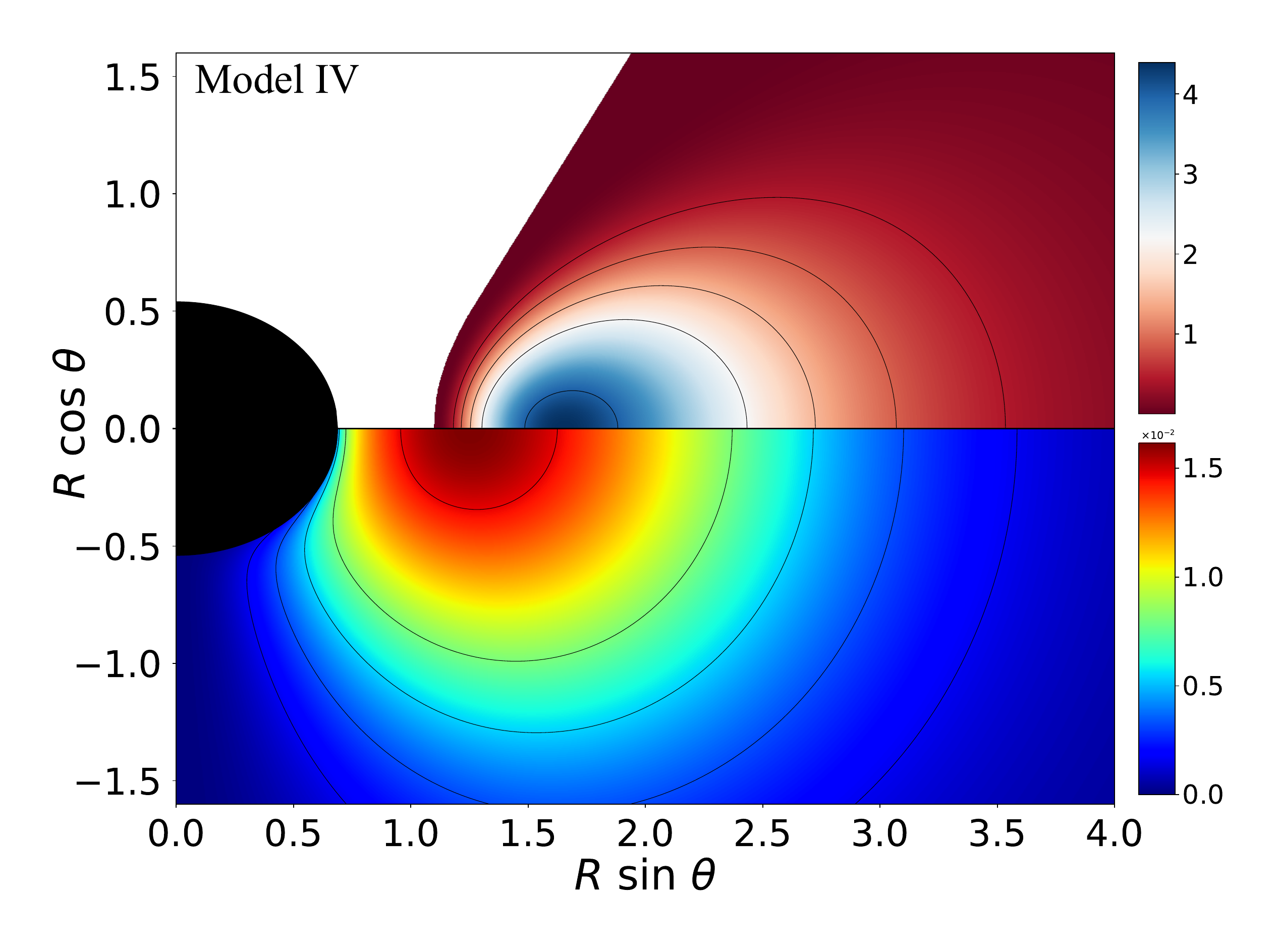}
\hspace{-0.45cm}
\includegraphics[scale=0.145]{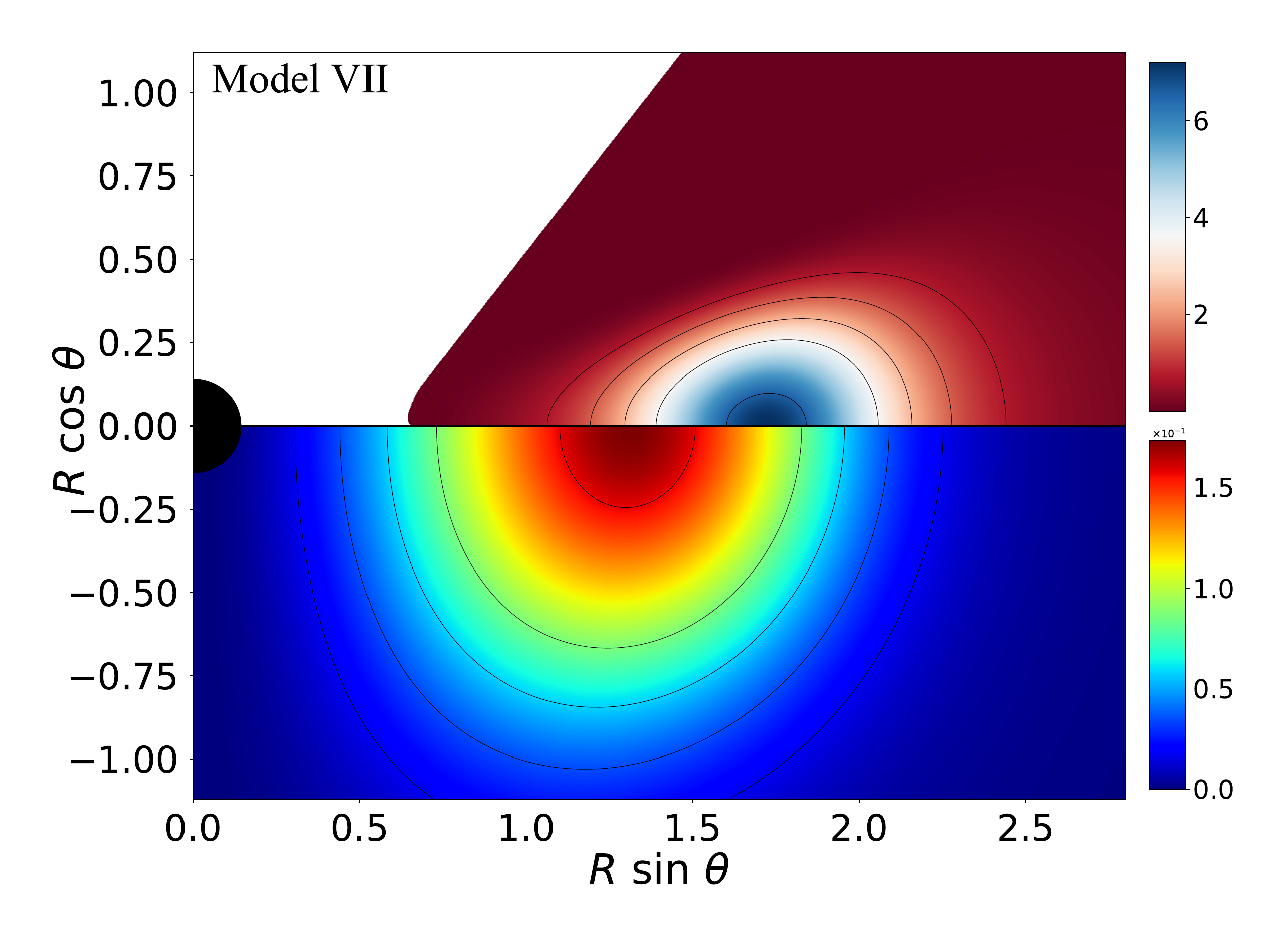}
\caption{Distribution of the gravitational energy density of the matter $\rho_{\mathrm{T}}$ (top half of each panel) and of the scalar field $\rho_{\mathrm{SF}}$ (bottom half of each panel) for $\alpha = 0.75$, $\beta_{\mathrm{m,c}} = 10^{-10}$, and $l_0 = l_{0, 2}$ (top row) and $l_0 = l_{0, 3}$ (bottom row). Spacetimes I, IV and VII are shown in the left, middle and right columns, respectively. Note that the spatial scale is not the same for all plots.}
\label{gravitational_energy_density_plots}
\end{figure*}

\begin{figure*}[t]
\includegraphics[scale=0.11]{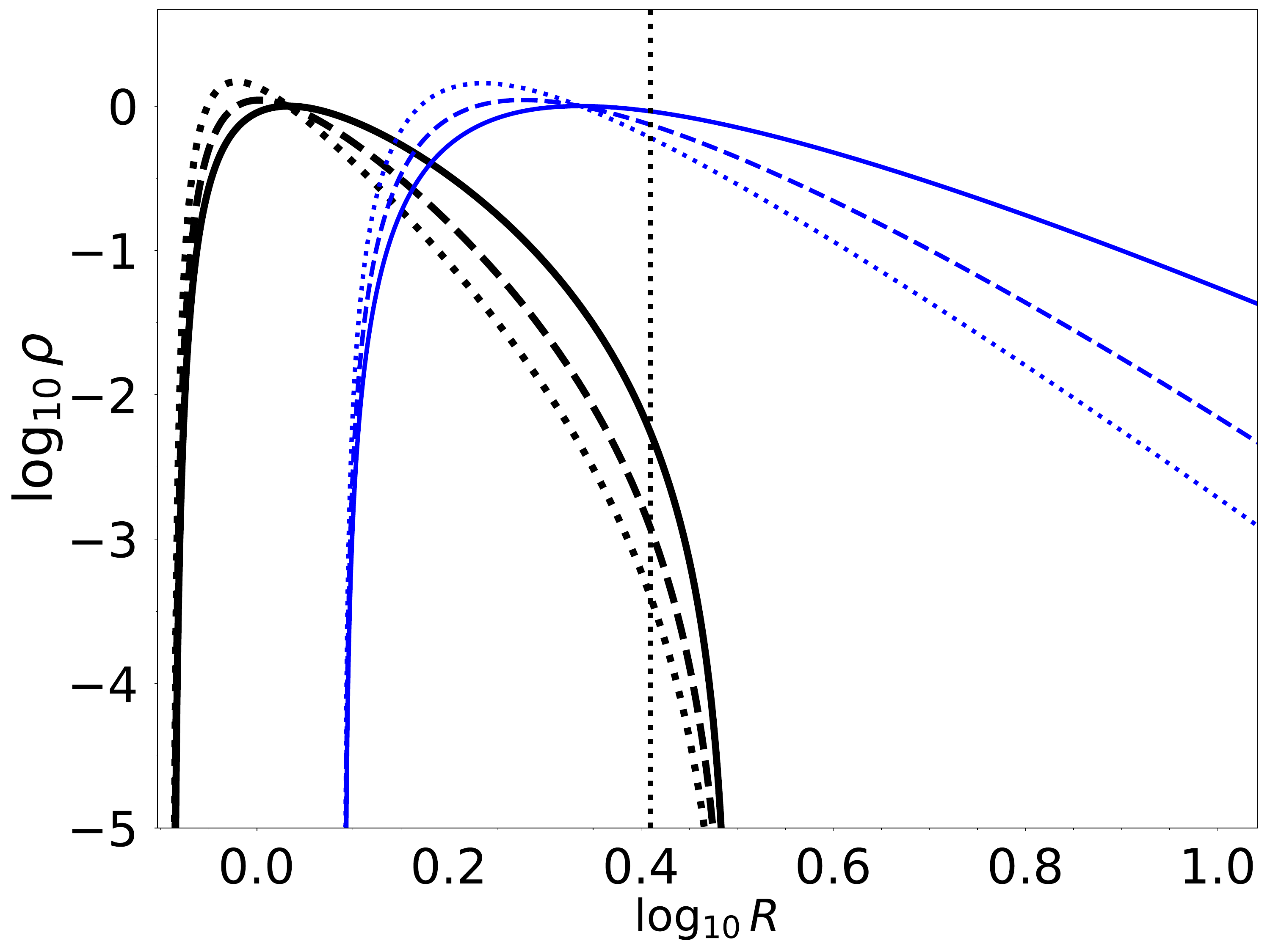}
\hspace{-0.1cm}
\includegraphics[scale=0.11]{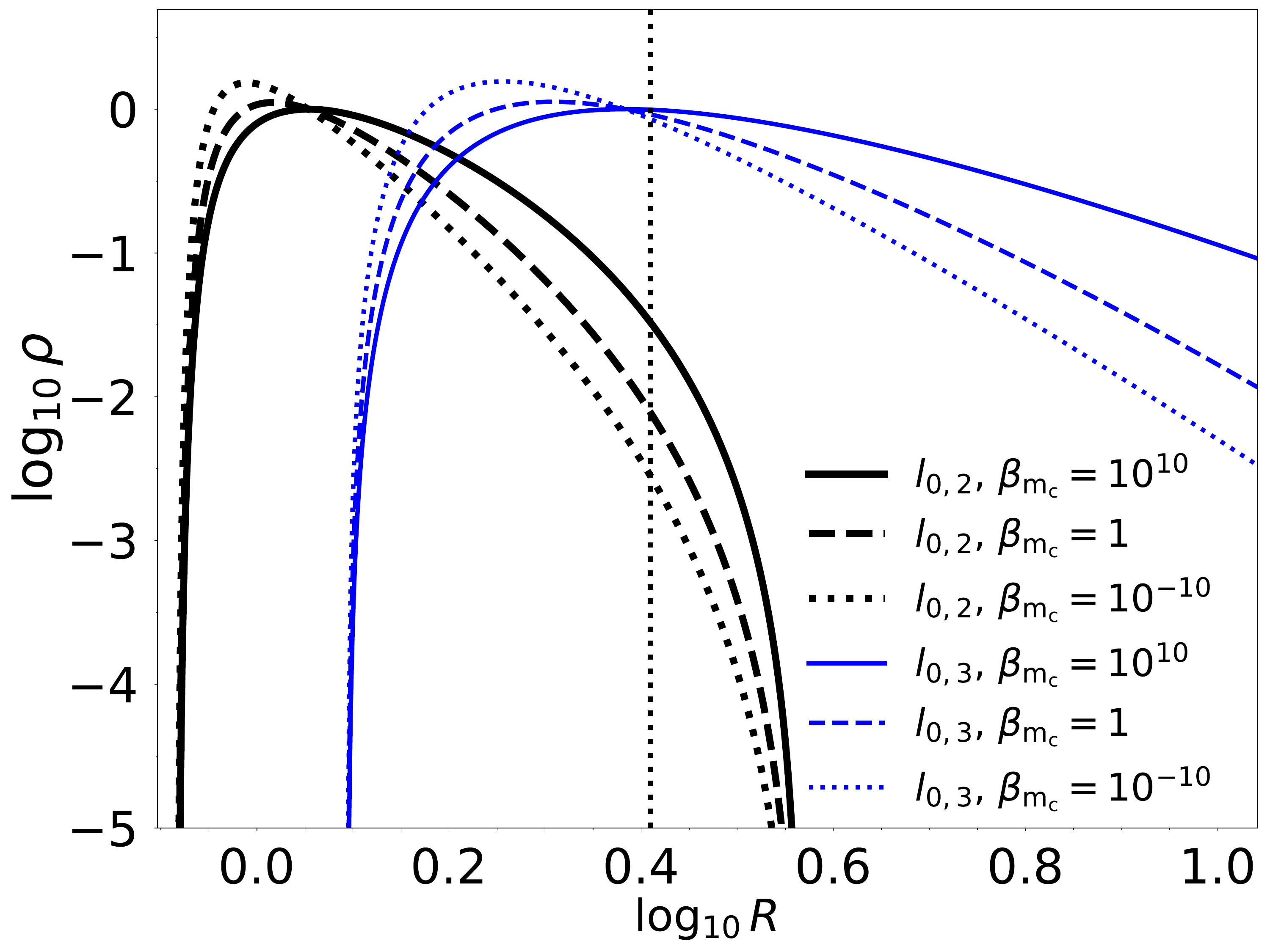}
\hspace{-0.1cm}
\includegraphics[scale=0.11]{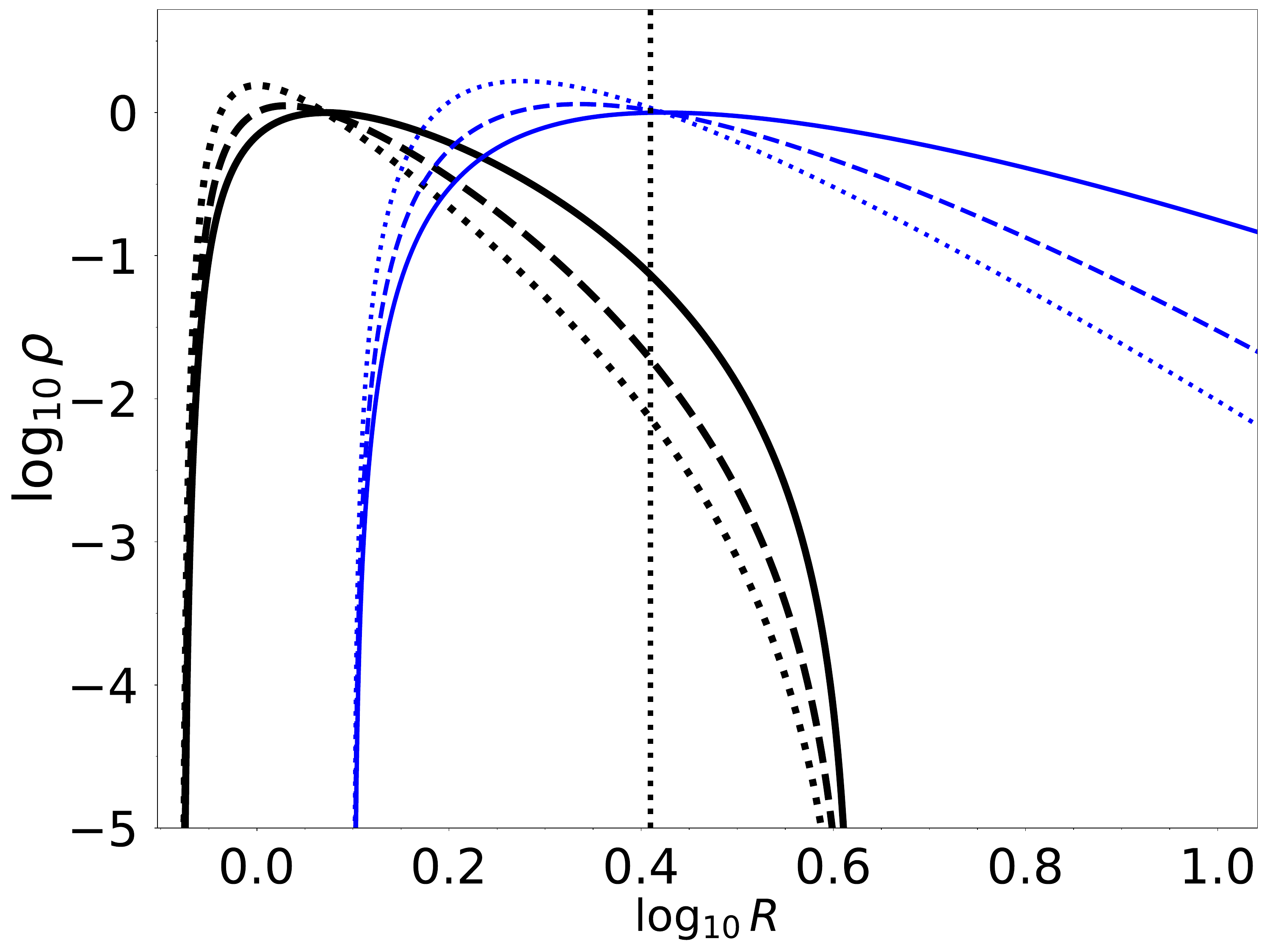}
\\
\vspace{0.0cm}
\includegraphics[scale=0.11]{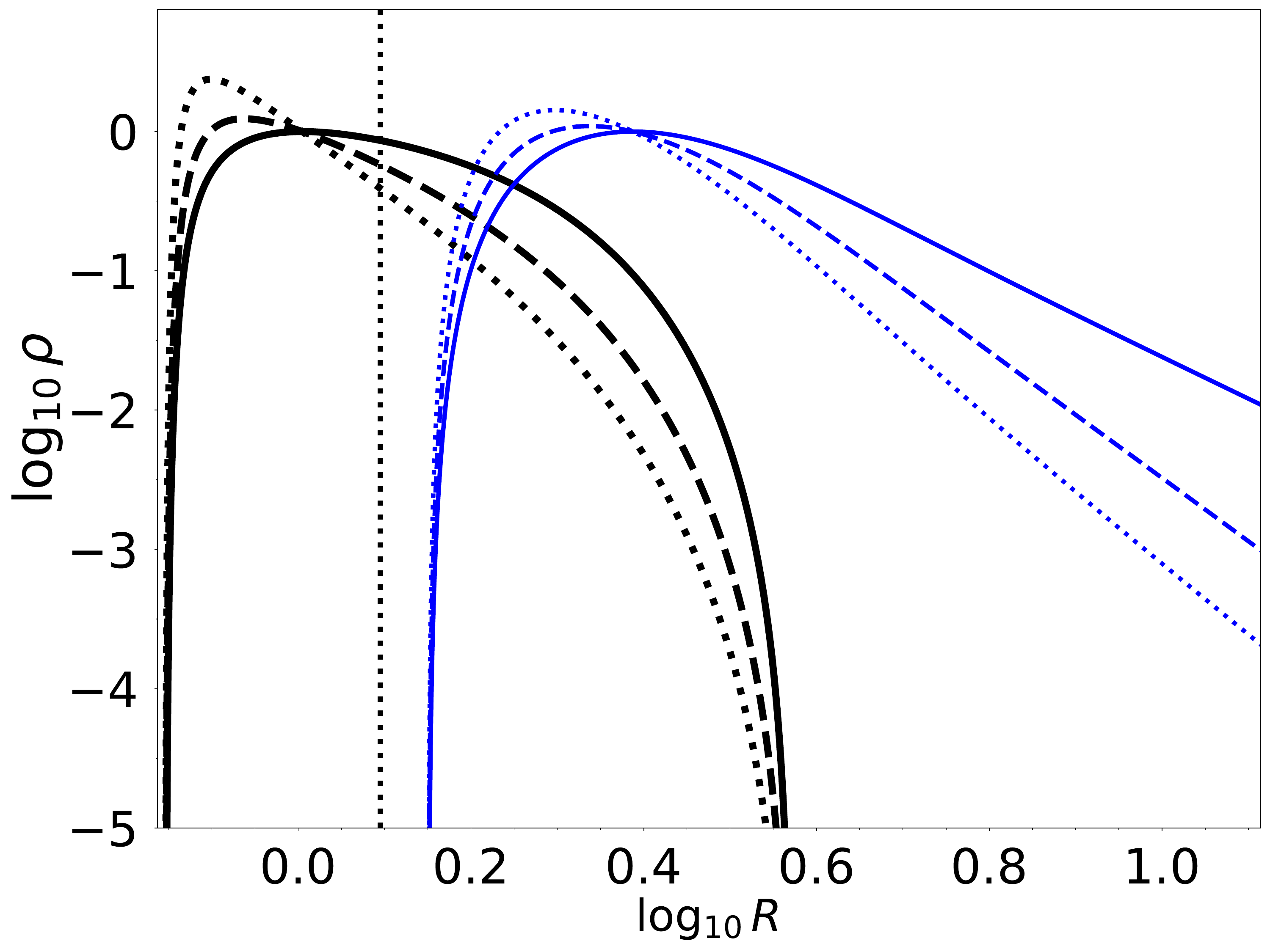}
\hspace{-0.1cm}
\includegraphics[scale=0.11]{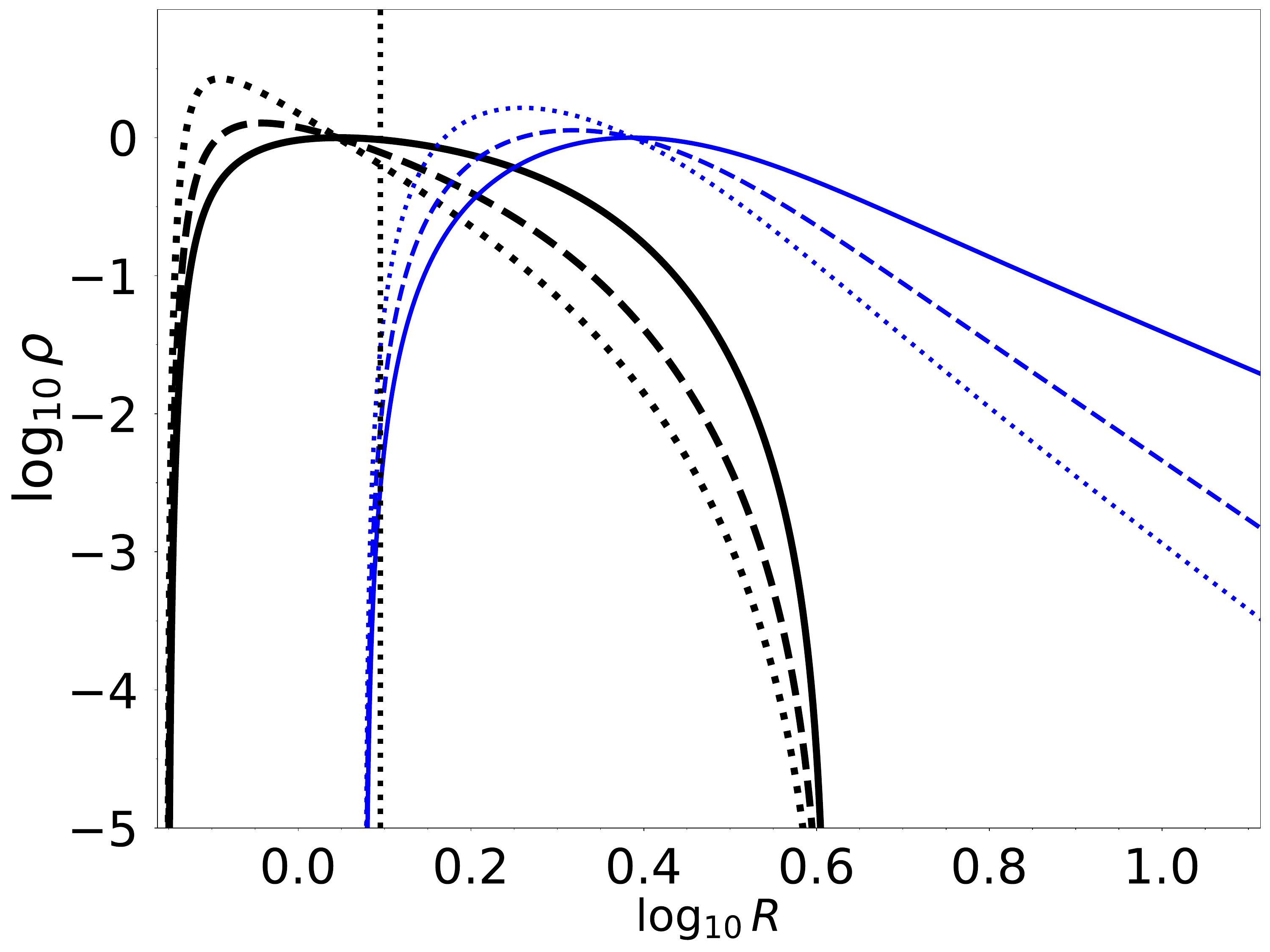}
\hspace{-0.1cm}
\includegraphics[scale=0.11]{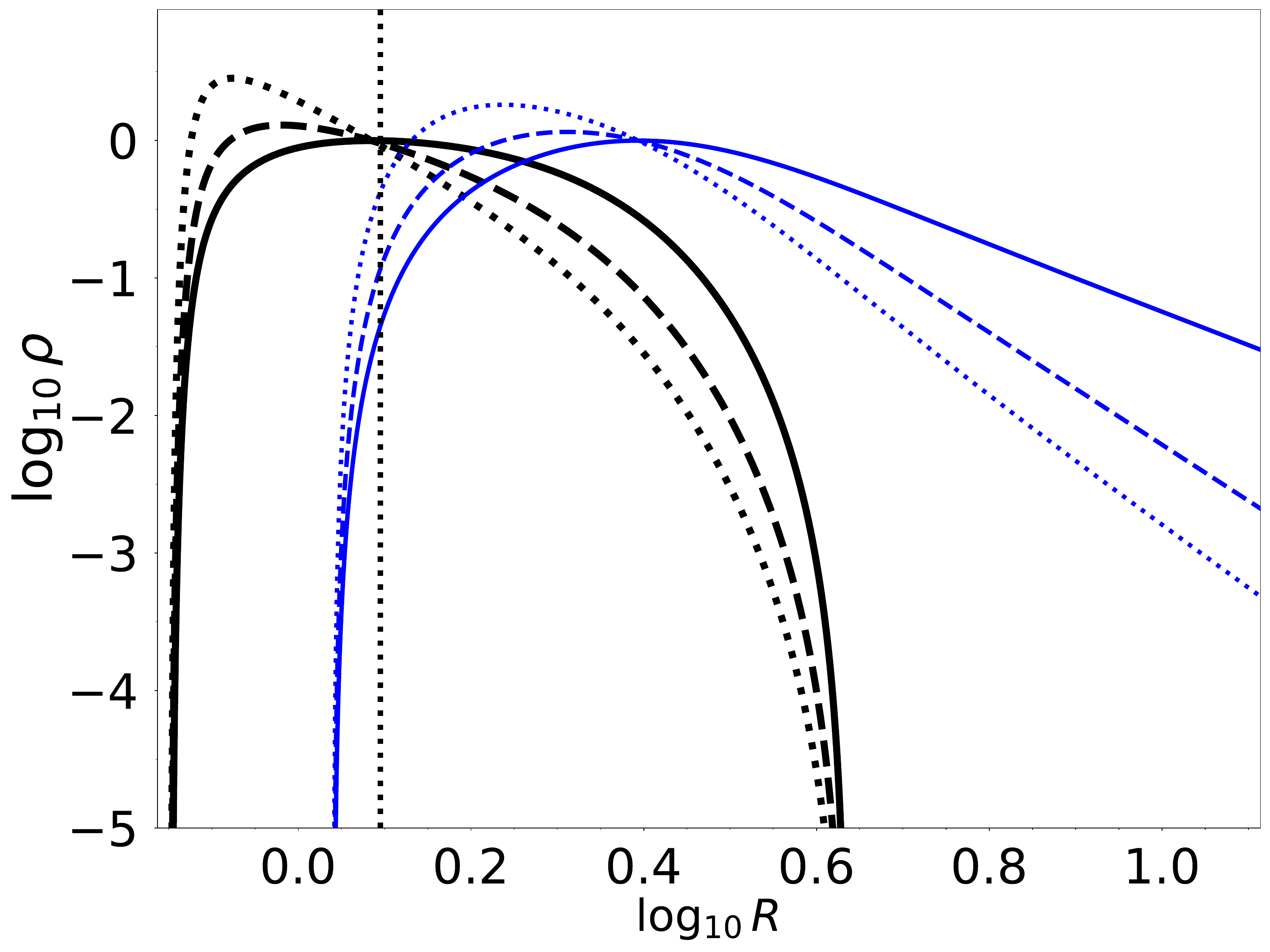}
\\
\vspace{0.0cm}
\includegraphics[scale=0.11]{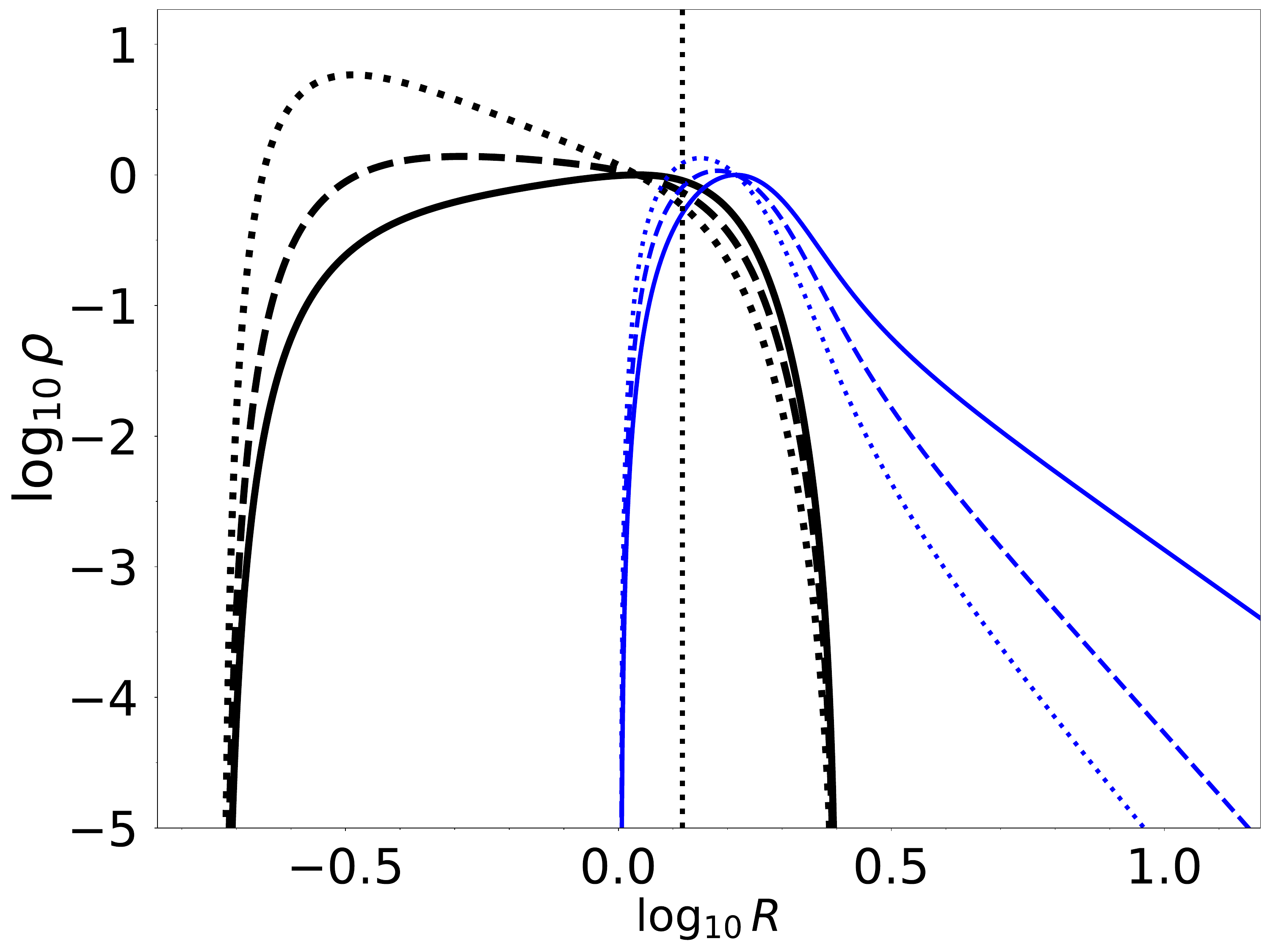}
\hspace{-0.1cm}
\includegraphics[scale=0.11]{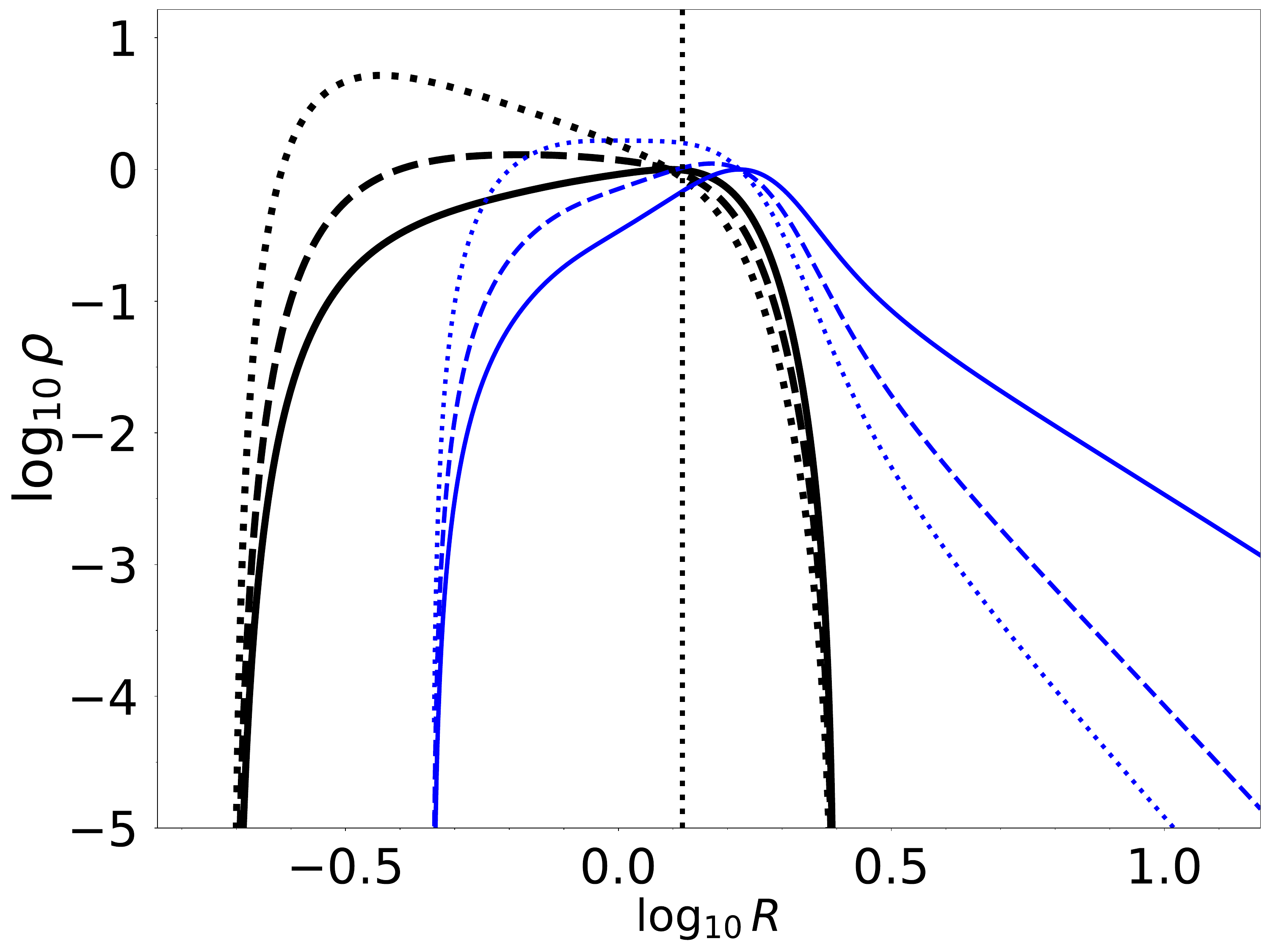}
\hspace{-0.1cm}
\includegraphics[scale=0.11]{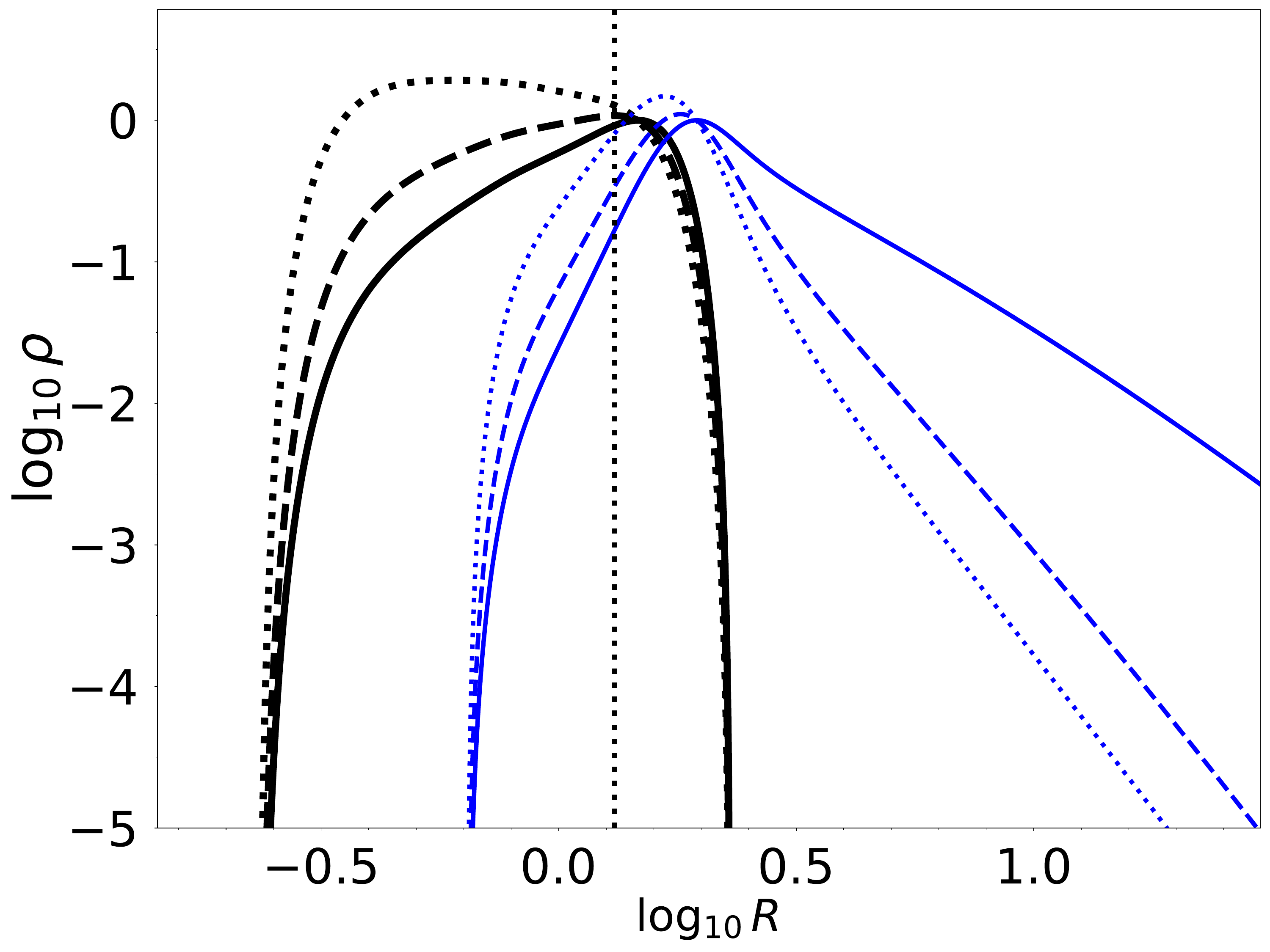}
\caption{Radial profiles of the logarithm of the rest-mass density at the equatorial plane. From top to bottom, the rows correspond to the three different spacetimes we are considering, namely Models, I, IV and VII. From left to right, the columns correspond to different values of the exponent of the angular momentum distribution $\alpha$, namely $0$, $0.5$ and $0.75$. In each panel, black and blue curves correspond to models with the constant part of the angular momentum distribution $l_0$ computed following the criterion 2 and 3, respectively. The solid, dashed and dotted lines correspond to different values of the magnetization parameter at the center of the disk $\beta_{\mathrm{m, c}}$, namely $10^{10}$, $1$, $10^{-10}$. (See  legend in the top-central panel.) The vertical black dotted lines denote the location of the maximum of the gravitational energy density of the scalar field $\rho_{\mathrm{SF, max}}$.}
\label{radial_log_density_plots}
\end{figure*}

\subsubsection{Angular momentum distribution outside the equatorial plane - von Zeipel's cylinders}

To obtain the specific angular momentum outside the equatorial plane ($\theta \neq \pi/2$) we take the same approach as in~\cite{Daigne:2004}. This approach  considers that $l$ is constant along curves of constant angular velocity $\Omega$ that cross the equatorial plane at a particular point $(r_0, \pi/2)$. The specific angular momentum distribution outside the equatorial plane $l(r,\theta)$ is hence obtained  by considering $\Omega(r, \theta) = \Omega(r_0, \pi/2)$. By replacing this condition in Eq.~\eqref{eq:ang_mon_ang_vel} we arrive at 
\begin{eqnarray}\label{eq:vonzeipel}
[g_{tt}(r, \theta)\tilde{g}_{t\phi}(r_0)-\tilde{g}_{tt}(r_0)g_{t\phi}(r, \theta)]l^2_{\mathrm{eq}}(r_0)
\nonumber \\ 
+[g_{tt}(r, \theta)\tilde{g}_{\phi\phi}(r_0)-\tilde{g}_{tt}(r_0)g_{\phi\phi}(r, \theta)]l_{\mathrm{eq}}(r_0)
\nonumber \\ 
+[g_{t\phi}(r, \theta)\tilde{g}_{\phi\phi}(r_0)-\tilde{g}_{t\phi}(r_0)g_{\phi\phi}(r, \theta)] = 0\,,
\end{eqnarray}
where $l_{\mathrm{eq}}(r_0)$ is the specific angular momentum at the point $(r_0, \pi/2)$ and the metric components $\tilde{g}_{\alpha\beta}(r_0)$ refer to quantities evaluated at the equatorial plane. Solving Eq.~\eqref{eq:vonzeipel} for different values $r_0$ yields the equation of the curves along which $l(r, \theta) = l_{\mathrm{eq}}(r_0)$, i.e.~the so-called von Zeipel cylinders. 

It is worth remarking that this approach to compute the angular momentum distribution outside the equatorial plane is a better choice for our case than the approach considered in~\cite{Gimeno-Soler:2017} where a set of equipotential surfaces were computed to map the disk.   
On the one hand, this approach is computationally cheaper when compared to the one followed in~\cite{Gimeno-Soler:2017}, where a large number of equipotential surfaces and a very small integration step were required to compute the physical quantities in the disk with an acceptable accuracy.
On the other hand, one could argue that this approach can be seen as a more natural way of building the angular momentum distribution, as it is built from the integrability conditions of Eq.~\eqref{eq:potential_equation} instead of an {\it ad-hoc} assumption about the form of the angular momentum distribution outside the equatorial plane.

\begin{figure*}[t]
\includegraphics[scale=0.11]{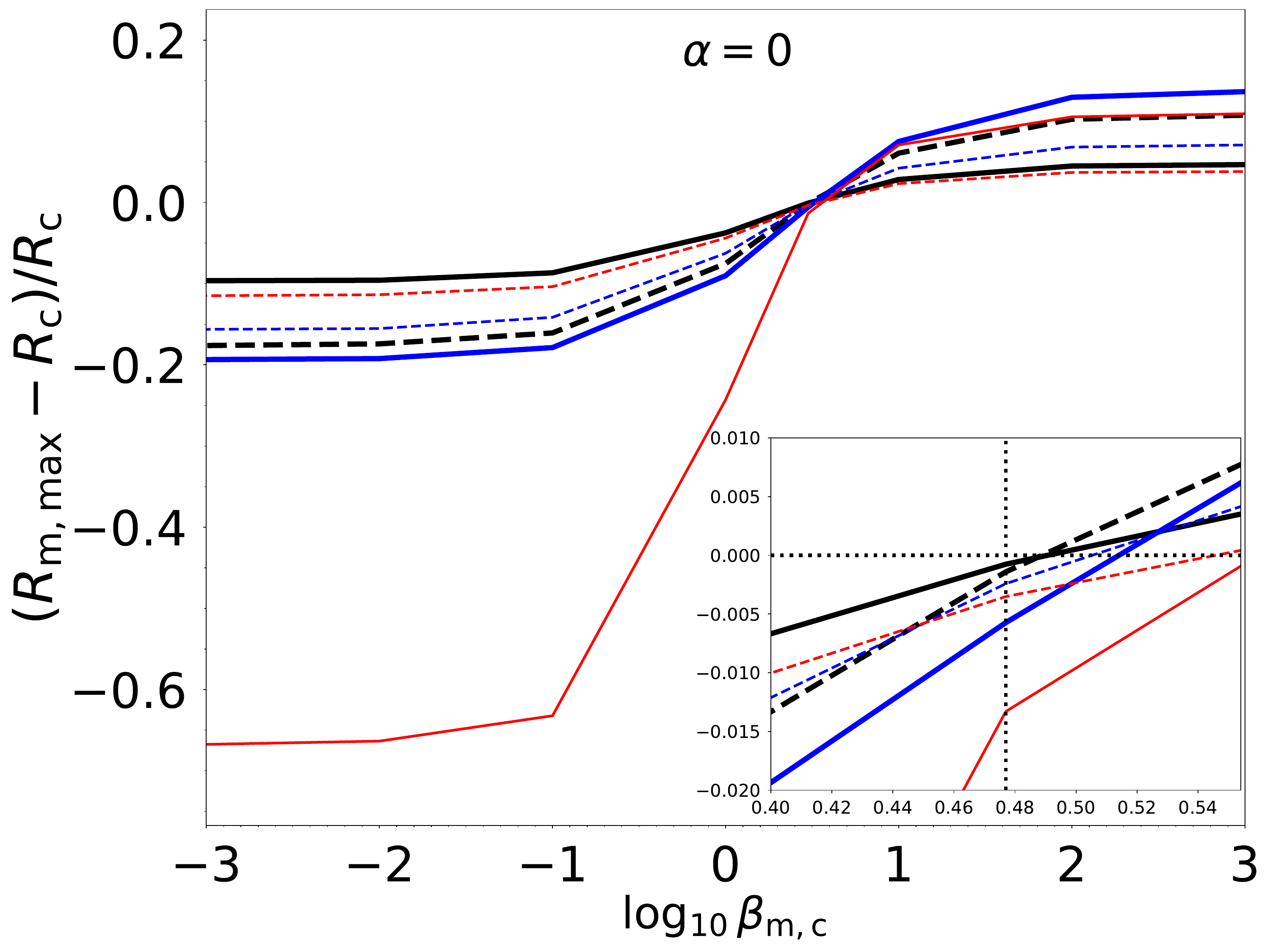}
\hspace{-0.1cm}
\includegraphics[scale=0.11]{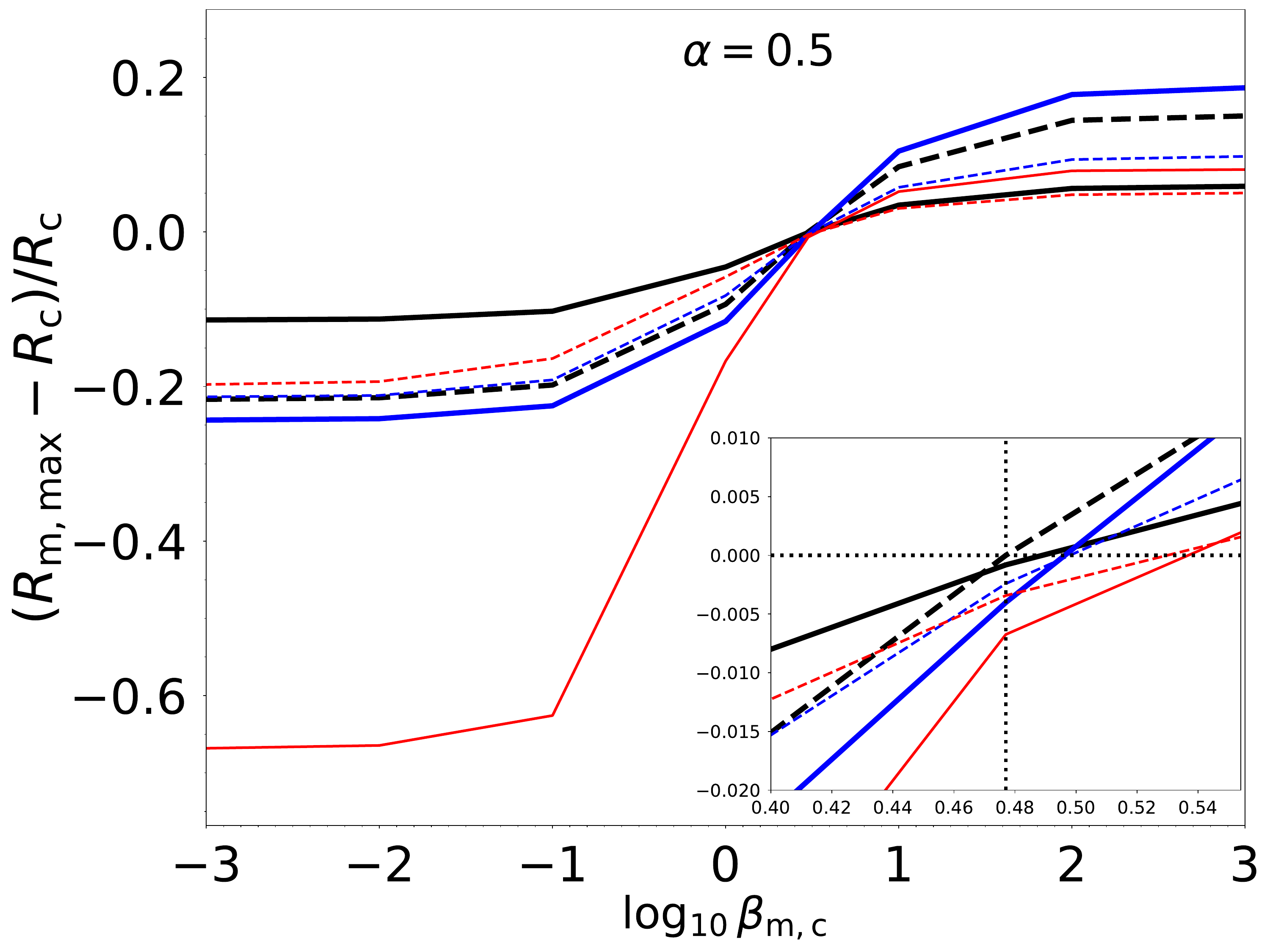}
\hspace{-0.1cm}
\includegraphics[scale=0.11]{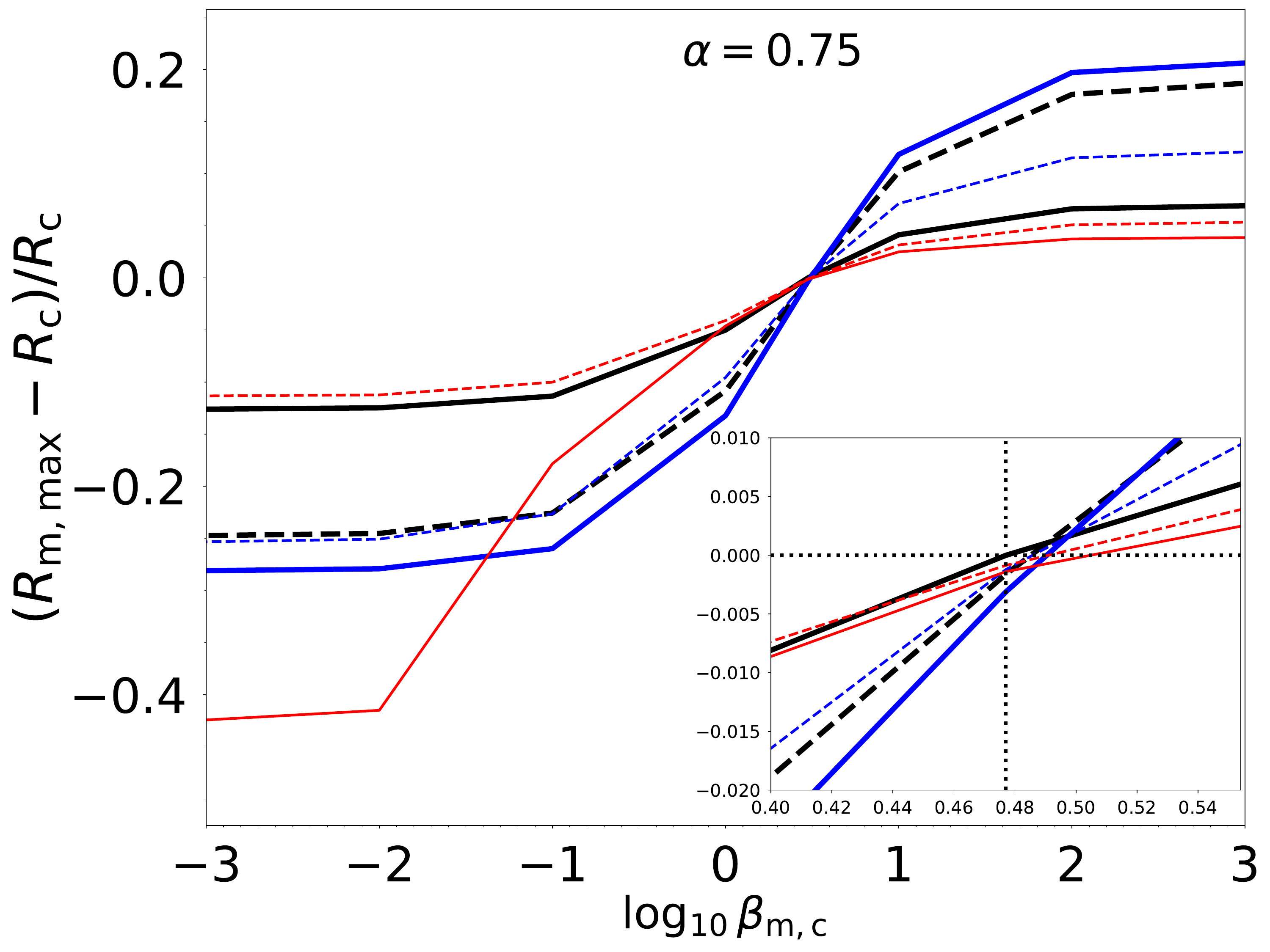}
\\
\vspace{0.0cm}
\includegraphics[scale=0.11]{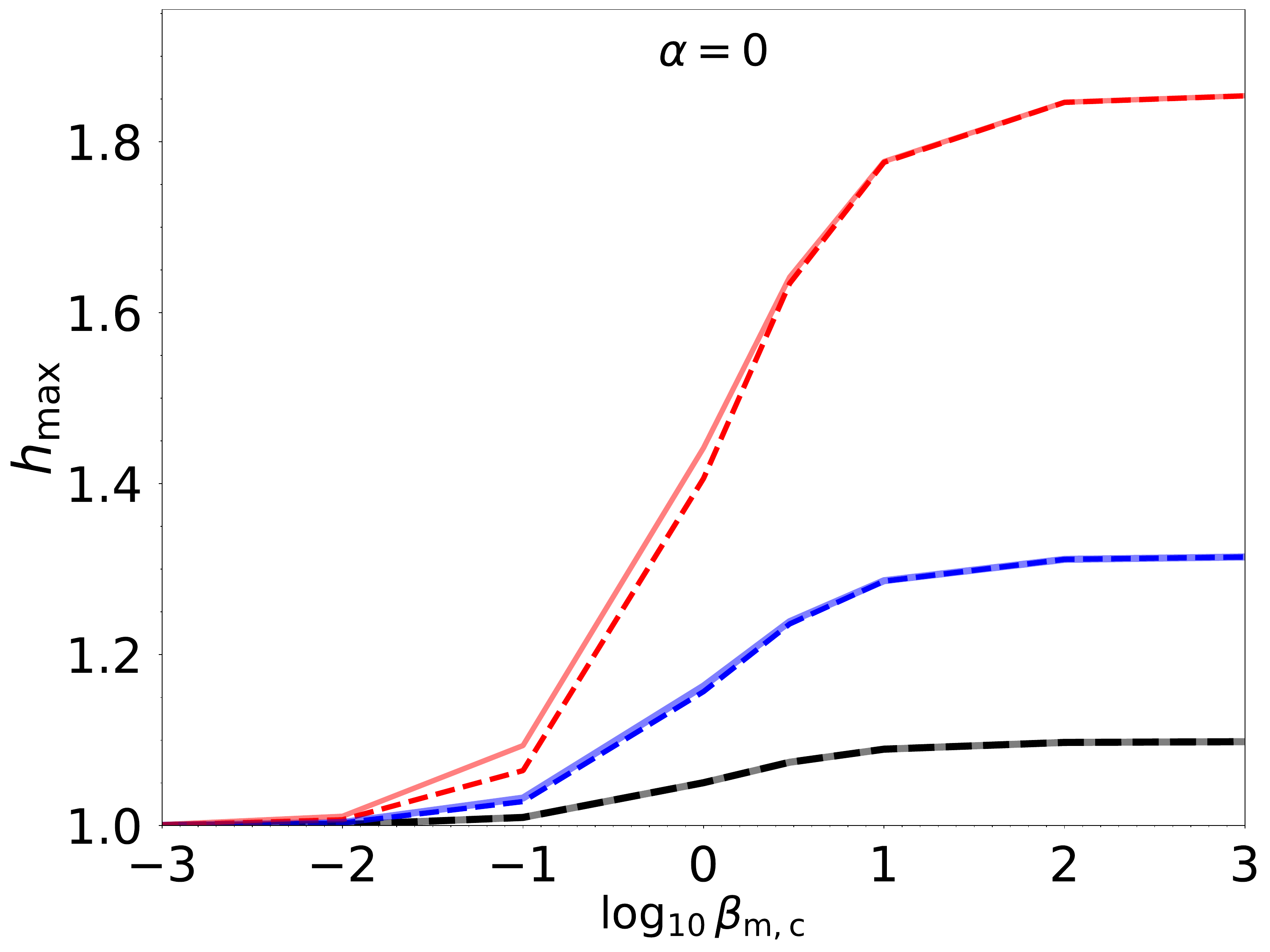}
\hspace{-0.1cm}
\includegraphics[scale=0.11]{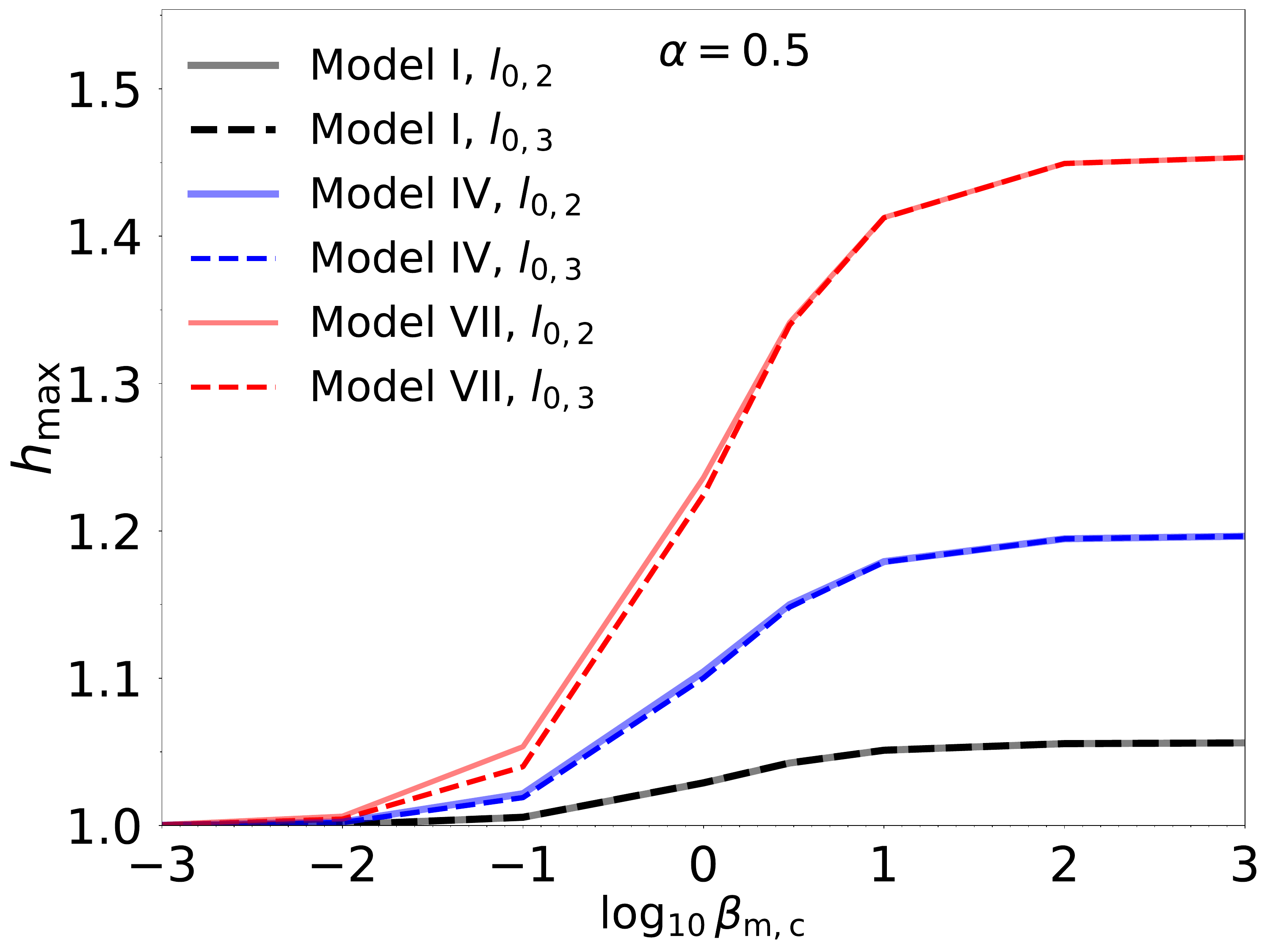}
\hspace{-0.1cm}
\includegraphics[scale=0.11]{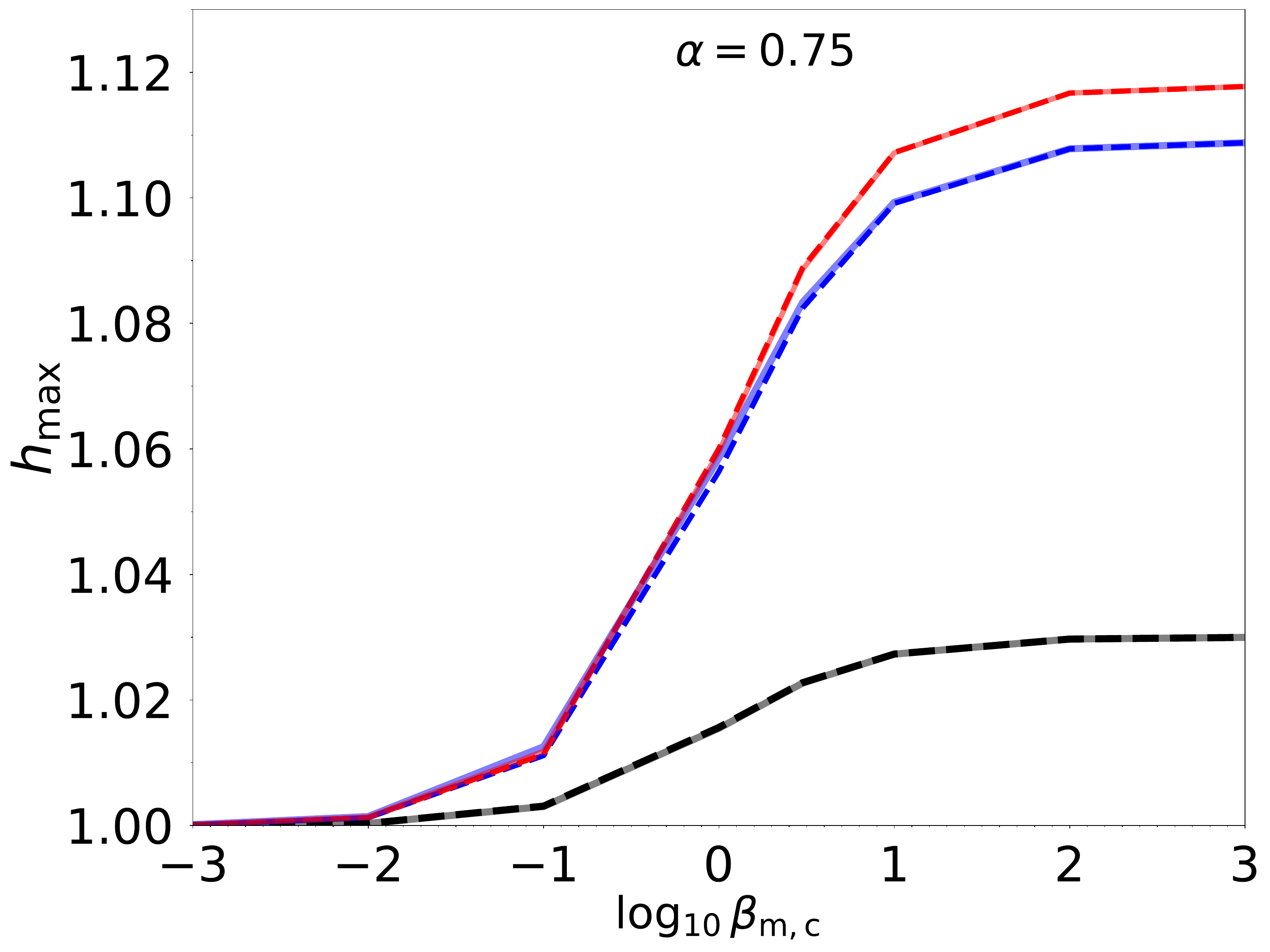}
\\
\vspace{0.0cm}
\includegraphics[scale=0.11]{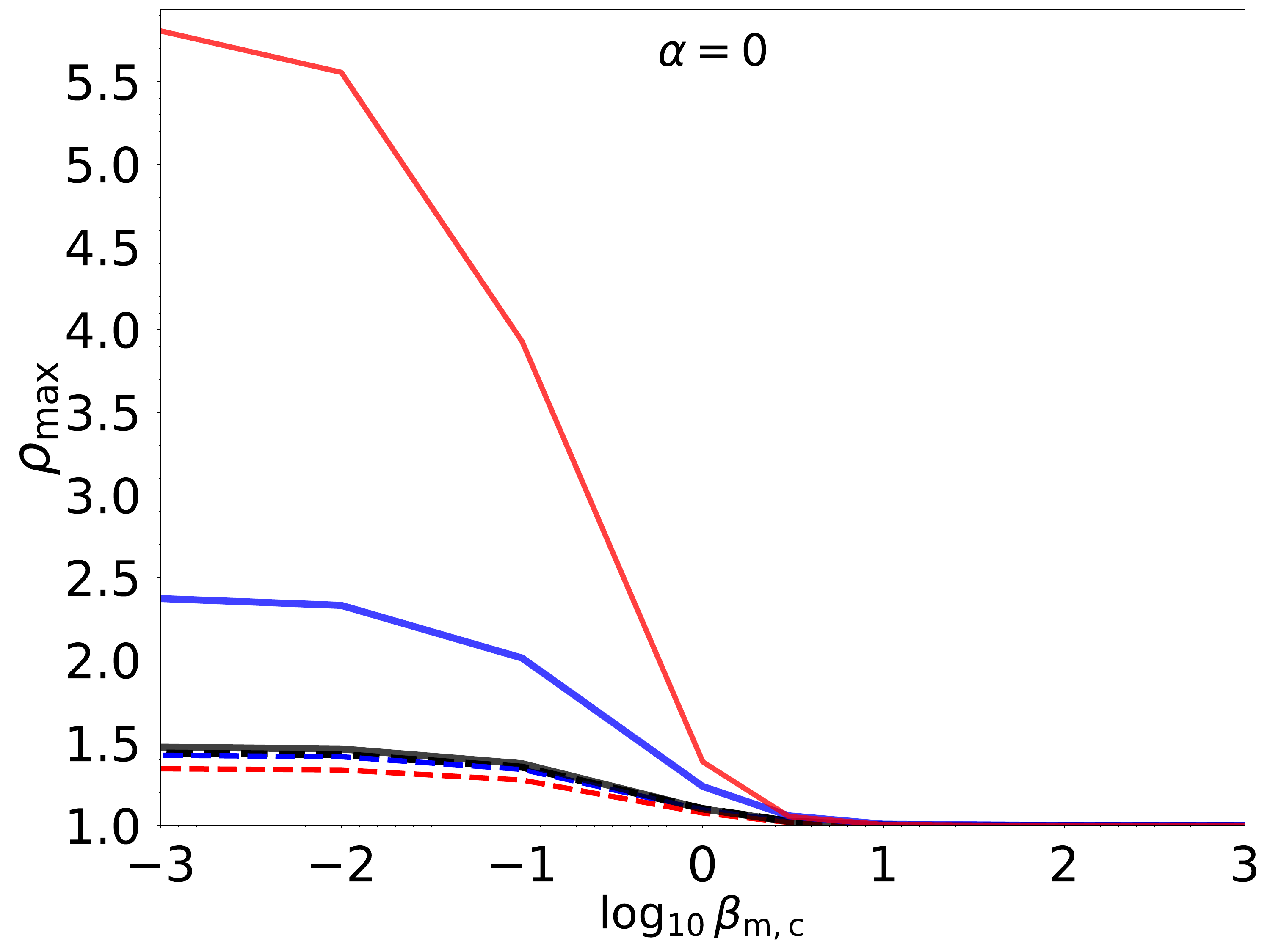}
\hspace{-0.1cm}
\includegraphics[scale=0.11]{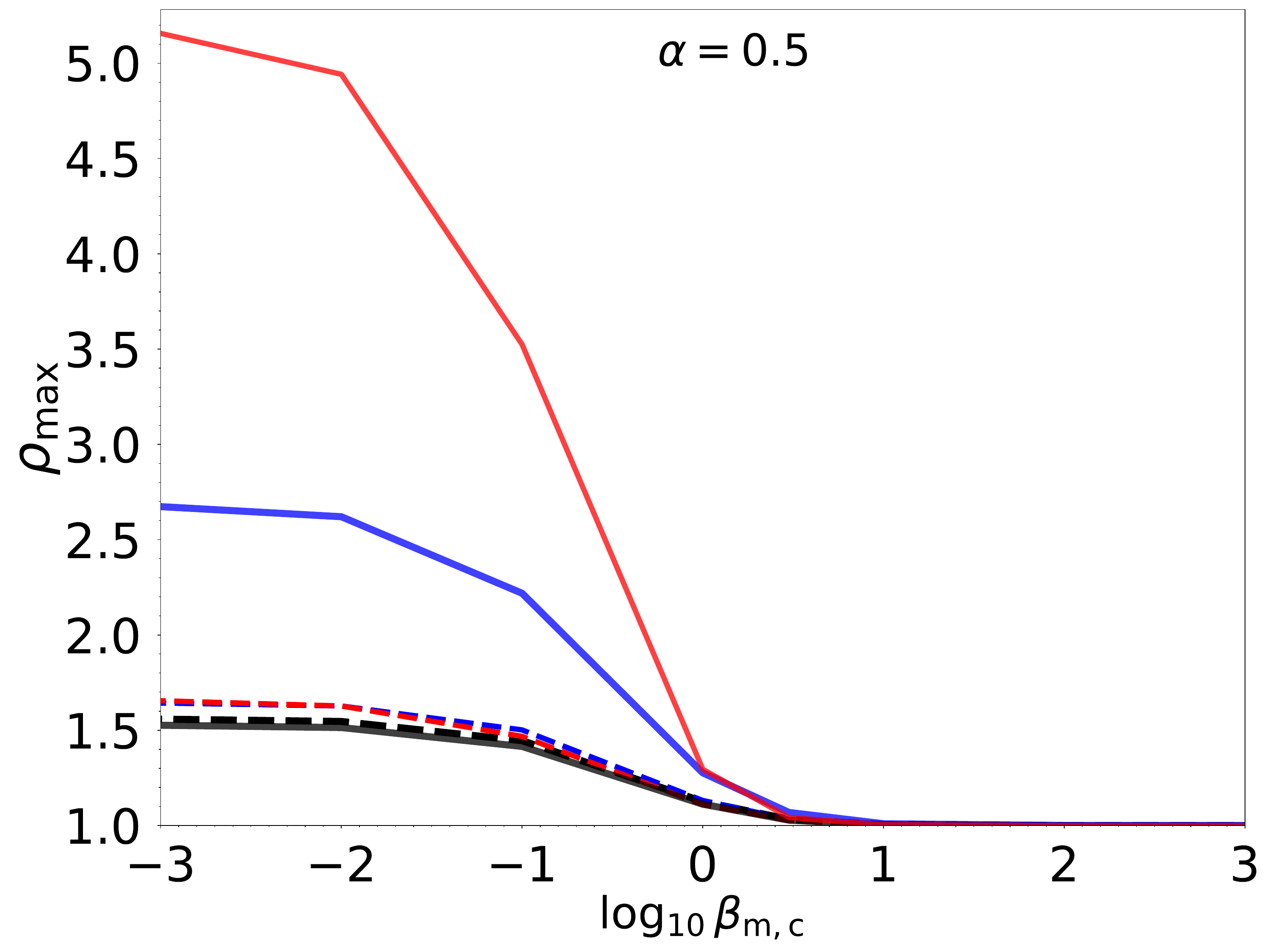}
\hspace{-0.1cm}
\includegraphics[scale=0.11]{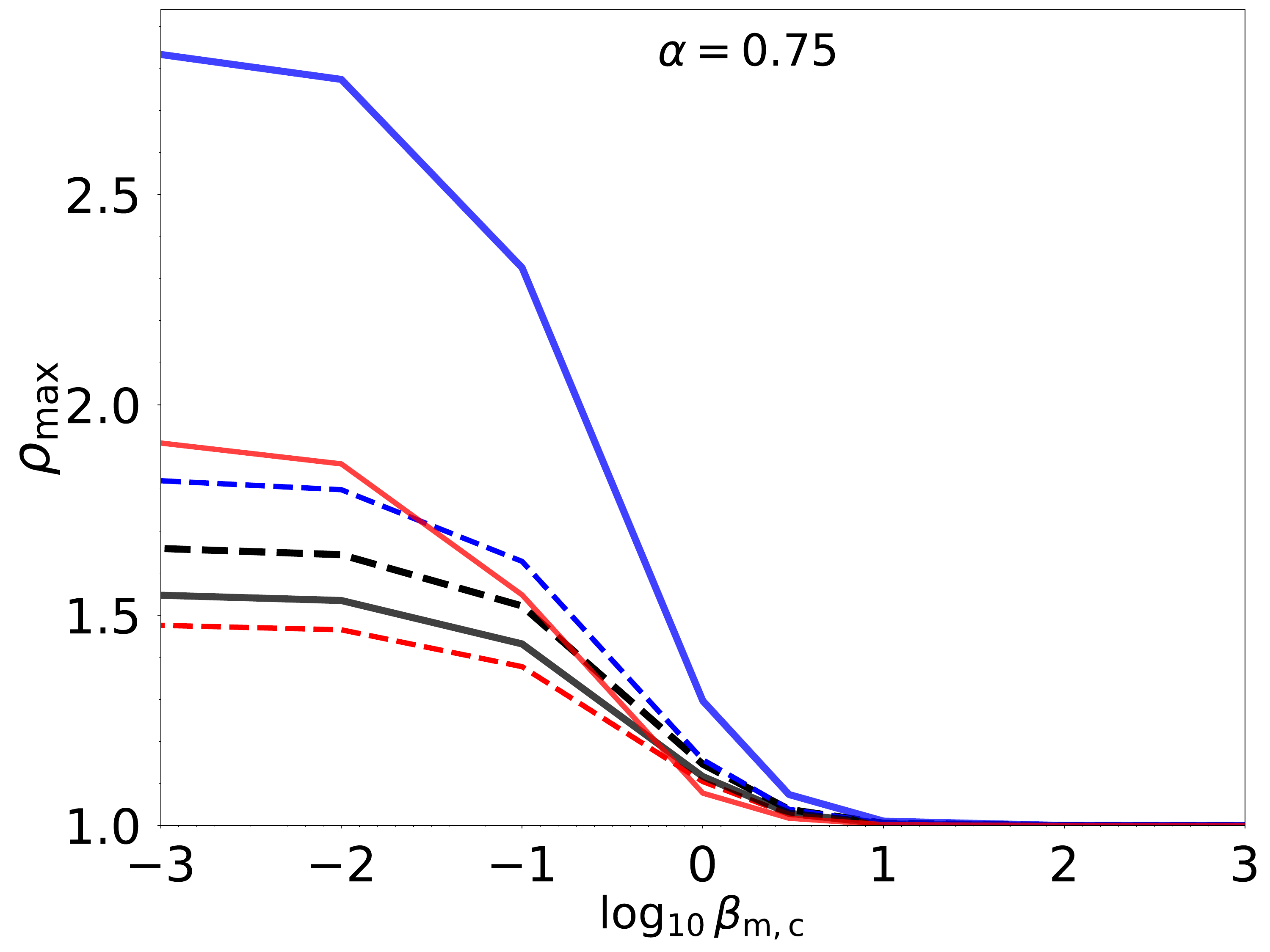}
\caption{Effects of the magnetization on the values of the relative variation of the maximum of the magnetic pressure $(R_{\mathrm{m, max}}- R_{\mathrm{c}})/R_{\mathrm{c}}$ (top panels), the maximum value of the specific enthalpy $h_{\mathrm{max}}$ (central panels) and the maximum value of the rest-mass density (bottom panels). From left to right, the columns correspond to different values of the exponent of the angular momentum distribution $\alpha$, indicated in the plots. In all the panels, black, blue and red lines correspond to the three KBHsSH spacetimes we are considering, namely I, IV and VII. Solid and dashed lines correspond to either criterion 2 or 3 employed to compute $l_0$, respectively. (See legend in the central plot.) The top panels also display an inset of the region around $\beta_{\mathrm{m, c}} = 3$. The horizontal and vertical dotted lines correspond to $R_{\mathrm{m, max}} = R_{\mathrm{c}}$ and $\beta_{\mathrm{m, c}} = 3$ respectively.
Note that in this figure we include additional results for values of $\beta_{\mathrm{m ,c}}$ that are not present in the tables, namely $\beta_{\mathrm{m ,c}} = \{10^{3}, 10^{2}, 10, 3, 10^{-1}, 10^{-2}, 10^{-3}\}$.}
\label{rmmax_rhomax_hmax_logbeta_fig}
\end{figure*}

\subsection{Magnetized disks}

As in {Paper I} we consider that the matter in the disk is described within the framework of  ideal, general relativistic MHD. Starting from the conservation laws $\nabla_{\mu} T^{\mu\nu} = 0$, $\nabla_{\mu} \,^\ast F^{\mu\nu} = 0$ and $\nabla_{\mu} (\rho u^{\mu}) = 0$, where $\nabla_{\mu}$ is the covariant derivative, $^\ast F^{\mu\nu} = b^{\mu}u^{\nu} - b^{\nu}u^{\mu}$ is the (dual of the) Faraday tensor, $b^{\mu}$ is the magnetic field 4-vector and
\begin{equation}\label{eq:e-m_tensor} 
T^{\mu\nu} = (\rho h + 2 p_{\mathrm{m}})u^{\mu}u^{\nu} + (p + p_{\mathrm{m}})g^{\mu\nu} - b^{\mu}b^{\nu},
\end{equation}
is the energy-momentum tensor of a  magnetized perfect fluid. In the latter  $h$, $\rho$, $p$, and $p_{\mathrm{m}} = b^{\mu}b_{\mu}/2$ are the fluid specific enthalpy, density, fluid pressure, and magnetic pressure, respectively. It is also convenient to define the magnetization parameter, that is the ratio of fluid pressure to magnetic pressure 
\begin{eqnarray}
\beta_{\mathrm{m}} = p/p_{\mathrm{m}}\,.
\end{eqnarray}
Assuming that the magnetic field is purely azimuthal i.e.~$b^r = b^{\theta} = 0$ and stationarity and axisymmetry of the flow, it immediately follows that the conservation equations of the current density and of the Faraday tensor are trivially satisfied.
Contracting the divergence of Eq.~\eqref{eq:e-m_tensor} with the projection tensor $h^{\alpha}_{\,\,\beta} = \delta^{\alpha}_{\,\,\beta} + u^{\alpha}u_{\beta}$ and rewriting the result in terms of the specific angular momentum $l$ and of the angular velocity $\Omega$, we arrive at
 \begin{equation}\label{eq:diff_ver}
 \partial_i(\ln |u_t|) - \frac{\Omega \partial_i l}{1-l\Omega} + \frac{\partial_i p}{\rho h} + \frac{\partial_i(\mathcal{L}b^2)}{2\mathcal{L}\rho h} = 0\,,
 \end{equation}
 where $i = r, \theta$ and $\mathcal{L} = g_{t\phi}^2 - g_{tt}g_{\phi\phi}$.
To integrate Eq.~\eqref{eq:diff_ver} we need to assume an equation of state (EoS). As in Paper I we assume a polytropic EoS of the form
\begin{equation}\label{eq:eos_fluid}
p = K \rho^{\Gamma},
\end{equation}
with $K$ and $\Gamma$ constants.
For the magnetic part, we can write an EoS equivalent to Eq.~\eqref{eq:eos_fluid}, but for $\tilde{p}_{\mathrm{m}} = \mathcal{L} p_{\mathrm{m}}$
\begin{equation}
\tilde{p}_{\mathrm{m}} = K_{\rm m} \tilde{w}^q\,,
\end{equation}
where $K_{\rm m}$ and $q$ are constants and $w =\rho h$. Thus, we can express the magnetic pressure $p_{\mathrm{m}}$ as
\begin{equation}\label{eq:eos_magnetic}
p_{\mathrm{m}} = K_{\rm m} \mathcal{L}^{q-1} (\rho h)^q\,.
\end{equation}
Now, we can integrate Eq.~\eqref{eq:diff_ver} to arrive at

\begin{equation}\label{eq:final}
W - W_{\mathrm{in}} + \ln \left(1 + \frac{\Gamma K}{\Gamma -1}\rho^{\Gamma -1}\right) + \frac{q}{q-1}K_{\rm m}(\mathcal{L}\rho h)^{q-1}=0,
\end{equation}
where $W$ stands for the (gravitational plus centrifugal) potential and is defined as
\begin{equation}\label{eq:potential_equation}
W(r, \theta) - W_{\mathrm{in}} = \ln |u_t| - \ln |u_{t\mathrm{, in}}| - \int^l_{l_\mathrm{in}}\frac{\Omega \mathop{dl}}{1-l\Omega}\,,
\end{equation}
where subscript `in' denotes that the corresponding quantity is evaluated at the inner edge of the disk i.e.~$(r_{\mathrm{in}}, \pi/2)$.
We also need to introduce the total gravitational energy density for the disk, $\rho_{\mathrm{T}}=-T^t_t + T^i_i$, and for the scalar field, $\rho_{\mathrm{SF}}=-(T_{\mathrm{SF}})^t_t + (T_{\mathrm{SF}})^i_i$. These are given by
\begin{eqnarray}\label{eq:torus_energy_density}
\rho_{\mathrm{T}} &=&  \frac{\rho h (g_{\phi\phi} - g_{tt} l^2)}{g_{\phi\phi} + 2 g_{t\phi} l + g_{tt} l^2} + 2 (p + p_{\mathrm{m}}),
\\
\rho_{\mathrm{SF}} &=&  2 \left(\frac{2 e^{-2 F_0} \omega (\omega-m W)}{N} - \mu^2\right) \varphi^2.
\end{eqnarray}
Using these expressions we can compute the total gravitational mass of the torus and of the scalar field as the following integral
\begin{equation}\label{eq:mass_integral}
\mathcal{M} = \int  \rho \sqrt{-g}\,\mathrm{d}^3x\,,
\end{equation}
where $g$ is the determinant of the metric tensor and $\rho\equiv \rho_{\mathrm{T}}, \rho_{\mathrm{SF}}$.

\section{Methodology}
\label{procedure}

We turn next to describe the numerical procedure to build the disks and our choice of parameter space. In this work, as mentioned before, we only consider a subset of the KBHsSH spacetimes considered in {Paper I} (namely, spacetimes I, IV and VII). This choice is made to keep the number of free parameters of our models reasonably tractable.
Likewise, as in Paper I we fix the mass of the scalar field to $\mu = 1$, the exponents of the polytropic EoS to $q = \Gamma = 4/3$, and the density at the center of the disk to $\rho_{\mathrm{c}} = 1$. We also consider only three representative values for the magnetization parameter at the center of the disk $\beta_{\mathrm{m,c}}$, namely $10^{10}$ (which effectively corresponds to an nonmagnetized disk), $1$ (mildly magnetized) and $10^{-10}$ (strongly magnetized).

\subsection{Angular momentum and potential at the equatorial plane}

From Eq.~\eqref{eq:ansatz} it is apparent that the parameter space in the angular momentum sector can be fairly large, i.e.~both the constant part of the angular momentum distribution $l_0$ and the exponent $\alpha$ are continuous parameters. To reduce this part of the parameter space, first we  restrict ourselves to four values of the exponent $\alpha$, namely $0$, $0.25$, $0.5$ and $0.75$. To obtain the constant part of the angular momentum distribution,  $l_0$, we consider three different criteria that yield three values of $l_0$ for each value of $\alpha$:
\begin{itemize}
    \item[1.- ] $l_0$ is such that $W_{\mathrm{cusp}} = 0$ and $r_{\mathrm{in}}$ is chosen such that $\Delta W_{\mathrm{c}} = 0.5 W_{\mathrm{c, 1}}$
    \item[2.- ] $l_0$ is such that $W_{\mathrm{cusp}} < 0$ and $\Delta W_{\mathrm{c}} = 0.5 W_{\mathrm{c, 1}}$
    \item[3.- ] $l_0$ is such that $W_{\mathrm{cusp}} > 0$ and $\Delta W_{\mathrm{c}} = 0.5 W_{\mathrm{c, 1}}$\,,
\end{itemize}
In the previous expressions  $W_{\mathrm{cusp}}$ is the value of the potential at the point where the isopotential surfaces cross (forming a cusp). That point corresponds to a maximum of the potential. In addition $\Delta W_{\mathrm{c}}$ is defined as
\begin{equation}
\Delta W_{\mathrm{c}} = \left\{ \label{eq:deltaWc} 
  \begin{array}{ll}
    W_{\mathrm{c}} - W_{\mathrm{in}} &  \text{if } W_{\mathrm{in}} < 0\\
    W_{\mathrm{c}} & \text{if } W_{\mathrm{in}} \geq 0\,,
  \end{array}
\right.
\end{equation}
where $W_{\mathrm{c}}$ is the potential at the center of the disk 
and $W_{\mathrm{c, 1}}$ is the value of the potential at the center when $W_{\mathrm{cusp}} = 0$. This value also corresponds to the maximum possible value of $|\Delta W_{\mathrm{c}}|$ for a specific choice of $\alpha$. 
Our choice for the three values of the constant part of the angular momentum distribution $l_0$ is particularly useful because it allows us to get rid of the dependence on $\Delta W_{\mathrm{c}}$ of the physical quantities in the disk computed with each criterion. As it can be seen when inspecting Eq~\eqref{eq:final}, the rest-mass density $\rho$ and the specific enthalpy $h$ (and the pressure $p$ and the magnetic pressure $p_{\mathrm{m}}$ which are computed from them) are only dependent on the potential distribution, the magnetization parameter $\beta_{\mathrm{m,c}}$ and the geometry of the spacetime (for fixed $\Gamma$ and $q$). Therefore, if we remove the dependence on $\Delta W_{\mathrm{c}}$, the disk morphology and the physical quantities in the disk only depend on the angular momentum distribution, $l(r, \theta)$, the magnetization parameter at the center of the disk $\beta_{\mathrm{m,c}}$ and the geometry of the spacetime. 
It is also worth to mention that this way of prescribing the angular momentum distribution only depends on the metric parameters and their derivatives (through the potential, the Keplerian angular momentum and the definition of the von Zeipel cylinders). Therefore, if we compare two solutions built in different spacetimes, but following the same criterion to prescribe the angular momentum distribution, we can be sure that the differences between these two solutions are a consequence of the fact that the two solutions correspond to different spacetimes.

Figure~\ref{radial_angmom_potential} displays radial profiles of the angular momentum along the equatorial plane, together with the corresponding profiles of the potential, for our three choices of $l_0$ and for $\alpha = 0.5$. From left to right each panel corresponds to one of the three different models of KBHsSHs we are considering, namely models I, IV and VII. The radial coordinate used in these plots (and in all figures in the paper) is the perimeteral radius $R$, related to the Boyer-Lindquist radial coordinate $r$ according to $R=e^{F_2}r$ (see Paper I for details on the geometrical meaning of this coordinate). 
It can be seen that for the three spacetimes, the profiles of $l_{\mathrm{eq}}(R)$ for criteria 1 and 2 are very similar. The deviations from the Kerr black hole case can be observed in the Keplerian angular momentum profile: in the first column, $l_{\mathrm{K}}(R)$ looks very similar to that of a rapidly rotating Kerr BH; some small deviations are visible in the profile plotted in the second column; finally, in the third column, a significant  deviation from what should be expected from any Kerr BH is noticeable.
The second row of Fig.~\ref{radial_angmom_potential} depicts the potential distribution at the equatorial plane,  $W_{\mathrm{eq}}(R)$. It becomes apparent that a very small variation in the value of $l_0$ affects significantly the value of $W_{\mathrm{cusp}}$ (e.g.~the bottom left panel shows that, when comparing the profiles from criteria 1 and 2, a difference between the values of $l_0$ of about $\sim 1\%$, yields a large  difference in the value of $\Delta W_{\mathrm{max}} = W_{\mathrm{c}} - W_{\mathrm{cusp}}$ such that $\Delta W_{\mathrm{max, 1}} = 2 \Delta W_{\mathrm{max, 2}}$).

To compute the potential at the equatorial plane we rewrite Eq.~\eqref{eq:potential_equation} as
\begin{equation}\label{eq:radial_potential}
W_{\mathrm{eq}}(r) = - \int^{+\infty}_{r}\left(\frac{\partial\ln |u_{t\mathrm{,in}}|(r)}{\partial r}  - \frac{\Omega_{\mathrm{eq}} \frac{\mathop{dl}}{\mathop{dr}}}{1-l_{\mathrm{eq}}\Omega_{\mathrm{eq}}}\right)\mathop{dr}\,,
\end{equation}
where we have used that $W_{\mathrm{eq}}(r) \rightarrow 0$ when $r \rightarrow \infty$ and $u_t$ can be written as
\begin{equation}\label{eq:ut}
u_t = - \sqrt{\frac{g_{t\phi}^2- g_{tt}g_{t\phi}}{g_{\phi\phi}+2g_{t\phi}l+g_{tt}l^2}}.
\end{equation}
Then, to obtain the values of $l_0$ we require, we choose the following procedure. First, we start by considering a constant distribution of angular momentum (i.e.~$\alpha = 0$) and $l_0 = l_{\mathrm{mb}}$ where $l_{\mathrm{mb}} = l_{\mathrm{K}}(r_{\mathrm{mb}})$ and $r_{\mathrm{mb}}$ is the radius of the marginally bound orbit. Notice that this choice of the parameters corresponds to the cases we considered in~\cite{paper1} (and implies that $W_{\mathrm{cusp}} =0$) and that we obtain $l_{\mathrm{mb}}$ and $r_{\mathrm{mb}}$ computing the minimum of Eq.(8) in Paper I. 
We also need to compute $l_{\mathrm{ms}}$ and $r_{\mathrm{ms}}$ as the minimum of the Keplerian angular momentum (Eq.~\eqref{eq:keplerian_ang_mom}).
In this way, Eq.~\eqref{eq:potential_equation} amounts to evaluate $W = \ln|u_t|$ and we only need to obtain the minimum of $W(r, \pi/2)$. The location of this minimum corresponds to the center of the disk $r_{\mathrm{c}}$ and the value of the potential there is $W_{\mathrm{c, 1}}$ ($r_{\mathrm{c}}$ also corresponds to the largest solution of $l_{\mathrm{K}}(r) = l_0$). Once we have the value of $W_{\mathrm{c, 1}}$ for $\alpha = 0$, we can compute the required quantities to build the three distributions of angular momentum we need.
For the first case we only need to find the value of $r_{\mathrm{in}}$ that fulfills the condition
\begin{equation}\label{eq:rin_1_solver}
W(r_{\mathrm{in}}, \pi/2) = 0.5 W_{\mathrm{c, 1}}.
\end{equation}
For the second case, we iteratively solve the following equation for $l_0$
\begin{equation}\label{eq:angmom_2_solver}
    W_{\mathrm{cusp}}(l_0) - W_{\mathrm{c}}(l_0) = 0.5  W_{\mathrm{c, 1}}\,,
\end{equation}
taking into account that $l_0$ must be in the interval $l_{\mathrm{K}}(r_{\mathrm{ms}}) < l_0 < l_{\mathrm{K}}(r_\mathrm{mb})$ so $W_{\mathrm{cusp}} < 0$.
And in the third case, we solve 
\begin{equation}\label{eq:angmom_3_solver}
W_{\mathrm{c}}(l_0) = 0.5  W_{\mathrm{c, 1}}\,,
\end{equation}
in the same way as in the second case, but taking into account that $l_0 > l_{\mathrm{K}}(r_{\mathrm{mb}})$, so that $W_{\mathrm{cusp}} > 0$.

To obtain the values of $l_0$ for $\alpha \neq 0$ we only have to take into account that the potential is defined by the integral Eq.~\eqref{eq:radial_potential}. As we do not have an easy way to compute $W_{\mathrm{c, 1}}$ (i.e.~to compute a value for $l_0$ such that the condition  $W_{\mathrm{cusp}} = 0$ is guaranteed), we have to solve iteratively the following equation for $l_0$
\begin{equation}\label{eq:alphanot0_Wc}
    W_{\mathrm{eq}}(r_{\mathrm{in}}; l_0) = 0\,,
\end{equation}
where the left-hand side of the equation is an integral and we know that, for $\alpha > 0$, the value of $l_0$ corresponding to this case will always be between $l_{\mathrm{K}}(r_{\mathrm{ms}}) < l_0 <  l_{\mathrm{K}}(r_{\mathrm{mb}})$. With this, we can obtain the value of $W_{\mathrm{c, 1}}$ for any $\alpha \neq 0$ and following the aforementioned three steps and taking into account that now the potential is defined by the integral \eqref{eq:radial_potential}, we can compute all angular momentum and potential distributions at the equatorial plane that we require.
It is worth to mention that the value of $W_{\mathrm{c, 1}}$ is very sensitive to small changes in $l_0$, due to the fact that the potential is very steep around the maximum, so we solve equations \eqref{eq:rin_1_solver}, \eqref{eq:angmom_2_solver}, \eqref{eq:angmom_3_solver} and \eqref{eq:alphanot0_Wc} using the bisection method with a tolerance that ensures  that the computed values of $l_0$ fulfill $|(\Delta W_{\mathrm{c}} - 0.5 W_{\mathrm{c, 1}})/- (0.5 W_{\mathrm{c, 1}})| < 10^{-8}$ for all the cases we have considered. The integral \eqref{eq:radial_potential} is solved using the trapezoidal rule with a radial grid 100 times denser than our regular grid (see below), to ensure the correct finding of $r_{\mathrm{cusp}}$, $r_{\mathrm{in}}$ and $r_{\mathrm{c}}$. 

\subsection{Angular momentum and potential outside the equatorial plane}

To extend the angular momentum and the potential distribution to the region outside the equatorial plane we need to solve Eq.~\eqref{eq:vonzeipel}. As we use a numerical grid, it is inconvenient to solve the curves starting from the equatorial plane, as in general the von Zeipel cylinders will not pass through the points in the grid.  Instead, we run through all the $(r, \theta)$ points in our grid and, for each point,  we solve Eq.~\eqref{eq:vonzeipel} to obtain the  crossing point of the corresponding von Zeipel cylinder in the equatorial plane, $(r_0, \pi/2)$. To improve the accuracy of the procedure, we interpolate the function $l_{\mathrm{eq}}(r)$ with a third-order spline and we solve Eq.~\eqref{eq:vonzeipel} with the bisection method and a tolerance of $\sim 10^{-8}$. A sample of the geometry of these cylinders is shown in Fig.~\ref{vonZeipel_cylinders} for the first criterion for the angular momentum at the equatorial plane and $\alpha = 0.5$.
To obtain the potential we follow~\cite{Daigne:2004} and  use the fact that the specific angular momentum is constant along the von Zeipel cylinders to recast Eq.~\eqref{eq:potential_equation} as
\begin{equation}
W(r, \theta) = W_{\mathrm{eq}}(r_0) + \ln\left[\frac{-u_t(r, \theta)}{-u_t(r_0, \pi/2)}\right]\,,
\end{equation}
which yields the potential everywhere.

\subsection{Building the magnetized disk}

To build the disk we follow the same procedure as in {Paper I}. First, we compute the polytropic constant $K$ by solving
\begin{multline}
\label{eq:to_solve_K}
W - W_{\mathrm{in}} + \ln \left(1 + \frac{\Gamma K}{\Gamma -1}\rho_{\mathrm{c}}^{\Gamma -1}\right) 
\\
+ \frac{q}{q-1} \frac{K\rho_{\mathrm{c}}^{\Gamma}}{\beta_{\mathrm{m_c}} \left(\rho_{\mathrm{c}} + \frac{K\Gamma\rho_{\mathrm{c}}^{\Gamma}}{\Gamma-1}\right)} =0\,,
\end{multline}
which is Eq.~\eqref{eq:final} evaluated at the center of the disk $r_{\mathrm{c}}$. Once  $K$ is computed we can obtain the remaining relevant quantities at the center, namely $p_{\mathrm{c}}$, $p_{\mathrm{m,c}}$ and $h_{\mathrm{c}}$ along with the polytropic constant of the magnetic EoS $K_{\mathrm{m}}$.
Then, to compute the distribution of the rest-mass density $\rho(r,\theta)$, we only have to solve 
\begin{multline}
\label{eq:to_solve_rho}
W - W_{\mathrm{in}} + \ln \left(1 + \frac{\Gamma K}{\Gamma -1}\rho^{\Gamma -1}\right) 
\\
+ \frac{q}{q-1}K_{\rm m}\left(\mathcal{L}\left(\rho + \frac{K\Gamma \rho^{\Gamma}}{\Gamma - 1}\right)\right)^{q-1}=0\,,
\end{multline}
if $W(r, \theta) < 0$. For $W(r, \theta) > 0$ we set $\rho = p = p_{\mathrm{m}} = 0$. Note that Eqs.\eqref{eq:to_solve_K} and \eqref{eq:to_solve_rho} are both trascendental equations, and must be solved numerically.
As in~\cite{paper1} we solve these equations using a non-uniform $(r, \theta)$ grid with a typical domain given by $[r_{\mathrm{H}}, 199.2] \times [0, \pi/2]$ and a typical number of points $N_r \times N_{\theta} = 2500 \times 300$. Those numbers are only illustrative as the actual values depend on the horizon radius $r_{\mathrm{H}}$ and on the specific model. The spacetime metric data on this grid is interpolated from the original data obtained by~\cite{Herdeiro:2015b}. The original grid in~\cite{Herdeiro:2015b} is a uniform ($x$, $\theta$) grid (where $x$ is a compactified radial coordinate) with a domain $[0, 1] \times [0, \pi/2]$ and a number of points of $N_x \times N_{\theta} = 251 \times 30$~\footnote{In particular, the three spacetimes which are presented here, are publicly available in~\cite{grav_web}}.

\begin{figure*}[t]
\includegraphics[scale=0.17]{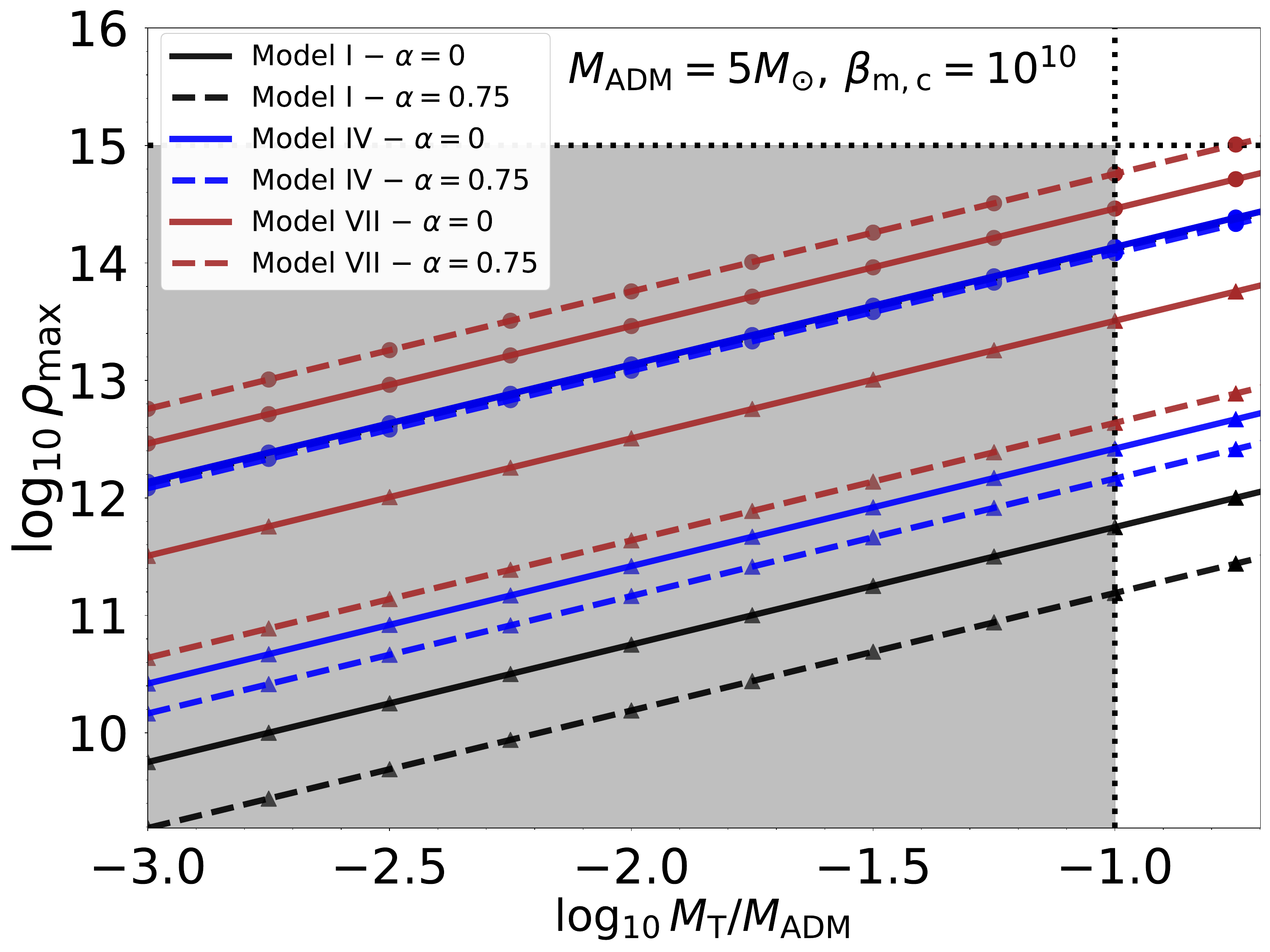}
\hspace{0.0cm}
\includegraphics[scale=0.17]{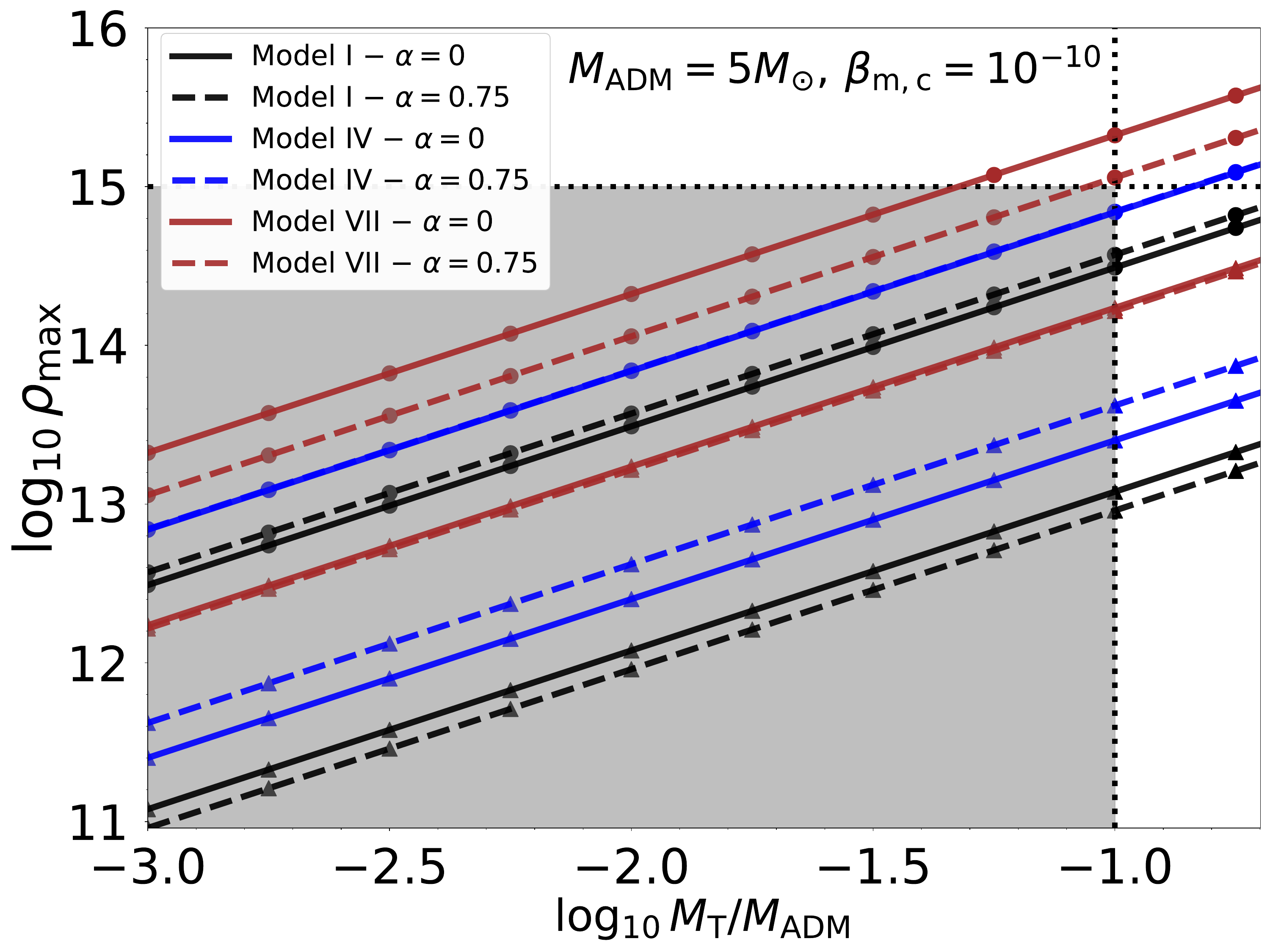}
\\
\vspace{0.0cm}
\includegraphics[scale=0.17]{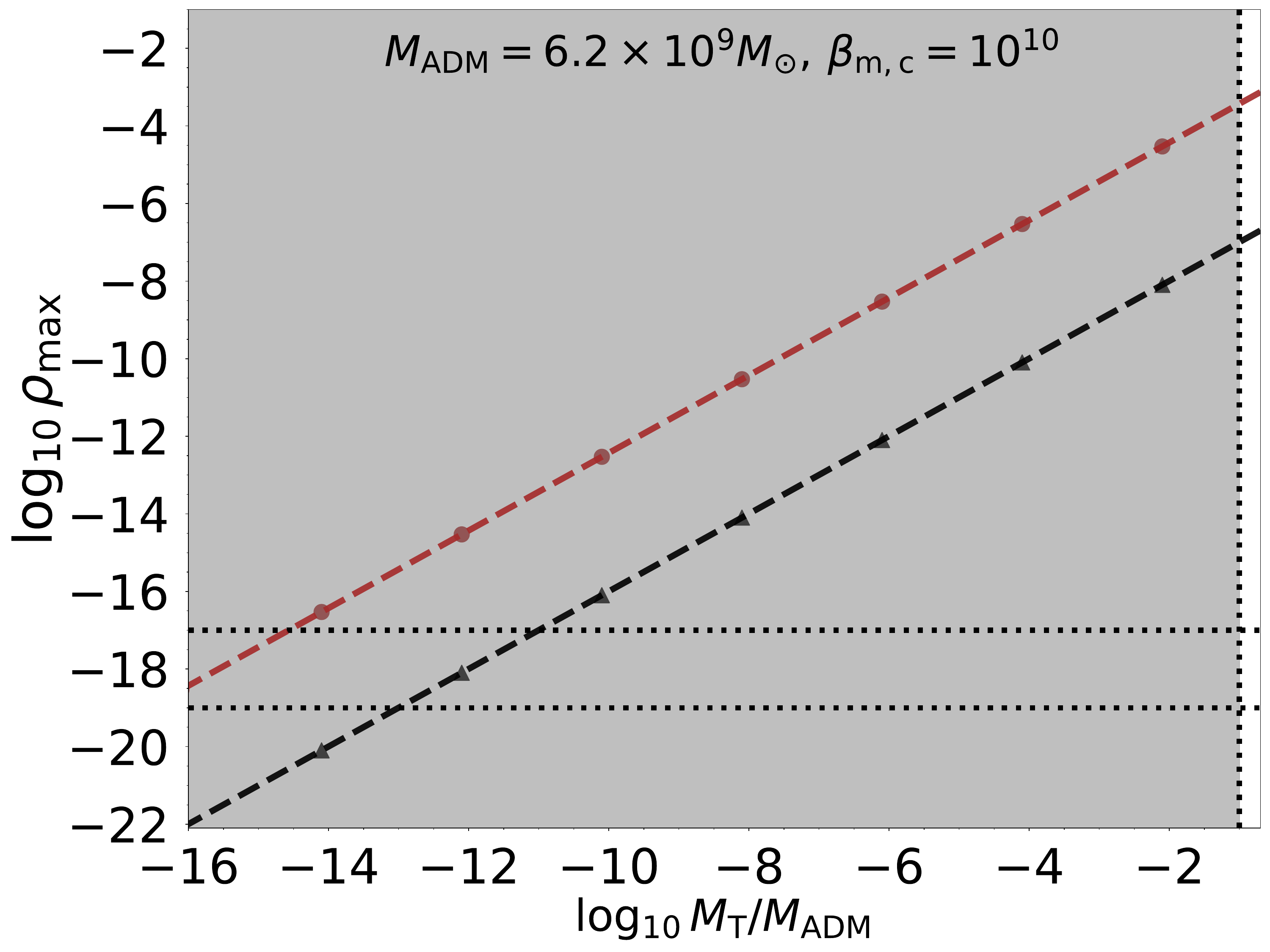}
\hspace{0.0cm}
\includegraphics[scale=0.17]{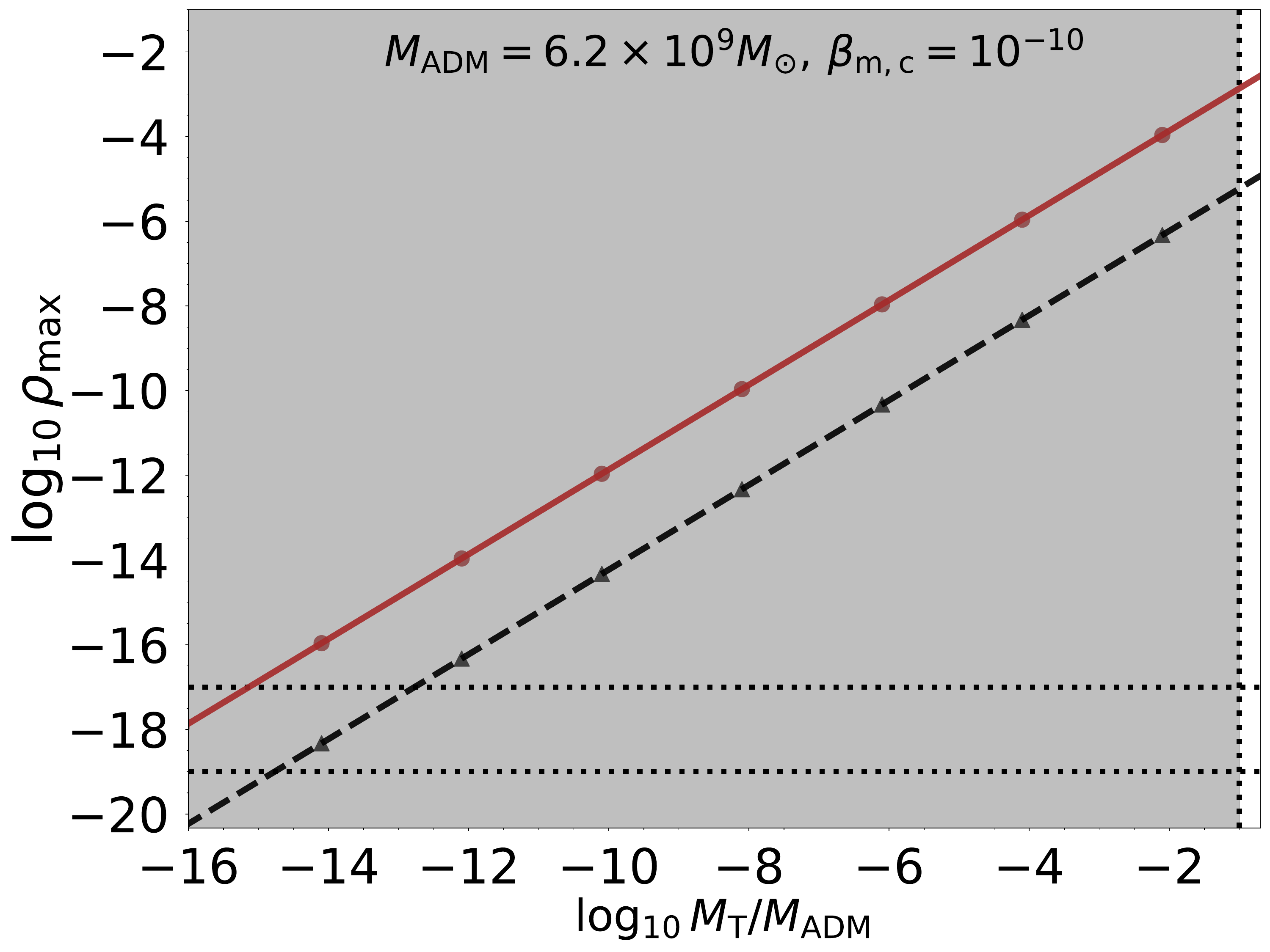}
\caption{Dependence of the logarithm of the maximum of the rest-mass density $\rho_{\mathrm{max}}$ in cgs units as a function of the logarithm of the mass of the accretion disk in units of the ADM mass of the spacetime $M_{\mathrm{T}}/M_{\mathrm{ADM}}$. The top panels show cases with $M_{\mathrm{ADM}} = 5 M_{\odot}$ and for the bottom panels $M_{\mathrm{ADM}} = 6.2 \times 10^{9} M_{\odot}$. The left column shows nonmagnetized disks ($\beta_{\mathrm{m, c}} = 10^3$) and the right column shows highly magnetized disks ($\beta_{\mathrm{m, c}}= 10^{-3}$). In the top panels, we show a subset of our parameter space. In particular we show, for our three KBHsSH spacetimes I, IV and VII (black, blue and brown lines, respectively), two values of the exponent $\alpha = 0, 0.75$ (solid and dashed lines respectively) and two ways of prescribing $l_0$, namely criteria 2 and 3 (circle and triangle markers, respectively). The shaded region of each plot shows the region where our results are physically acceptable. The vertical and horizontal black dotted lines represent $M_{\mathrm{T}}/M_{\mathrm{ADM}} = 0.1$ and  $\rho_{\mathrm{max}} = 10^{15} \mathrm{g} \mathrm{cm^{-3}}$ respectively.}
\label{mass_density_plots}
\end{figure*}

\section{Results}
\label{results}

Taking into account the different parameters that characterize our problem setup, we build a total of 108 models of thick disks around KBHsSH, 36 for each of the three hairy BH spacetimes we consider. The main thermodynamical and geometrical characteristics of the models are reported in tables \ref{results_I}, \ref{results_IV} and \ref{results_VII}, for spacetimes I, IV, and VII, respectively.  For all models the various physical quantities listed in the tables follow  qualitatively similar trends when compared to the results in {Paper I} for constant angular-momentum tori. In particular, for increasing magnetization the maximum of the rest-mass density increases, and the location of the fluid and magnetic pressure maximum shifts towards the black hole. This is accompanied by a global reduction in the size of the disks, only visible for the finite-size disks corresponding to the $l_0,1$ and $l_0,2$ cases. Moreover, since the value of $\Delta W_{\mathrm{c}}$ is the same for the three values of $l_0$, for each spacetime and value of the exponent $\alpha$   the maximum specific enthalpy $h_{\mathrm{max}}$ and the fluid pressure maximum $p_{\mathrm{max}}$ are equal when the disk is unmagnetized ($\beta_{\mathrm{m, c}} = 10^{10}$). When the magnetization increases, $h_{\mathrm{max}}\rightarrow 1$ but at a different rate for each value of $l_0$ (although the differences can be very small depending on the spacetime and the value of $\alpha$). We also observe that increasing the magnetization also increases the value of $\rho_{\mathrm{max}}$ in a different way depending on the value of $l_0$. We conclude that the specific value of $\rho_{\mathrm{max}}$ achieved when $\beta_{\mathrm{m, c}} \rightarrow 0$ does not depend on the value of $\Delta W_{\mathrm{c}}$, but depends only on the disk model and on the spacetime. Quantitative features and differences between the models are discussed below. 

\subsection{Morphology of the disks}

The morphology of the disks in the $(R\sin\theta,R\cos\theta)$ plane is shown in figures \ref{Model_I_2Dplots}, \ref{Model_IV_2Dplots}, and \ref{Model_VII_2Dplots}. Respectively, they correspond to spacetimes I, IV, and VII. These figures depict the distribution of the rest-mass density $\rho$ for our three values of the central magnetization parameter $\beta_{\rm m,c}$ (one per column) and for our three values of the constant part of the specific angular momentum $l_0$ (one per row). In all three figures the exponent $\alpha$ of the specific angular momentum law is fixed to $\alpha=0.5$, as an illustrative example. The morphological trends observed in this case also apply to the other values of $\alpha$ we scanned. Specific information about the radial size of the disks for all $\alpha$ values are reported in the tables.

The figures reveal that the size of the disks is similar for the two cases $l_{0,1}$ and $l_{0,2}$ and it is remarkably different for $l_{0,3}$. In the latter, disks are significantly larger (in fact the outermost isodensity contour closes at infinity, as shown in the value of $R_{\rm out}$ in the tables). This trend applies to all values of the magnetization parameter, to all three spacetimes, and to all values of $\alpha$, as can be determined from the tables. The fact that the morphological differences for $l_{0,1}$ and $l_{0,2}$ are minor is related to the fact that the angular momentum profiles along the equatorial plane are fairly similar for those two cases (as shown in Fig.~\ref{radial_angmom_potential}). Actually, the  $l_{0,2}$ models resemble a slightly smaller version of the $l_{0,1}$ disks, attaining larger values of $\rho_{\mathrm{max}}$ and $p_{\mathrm{max}}$ when the magnetization starts becoming relevant.

As in the constant angular momentum models of Paper I the 
location  of  the  centre  of  the  disk moves  closer  to the black hole as the magnetization increases, and the upper values of the isodensity contours also become larger. Moreover, the inner radius of the disks also shift closer to the black hole for $l_{0,1}$ and $l_{0,2}$ than for $l_{0,3}$. Both of these trends are observed for all three spacetimes. Specific values of those radii are reported in the tables.

Fig.~\ref{gravitational_energy_density_plots} shows the gravitational energy density, Eqs.~(\ref{eq:torus_energy_density}) and (20), for both the fluid matter (top half of each panel) and for the scalar field (bottom half). We compare the distribution of the energy density in the three KBHsSH spacetimes for the particular case $\alpha = 0.75$, $\beta_{\mathrm{m, c}} = 10^{-10}$ as an illustrative example. The top panels correspond to $l_0 = l_{0, 2}$ and the bottom panels to $l_0 = l_{0, 3}$. We note that, in general, the location of the area where the maximum values for the energy density for the fluid and for the scalar field are attained do not coincide. The most striking morphological difference appears in spacetime VII where, in some cases, a second maximum in the gravitational energy density distribution of the fluid appears (see top-right plot of Fig.~\ref{gravitational_energy_density_plots}). The region of the parameter space in which this situation occurs is discussed below. 

In Fig.~\ref{radial_log_density_plots} we plot the radial profiles of the rest-mass density at the equatorial plane in double logarithmic scale. Models built for spacetimes I and IV (top and central rows) show very similar qualitative profiles in all cases (i.e.~different values of $\alpha$ and $\beta_{{\rm m,c}}$). Those profiles are also fairly similar to those expected of a constant angular momentum torus around a Kerr black hole (see~\cite{paper1}). Again, the most prominent differences are apparent for spacetime VII shown in the bottom row: for spacetime I and IV, the rest-mass density maximum is close to the inner edge of the disk while for spacetime VII and for unmagnetized disks (solid lines), this maximum is significantly further away. This is related to the fact that most of the mass and angular momentum of spacetime VII are stored in the scalar field. Moreover, compared to spacetimes I and IV, for spacetime VII the location of the density maximum for models $l_0 = l_{0, 2}$ and $l_0 = l_{0, 3}$ in the
unmagnetized case ($\beta_{\mathrm{m, c}} = 10^{10}$) are very close to each other (see solid black and blue curves).

Focusing on the unmagnetized case, the bottom row of the figure shows that the rest-mass density is higher in the region where the hair has most of its gravitational energy density ($\log_{10} R\sim 0.1$; see vertical line), irrespective of $l_0$. The central and right panel reveal an interesting effect. Compared to the left panel, the profile of the $l_0 = l_{0, 2}$ case in the central panel is similar but that of the $l_0 = l_{0, 3}$ case develops a low-density inner region (notice the change in slope in the blue solid curves).
When the magnetization increases the maximum of the distribution shifts towards the black hole (central panel, blue dotted curve) but the profile flattens and a significant fraction of the mass is left around $R_{\mathrm{c}}$ (signalled by the maximum of the solid lines).

If we now focus on the right panel, we observe that what we have just discussed  for the $l_{0, 3}$ case for $\alpha = 0.5$ (i.e.~the flattening of the profile) occurs for the $l_{0, 2}$ case in the $\alpha = 0.75$ case. The flattening in the distribution implies the appearance of a second maximum in the gravitational energy density of the torus, $\rho_{\rm T}$, which is roughly located in the same region where the $\rho_{\rm SF}$ maximum is attained (see top-right panel in
 Fig.~\ref{gravitational_energy_density_plots}). Correspondingly, for the $l_{0, 3}$ case (blue dashed and dotted curves) we see that the location of $\rho_{\mathrm{max}}$ does not move all the way down to the inner edge of the disk and a low density region is left even in the highly magnetized case.

We note that this trend is expected to happen also for the $l_0 = l_{0, 2}$ (black curves) if we increase the value of $\alpha$. We have tested this by building models with  $\alpha = 0.8$. It seems that large enough values of $\alpha$ the gravitational well of the hair can act as a barrier preventing the maximum of the rest-mass density to reach the inner edge of the disk. This effect seems to appear first (i.e.~for smaller values of $\alpha$) for $l_{0, 3}$, then for $l_{0, 1}$ (radial profiles not shown in Fig.~\ref{radial_log_density_plots}) and lastly for $l_{0, 2}$.

We close this section by noting that some models for spacetime VII bear some morphological resemblance with the findings of~\cite{Dyba:2020, Dyba:2021} for self-gravitating massive tori. In particular, this similarity is found for nonmagnetized disks and when the maximum of the rest-mass density is close to the maximum of gravitational energy density of the field. Some of their massive models also present a second ergoregion as in our spacetime VII (see~\citep{Herdeiro:2015b}) but due to the self-gravity of the disk. This resemblance can be explained by the fact that in our case, the scalar field distribution can mimic the self-gravity of the disk.


\subsection{Effects of the magnetization}

In Fig.~\ref{rmmax_rhomax_hmax_logbeta_fig} we discuss the effects of the magnetization on the disk properties, for a subset of the models reported in Tables~\ref{results_I}, \ref{results_IV} and \ref{results_VII}.
The top row of Fig.~\ref{rmmax_rhomax_hmax_logbeta_fig} shows the deviation in the location of the maximum of the magnetic pressure $R_{\mathrm{m,c}}$ (reached at the equatorial plane) with respect the location of the center of the disk $R_{\mathrm{c}}$. This is a relevant quantity to analyze because our previous results in~\cite{Gimeno-Soler:2017} and {Paper I} showed that $R_{\mathrm{m,c}} > R_{\mathrm{c}}$ for weakly magnetized disks and $R_{\mathrm{m,c}} < R_{\mathrm{c}}$ for strongly magnetized disks. The exact value of $\beta_{\mathrm{m, c}}$ for which $R_{\mathrm{m,c}} = R_{\mathrm{c}}$ is related to the exponent of the EoS $\Gamma$ and to the value of the potential gap at the center of the disk $\Delta W_{\mathrm{c}}$ (or to the maximum value of the specific enthalpy $h_{\mathrm{max}}$ when $\beta_{\mathrm{m, c}} \rightarrow \infty$). In particular, in reference~\cite{Gimeno-Soler:2017} it was shown that, if $\Delta W_{\mathrm{c}}$ is sufficiently small, then $h \rightarrow 1$ and the value of the magnetization parameter such that $R_{\mathrm{m,c}} = R_{\mathrm{c}}$ is $\beta_{\mathrm{m, c}} = 1/(\Gamma - 1)$.
In the rightmost part of the top panels (which correspond to cases increasingly less magnetized) we observe that most models can be ordered by their value of $(R_{\mathrm{m,c}} - R_{\mathrm{c}}/R_{\mathrm{c}}$ irrespective of $\alpha$, the greatest deviation being observed for spacetime IV and $l_0 = l_{0, 2}$ (blue solid curve) and the smallest for spacetime VII and $l_0 = l_{0, 3}$ (red dashed curve). The only exception to this trend is spacetime VII for $l_0 = l_{0, 2}$ where the value of $(R_{\mathrm{m,c}} - R_{\mathrm{c}}/R_{\mathrm{c}}$ goes from the second highest for $\alpha = 0$ (left column) to the smallest for $\alpha = 0.75$ (right column). 
In the inset of all three plots in the top row we display the region around $R_{\mathrm{m,c}} = R_{\mathrm{c}}$ and $\beta_{\mathrm{m, c}} = 3$. In particular, we find that, as expected, models with a smaller value of $\Delta W_{\mathrm{c}}$ pass closer to the point $(\log_10 3, 0)$. This can be seen both for each spacetime with constant $\alpha$ and for each model when changing the value of $\alpha$. Moreover, we also observe that in general, the models with $l_0 = l_{0, 3}$ pass closer to the point $(\log_{10} 3, 0)$ when compared to their counterparts with $l_0 = l_{0, 3}$, with the exceptions of model I for $\alpha = 0$ where they almost coincide (see the black curves in the top left panel) and of model I for $\alpha = 0.75$, where this behaviour is inverted (top right panel).

On the other hand, in the leftmost part of each plot in the top row (which correspond to highly magnetized cases) we find that for $\alpha = 0$ and $0.75$ (left and central panels), the value of $(R_{\mathrm{m,c}} - R_{\mathrm{c}}/R_{\mathrm{c}}$ also provides a neat ordering of the models, from spacetime VII, $l_0 = l_{0, 2}$ (red solid line) with the highest deviation, to spacetime I, $l_0 = l_{0, 2}$ (black solid curve) with the smallest. However, for $\alpha = 0.75$ (top left panel) we find that spacetime IV, $l_{0, 3}$ now has a slightly larger deviation than spacetime I, $l_{0, 3}$, and for spacetime VII it is found that $l = l_{0, 3}$ has the smaller deviation and that for $l = l_{0, 2}$ the behavior of $(R_{\mathrm{m,c}} - R_{\mathrm{c}}/R_{\mathrm{c}}$ with respect $\log_{10} \beta_{\mathrm{m, c}}$ is abnormal when compared to the other models. The discrepancies observed for the highly magnetized models for spacetime VII can be related to the peculiarities we discussed in the radial profiles in Fig.~\ref{radial_log_density_plots}.

In the central row of Fig.~\ref{rmmax_rhomax_hmax_logbeta_fig} we show the dependence of the maximum of the specific enthalpy $h_{\mathrm{max}}$ on the magnetization of the disk. For each model the value of $h_{\mathrm{max}}$ goes from $e^{\Delta W_{\mathrm{c}}}$ for unmagnetized disks ($\beta_{\mathrm{m, c}} \rightarrow \infty$) to $1$ for extremely magnetized disks ($\beta_{\mathrm{m, c}} \rightarrow 0$) (for a discussion on this topic see, for instance~\citep{paper1,Cruz2020}). It is apparent that, as expected, the value of $h_{\mathrm{max}}$ is consistently higher for models with a higher value of $\Delta W_{\mathrm{c}}$ up to small values of $\beta_{\mathrm{m, c}}$. Moreover, we can see that models with $l_0 = l_{0, 2}$ (solid curves) have slightly higher values of $h_{\mathrm{max}}$ for values of $\log_{10}\beta_{\mathrm{m, c}}$ between $0$ and $-2$. This difference could be related to the fact that, even though the value of $\Delta W_{\mathrm{c}}$ is the same for both $l_{0, 2}$ and $l_{0, 3}$ models, the gravitational potential distribution that they feel is quite different (see Fig.~\ref{radial_angmom_potential}).

Finally, the bottom row of Fig.~\ref{rmmax_rhomax_hmax_logbeta_fig} depicts the dependence of the maximum of the rest-mass density $\rho_{\mathrm{max}}$ on the magnetization of the disk. The observed behaviour is related, when the magnetization begins to be relevant in the disk, to two factors, namely, the shift of the maximum of the rest-mass density with respect the center of the disk and the radial extent of the high density region of the disk. We find that, in general, tori with a value of $R_{\mathrm{in}}$ ($l_{0, 2}$) closer to the horizon of the black hole exhibit larger values of $\rho_{\mathrm{max}}$. This is related to the fact that this kind of disks tend to have $\rho_{\mathrm{max}}$ closer to the inner edge of the disk and lesser radial extent of the high density region, as it can be noticed from the central rows of Figs.\ref{Model_I_2Dplots}, \ref{Model_IV_2Dplots} and \ref{Model_VII_2Dplots}. When comparing between KBHsSH spacetimes, we find that larger values of $\rho_{\mathrm{max}}$ are attained for larger values of $\Delta W_{\mathrm{c}}$. However, there are particular models that do not obey this trend. In the case of spacetime I, (black lines),  the solid and dashed lines are almost coincident for $\alpha = 0$ and $\alpha = 0.5$ (left and central panels), and for $\alpha = 0.75$ (right panel) the value of $\rho_{\mathrm{max}}$ is larger for $l_{0, 3}$. This can be explained taking into account that the size of the high-density region for spacetime I, $l_0 = l_{0, 2}$ does not change much for increasing magnetization.  Other cases that behave in a different way are spacetime VII, $l_0 = l_{0, 2}$ (red solid curve) for $\alpha = 0.75$ and $l_0 = l_{0, 3}$ (red dashed curve) for $\alpha = 0.5$. In the first case,  the value of $\rho_{\mathrm{max}}$ is the smallest for $\beta_{\mathrm{m, c}} = 0$. However, when the magnetization increases, $\rho_{\mathrm{max}}$ grows faster and it ends up reaching the second highest value for $\beta_{\mathrm{m, c}} = 10^{3}$. A similar effect (but in a smaller scale) can be observed for spacetime VII, $l_0 = l_{0, 3}$, $\alpha = 0.5$. In this case the effect is due to the flattening of the rest-mass density distribution that we described in the preceding section, where a significant fraction of the mass is left around $R_{\mathrm{c}}$, thus reducing the value of $\rho_{\mathrm{max}}/\rho_{\mathrm{c}}$ (i.e.~ $\rho_{\mathrm{max}}$ as $\rho_{\mathrm{c}} = 1$ by construction). It is also worth  remarking that with the exception of spacetime VII, if we fix the spacetime and $l_0$, the value of $\rho_{\mathrm{max}}$ in the extremely magnetized case is larger for increasing $\alpha$, in agreement with what was found for purely Kerr black holes in~\cite{Gimeno-Soler:2017}.

\subsection{Astrophysical implications} 

We turn next to discuss possible astrophysical implications of our models. To do so we compute the maximum value of the rest-mass density and of the mass of the disk in physical units. To this end, we recall (see {Paper I}) that the density in cgs units is related to the density in geometrized units by
\begin{equation}\label{eq:cgs_density}
\rho_{\mathrm{cgs}} = 6.17714 \times 10^{17} \left(\frac{G}{c^2}\right)\left(\frac{M_{\odot}}{M}\right)^{2}\rho_{\mathrm{geo}}\,.
\end{equation}
We can rewrite this equation in a more convenient way making the following considerations: The ADM mass of the spacetime is expressed in solar-mass units $M_{\mathrm{ADM}} = n M_{\odot}$. The mass of the accretion disk is expressed as a fraction of the ADM mass $M_{\mathrm{T}} = q M_{\mathrm{ADM}}$. 
Now, we define the function $\overline{\rho_{\mathrm{T}}}$ such that we can rewrite Eq.~\eqref{eq:mass_integral} as
\begin{equation}\label{eq:mass_density_factor}
M_{\mathrm{T}} = \rho_{\mathrm{max}}\int \overline{\rho_{\mathrm{T}}} \, \sqrt{-g}\, \mathrm{d}  x^3\,,
\end{equation}
where $\rho_{\mathrm{max}}$ is the maximum value of the rest-mass density in the disk. It is relevant to note that, as $\rho_{\mathrm{T}}$ in Eq.~\eqref{eq:torus_energy_density} does not depend linearly in $\rho$, some dependence on $\rho_{\mathrm{max}}$ is left in $\overline{\rho_{\mathrm{T}}}$, but the contribution of the nonlinear terms is very small for all the cases we are considering (the deviation between the exact formula and Eq.~\eqref{eq:mass_density_factor} is $< 10^{-6}$ for all our cases). Then, one can see that the ratio $M_{\mathrm{T}}/\rho_{\mathrm{max}}$ is constant if all the parameters but $\rho_{\mathrm{c}}$ are kept constant.
This fact allows us to write the value of the maximum rest-mass density in geometrized units for an accretion torus of mass $M_{\mathrm{T}}$ as
\begin{equation}\label{eq:rho_geo_max}
\rho_{\mathrm{geo}}^{\mathrm{max}} = \frac{M_{\mathrm{T}}\rho_{\mathrm{max}}}{M_{\mathrm{T}}(\rho_{\mathrm{c}}=1)} =  \frac{q M_{\mathrm{ADM}}^{\mathrm{geo}}\rho_{\mathrm{max}}}{M_{\mathrm{T} }(\rho_{\mathrm{c}}=1)} \,,
\end{equation}
where we have used that $M_{\mathrm{T}} = q M_{\mathrm{ADM}}$, $M_{\mathrm{T}}(\rho_{\mathrm{c}}=1)$ and $\rho_{\mathrm{max}}$ are the mass and the maximum rest-mass of the torus when $\rho_{\mathrm{c}} = 1$ and $M_{\mathrm{ADM}}^{\mathrm{geo}}$ is the ADM mass of the spacetime in geometrized units (i.e.~the ADM mass as is reported in Table~\ref{models_list}).
Now, we can rewrite Eq.~\eqref{eq:cgs_density} as
\begin{equation}\label{eq:cgs_max_density}
\rho_{\mathrm{cgs}}^{\mathrm{max}} = 6.17714 \times 10^{17} \left(\frac{1}{n}\right)^{2}\frac{q M_{\mathrm{ADM}}^{\mathrm{geo}}\rho_{\mathrm{max}}}{M_{\mathrm{T} }(\rho_{\mathrm{c}}=1)}\,.
\end{equation}
This equation allows us to compute the maximum value of the rest-mass density in cgs units in terms of the disk mass fraction $q$ provided that we know $n$, $M_{\mathrm{ADM}}^{\mathrm{geo}}$ (parameters of the model),  $\rho_{\mathrm{max}}$, and $M_{\mathrm{T} }(\rho_{\mathrm{c}}=1)$ (results of our computations). 

Figure \ref{mass_density_plots} depicts double logarithmic plots of Eq.~\eqref{eq:cgs_max_density}  showing the relation between the maximum value of the rest-mass density and $M_{\mathrm{T}}/M_{\mathrm{ADM}}$ for a subset of our parameter space and two different ADM masses for each KBHsSH spacetime. One is in the stellar mass regime ($M_{\mathrm{ADM}} = 5 M_{\odot}$; top panels) and another one is in the supermassive range ($M_{\mathrm{ADM}} = 6.2 \times 10^{9}M_{\odot}$, i.e.~the mass of the central black hole in M87; bottom panels). In the top panels we explore the limits of both our disk models and our approach to build them. The shaded region corresponds to the physically admissible solution space, and it is bounded by a horizontal line that represents unrealistically dense solutions ($\rho_{\mathrm{max}} = 10^{15} \mathrm{g} \mathrm{cm^{-3}}$) and by a vertical line that represents the point when the test fluid approximation for the disk begins to break down ($M_{\mathrm{T}} = 0.1M_{\mathrm{ADM}}$) and our approach becomes unsuitable to construct accretion tori. These top panels show interesting properties of our models that we should highlight here:
First, it can be seen that, irrespective of the value of the magnetization parameter at the center $\beta_{\mathrm{m, c}}$, for a given value of $q = M_{\mathrm{T}}/M_{\mathrm{ADM}}$, the models with $l_0 = l_{0, 3}$ (triangle markers) have smaller values of $\rho_{\mathrm{max}}$ when compared to the models with $l_0 = l_{0, 2}$ (circle markers). This is due to the fact that the models that are constructed following criterion 3 are significantly more radially extended than the ones built using criterion 2. It can be seen as well that increasing the magnetization parameter increases the value of $\rho_{\mathrm{max}}$ for constant $q$. This is caused by the change of morphology of the disk (the higher rest-mass density region moves towards the black hole and then its volume decreases), but this effect does not change the value of $\rho_{\mathrm{max}}$ in the same way for all the models. In particular, we observe that, in general, models with $\alpha = 0.75$ suffer a greater increase of  $\rho_{\mathrm{max}}$. In some cases, the difference is very small (e.g.~ Models I and IV for $l_0 = l_{0, 2}$) but it can also be considerably large (e.g.~Model VII for $l_0 = l_{0, 3}$). This is due to the fact that models with $\alpha = 0.75$ have a greater value of $R_{\mathrm{in}}$ than their counterparts with $\alpha = 0$, and then, the decrease of volume of the high rest-mass density region is bigger. However, the difference in magnitude of these changes are caused by the particular features of each spacetime. In particular, Model VII for $l_0 = l_{0, 2}$ and $\alpha = 0.75$ is the only case that deviates from the behavior described above. The reason for this deviation is the presence of a second maximum of the gravitational energy density $\rho_{\mathrm{T}}$ (see top right panel of Fig.~\ref{gravitational_energy_density_plots}). This second maximum suppresses the increase of $\rho_{\mathrm{T}}$ that would be present due to the high rest-mass density region moving toward the black hole. We also note that the most dense models should affect the hair distribution. In particular in cases I and IV, where less mass and angular momentum are stored in the field.

We conclude that, for a stellar-mass black hole, the values spanned by $\rho_{\mathrm{max}}$ are consistent with the maximum densities found in disks firmed in numerical-relativity simulations of binary neutron star mergers (see~\cite{Rezzolla:2010,Rezzolla:2017, Most:2021}). This result, which had already been found in the constant angular momentum models of {Paper I}, is corroborated when using the improved angular momentum distributions analyzed in the present work.

\begin{table}[t]
\caption{Mass parameter of the scalar field $\mu$ in eV for our three KBHsSH models if the ADM mass of each spacetime is $M_{\mathrm{ADM}}=5 M_{\odot}$ (first row) and $M_{\mathrm{ADM}}=6.2 \times 10^{9} M_{\odot}$ (i.e.~the mass of the central black hole in M87; second row).}        
\label{scalar_field_mass_tab}
\centering          
\begin{tabular}{c c c c}
\hline\hline
 & Model I & Model IV & Model VII \\
 \hline
\multicolumn{4}{c}{$M_{\mathrm{ADM}} = 5 M_{\odot}$}\\
\hline
 $\mu[eV]$ & $8.33 \times 10^{-11}$ & $1.87 \times 10^{-11}$ & $1.95 \times 10^{-11}$\\
\hline\hline
\multicolumn{4}{c}{$M_{\mathrm{ADM}} = 6.2 \times 10^{9} M_{\odot}$}\\
\hline
 $\mu[eV]$ & $6.69 \times 10^{-21}$ & $1.50 \times 10^{-20}$ & $1.57 \times 10^{-20}$\\
\hline\hline
\end{tabular}
\end{table}

In the bottom panels of Fig.~\ref{mass_density_plots} we consider the case of a supermassive black hole and only show the two lines (the top line and the bottom line for each case) that bound the parameter space spanned by our results. We also expand the horizontal axis to take into account the extremely low rest-mass densities (between $\sim 10^{-17}$ and $\sim 10^{-19} \mathrm{g} \, \mathrm{cm^{-3}}$) in the disk inferred by matching the results of general relativistic magneto-hydrodynamic (GRMHD) simulations with observations (see~\cite{EHT8} and also \cite{Chael:2019}). As we can see in this figure these values of $\rho_{\mathrm{max}}$ correspond to extremely low values of $q$ (from $< 10^{-11}$ for $\beta_{\mathrm{m,c}} = 10^{10}$ to $\lesssim 10^{-13}$ for $\beta_{\mathrm{m,c}} = 10^{-10}$). However, it is important to note that the disks in the aforementioned references are not stationary solutions (unlike ours) but are evolved dynamically instead, which means that they are subject to various processes that cause matter redistribution, angular momentum transport and magnetic field amplification -for low magnetized disks- or suppression -for strongly magnetized disks. (Some instances of these processes can be seen in~\cite{Cruz2020}.). These effects would change the value of the integral Eq.~(\ref{eq:mass_density_factor}) in a non trivial way, as the exact form the evolution affects the disk can depend on the characteristics of the spacetime.

It is also relevant to recall the formula that relates the maximum ADM mass of the KBHsSH with the mass parameter of the scalar field $\mu$  (see~\cite{Herdeiro:2015a} and references therein),
\begin{equation}\label{eq:max_mass_HBH}
M^{\rm max}_{\rm ADM} \simeq \alpha_{\mathrm{BS}} 10^{-19} M_{\odot} \left(\frac{\mathrm{GeV}}{\mu}\right)\,,
\end{equation}
with $\alpha_{\mathrm{BS}} = 1.315$ (corresponding to a value of the azimuthal harmonic index $m = 1$). Using the previous definitions, we can rewrite this formula as
\begin{equation}\label{eq:mu_eq}
\mu [\mathrm{eV}] = 10^{-10} \frac{M_{\mathrm{ADM}}^{\mathrm{geo}}}{n}\,.
\end{equation}
The values of $\mu$ for the two astrophysical scenarios we have considered in this section are reported in Table~\ref{scalar_field_mass_tab}. These values of $\mu$ are within the mass range suggested by the {\it axiverse} of string theory (see~\cite{Arvanitaki:2010}) portraying a large number of scalar fields in a mass range from $10^{-33}$ eV to $10^{-10}$ eV.

\section{Conclusions}
\label{conclusions}

Recent observational data from the LIGO-Virgo-KAGRA Collaboration and from the EHT Collaboration is allowing to probe the black hole hypothesis -- black holes apparently populate the Cosmos in large numbers and are regarded as the canonical dark compact objects. 
While this  hypothesis is thus far supported by current data, the ongoing efforts also place within observational reach the exploration of additional proposals for alternative, and {\it exotic}, compact objects. Indeed, possible model degeneracies have been already pointed out in~\cite{juan1,imitation}. In this paper we have considered a particular class of ECOs, namely Kerr black holes with synchronised hair resulting from minimally coupling Einstein’s gravity to bosonic matter fields~\cite{Herdeiro:2014a,Herdeiro:2016}. Such hairy black holes provide a counterexample to the no-hair conjecture and they have been shown to form dynamically (in the vector case) as the end-product of the superradiant instability~\cite{East:2017} (but see also~\cite{sanchis-gual:2020} for an alternative formation channel through the post-merger dynamics of bosonic star binaries) and to be effectively stable themselves against superradiance~\cite{Degollado:2018}.
In this work we have presented new equilibrium solutions of stationary models of magnetized thick disks (or tori) around Kerr black holes with synchronised scalar hair. The models reported are based on ideas put forward in our previous work~\cite{paper1} which focused on models following a constant radial distribution of the specific angular momentum along the equatorial plane. The models reported in the present paper, however, greatly extend those of~\cite{paper1} by accounting for fairly general and astrophysically motivated  distributions of the specific angular momentum. In particular, we have  
introduced a new way to prescribe the distribution of the disk's angular momentum based on a combination of two previous proposals discussed in~\cite{Daigne:2004} and \cite{Qian:2009}. 
Due to the intrinsic higher complexity of the new models, the methodology employed for their construction is markedly different to that employed in~\cite{paper1}. Following~\cite{Daigne:2004}, our approach has been based on the use of the so-called von Zeipel cylinders as a suitable (and computationally efficient) means to compute the angular momentum distribution outside the equatorial plane. Within this framework, we have chosen a fairly large parameter space (amounting to a total of 108 models) that has  allowed us to directly compare among different spacetimes with the same choice of specific angular momentum distribution, and to compare between different rotation profiles in the same spacetime.

While our models show some similarities to the constant angular momentum disks of {Paper I} (which we recover here as a particular limiting case of our improved distributions) important morphological differences also arise. We have found that, due to the scalar hair effect on the spacetime, the disk morphology and physical properties can be quite different than expected if the spacetime was purely Kerr. This has been revealed quite dramatically for KBHsSH spacetime VII which most deviates from the Kerr spacetime (as most of the mass and angular momentum of this spacetime is actually stored in the scalar field). Some of the tori built within this spacetime exhibit the appearance of a secondary maximum in the gravitational energy density with implications in the radial profile distributions of the thermodynamical quantities of the disks. We have also discussed possible astrophysical implications of our models, computing the maximum value of the rest-mass density and of the mass of the disk in physical units for the case of a stellar-mass black hole and a supermassive black hole. Comparisons with the results from mergers of compact binaries and GRMHD simulations performed by the EHT collaboration yield values compatible with our numbers, again pointing out possible model degeneracies. Finally, our study has also allowed us to provide estimates for the mass of the bosonic particle.

The two-parameter specific angular momentum prescription we have discussed here could be particularly useful for further studies, possibly including time-dependent evolutions, as it allows to build disks with different morphological features (different degrees of thickness and radial extent of the disk). Our models could be used as initial data for numerical evolutions of GRMHD codes to study their dynamics and  stability properties. In addition, perhaps most importantly, these disks could be used as illuminating sources to build shadows of Kerr black holes with scalar hair which might further constrain the no-hair hypothesis as new observational data is collected,~{following up on~\cite{Cunha:2015yba,Cunha:2016bjh,Cunha:2019ikd}}. Those aspects are left for future research and will be presented elsewhere.


\section*{Acknowledgements}
We thank Alejandro Cruz-Osorio for useful comments.
This work has been supported by the Spanish Agencia Estatal de Investigaci\'on (Grant No. PGC2018-095984-B-I00), by the Generalitat Valenciana (Grant No. PROMETEO/2019/071), by the {Center for Research and Development in Mathematics and Applications (CIDMA) through the Portuguese Foundation for Science and Technology (FCT - Funda\c c\~ao para a Ci\^encia e a Tecnologia), references UIDB/04106/2020 and UIDP/04106/2020 and by national funds (OE), through FCT, I.P., in the scope of the framework contract foreseen in the numbers 4, 5 and 6 of the article 23, of the Decree-Law 57/2016, of August 29, changed by Law 57/2017, of July 19. We acknowledge support from the projects PTDC/FIS-OUT/28407/2017,  CERN/FIS-PAR/0027/2019 and PTDC/FIS-AST/3041/2020}, and by the European Union's Horizon 2020 Research and Innovation (RISE) programme H2020-MSCA-RISE-2017  Grant  No. FunFiCO-777740.
The authors also acknowledge networking support by the COST Action CA16104.

\bibliographystyle{apsrev4-1}
\bibliography{references}

\end{document}